\documentclass[opre,nonblindrev]{informs3_arXiv}

\newif\ifarXiv

\arXivtrue 

\newif\ifadditional
\additionalfalse 

\ifarXiv
\OneAndAHalfSpacedXI
\newcommand{\eqspace}{}
\def \spacingalign {-0.3em}

\def \spacingtable {1.5}
\def \spacingfigure {-0.4cm}
\def \spacingcomput {1.3}
\def \spacinginfluence {1.3}

\else
\DoubleSpacedXI
\newcommand{\eqspace}{\setlength{\abovedisplayskip}{8pt} \setlength{\belowdisplayskip}{8pt}}
\def \spacingalign {-0.6em}

\def \spacingtable {1}
\def \spacingfigure {-1cm}
\def \spacingcomput {0.7}
\def \spacinginfluence {0.7}
\fi

\usepackage{endnotes}
\let\footnote=\endnote
\let\enotesize=\normalsize
\def\notesname{Endnotes}%
\def\makeenmark{$^{\theenmark}$}
\def\enoteformat{\rightskip0pt\leftskip0pt\parindent=1.275em
  \leavevmode\llap{\makeenmark}}

\usepackage[colorlinks=true,breaklinks=true,bookmarks=true,urlcolor=blue,
     citecolor=blue,linkcolor=blue,bookmarksopen=false,draft=false]{hyperref}

\usepackage{stmaryrd}
\usepackage[dvipsnames]{xcolor}
\usepackage{tikz}
\usepackage{dsfont}
\usetikzlibrary{arrows,positioning}
\usepackage{caption}
\usepackage{subcaption}
\usepackage{xspace}
\usepackage{multirow}
\usepackage{rotating}
\usepackage{framed}
\usepackage{color, colortbl}
\usepackage{rotating}
\usepackage{multirow}
\usepackage{caption}
\definecolor{LRed}{rgb}{1,.8,.8}
\definecolor{MRed}{rgb}{1,.6,.6}
\definecolor{HRed}{rgb}{1,.2,.2}
\definecolor{LYellow}{rgb}{1,1,0.7}

\usepackage{xspace}
\usepackage{tcolorbox}
\usepackage{kbordermatrix}
\usepackage{mathcomp}
\usepackage[inline]{trackchanges}
\usepackage{algorithm}
\usepackage{algpseudocode}
\usepackage{mathtools}

\newcolumntype{L}[1]{>{\raggedright\let\newline\\\arraybackslash\hspace{0pt}}m{#1}}
\newcolumntype{C}[1]{>{\centering\let\newline\\\arraybackslash\hspace{0pt}}m{#1}}
\newcolumntype{R}[1]{>{\raggedleft\let\newline\\\arraybackslash\hspace{0pt}}m{#1}}
\usepackage{hhline}

\renewcommand{\kbldelim}{(}
\renewcommand{\kbrdelim}{)}

\usepackage{pgfplots}

\usepackage{natbib}
 \bibpunct[, ]{(}{)}{,}{a}{}{,}%
 \def\bibfont{\small}%
 \def\bibsep{1.4pt}
 \def\bibhang{24pt}%
\TheoremsNumberedThrough     
\ECRepeatTheorems

\EquationsNumberedThrough    

\MANUSCRIPTNO{OPRE-2018-10-575} 

{\theoremstyle{THkey}}

\begin{document}

\newcommand{\opt}[1]{\alpha_{#1}}
\newcommand{\defender}{\textbf{P1}\xspace}
\newcommand{\attacker}{\textbf{P2}\xspace}
\newcommand{\aref}[1]{(\hyperref[#1]{\text{A}\ref{#1}})}
\newcommand{\lref}[1]{(\hyperref[#1]{\text{L}\ref{#1}})}
\newcommand{\supp}{\operatorname{supp}}
\newcommand{\Val}{\operatorname{F}}
\newcommand{\Costop}{\operatorname{C_2}}
\newcommand{\Costflow}{\operatorname{C_1}}
\newcommand{\flow}{x}
\newcommand{\Def}{S}
\newcommand{\att}{T}
\newcommand{\detect}{\operatorname{F}}
\newcommand{\detdef}{\widetilde{\operatorname{F}}}
\newcommand{\nprob}{\rho_{\sigma^1}(i)}
\newcommand{\eprob}{\rho_{\sigma^2}(e)}
\newcommand{\rdet}{r^N_{\Def}}
\newcommand{\rdetmixed}{r^N_{\sigma^1}}
\newcommand{\sdet}[1]{r({\sigma^{#1}})}
\newcommand{\ropt}{{r^N}^*}
\newcommand{\tr}[1]{\textcolor{red}{#1}}
\newcommand{\tb}[1]{\textcolor{blue}{#1}}
\newcommand{\tg}[1]{\textcolor{Green!75!black}{#1}}
\newcommand{\set}[1]{\mathcal{C}_{#1}}
\newcommand{\valdet}[2]{\operatorname{F}(#1,#2)}
\newcommand{\natt}[1]{|#1|}
\newcommand{\nash}{\Sigma}
\newcommand{\mm}{\text{(MSP)}\xspace}
\newcommand{\nodes}{\mathcal{V}}
\newcommand{\edges}{\mathcal{E}}
\newcommand{\smsc}{n^*}
\newcommand{\semm}{m^*}
\newcommand{\spm}{|\nodes|}
\newcommand{\msc}{\mathcal{S}}
\newcommand{\emm}{\mathcal{M}}
\newcommand{\bs}[1]{\boldsymbol{#1}}
\newcommand{\nbasis}[1]{\nodes_{#1}}
\newcommand{\ebasis}[1]{\edges_{#1}}
\newcommand{\lpo}{(\hyperlink{LPs}{$\text{LP}_1$})\xspace}
\newcommand{\lpot}{(\hyperlink{LP1tilde}{$\widetilde{\text{LP}_1}$})\xspace}
\newcommand{\lpom}{(\hyperlink{LPs}{\text{LP}_1})\xspace}
\newcommand{\lpt}{(\hyperlink{LPs}{$\text{LP}_2$})\xspace}
\newcommand{\lptt}{(\hyperlink{LP2tilde}{$\widetilde{\text{LP}_2}$})\xspace}
\newcommand{\lptm}{(\hyperlink{LPs}{\text{LP}_2})\xspace}
\newcommand{\lpob}{(\hyperlink{LPsb}{$\overline{\text{LP}_1}$})\xspace}
\newcommand{\lpobm}{(\hyperlink{LPsb}{\overline{\text{LP}_1}})\xspace}
\newcommand{\lptb}{(\hyperlink{LPsb}{$\overline{\text{LP}_2}$})\xspace}
\newcommand{\lptbm}{(\hyperlink{LPsb}{\overline{\text{LP}_2}})\xspace}
\newcommand{\lpop}{(\hyperlink{LPsb}{$\overline{\text{LP}_1}$})\xspace}
\newcommand{\lptp}{(\hyperlink{LPsb}{$\overline{\text{LP}_2}$})\xspace}
\newcommand{\rlpo}[1]{(\hyperlink{LP_S}{$\text{LP}_{#1}$})\xspace}
\newcommand{\rlpom}[1]{(\hyperlink{LP_S}{\text{LP}_{#1}})\xspace}
\newcommand{\rlpt}[1]{(\hyperlink{LP_S}{$\text{LP}_{#1}$})\xspace}
\newcommand{\rlptm}[1]{(\hyperlink{LP_S}{\text{LP}_{#1}})\xspace}
\newcommand{\pcg}{(\hyperlink{Pcg}{$M_{\text{CG}}$}$(\mathcal{I})$)\xspace}
\newcommand{\dcg}{(\hyperlink{Dcg}{$S_{\text{CG}}$}$(\rho^*)$)\xspace}
\newcommand{\sold}[3]{\sigma^1(#1,#2)_{#3}}
\newcommand{\sola}[3]{\sigma^2(#1,#2)_{#3}}
\newcommand{\ECOP}{(\hyperlink{(P)}{$\mathcal{P}$})\xspace}
\newcommand{\Fea}{(\hyperlink{(F)}{$\mathcal{F}$}$(b_2\rho^*)$)\xspace}
\newcommand{\order}[2]{#1}
\newcommand{\antiorder}[2]{#2}
\newcommand{\OPT}[1]{z{#1}}

\newcommand{\Value}[1]{\Val\left(#1\right)}
\newcommand{\Eff}[2]{\Val\left({#1}^{#2}\right)}
\newcommand{\Effp}[2]{\Val\left((#1)^{#2}\right)}
\newcommand{\Loss}[2]{\Val\left(#1 - {#1}^{#2}\right)}
\newcommand{\Lossp}[2]{\Val\left(#1 - (#1)^{#2}\right)}
\newcommand{\Cost}[1]{\Costop\left(#1\right)}
\newcommand{\Costf}[1]{\Costflow(#1)}
\newcommand{\Exponebis}[2]{\mathbb{E}_{#1}\left[#2\right]}
\newcommand{\Exptwobis}[2]{\mathbb{E}_{#1}\left[#2\right]}
\newcommand{\Exps}[3]{\mathbb{E}_{(#1,#2)}\left[#3\right]}
\newcommand{\Expone}[2][\empty]{\mathbb{E}_{{\sigma}^{#1}}\left[#2 \right]}
\newcommand{\Exptwo}[2][\empty]{\mathbb{E}_{{\sigma}^{#1}}\left[#2 \right]}
\newcommand{\Expboth}[3]{\mathbb{E}_{\sigma^{#1}}\left[#3\right]}

\newcommand{\Cross}{$\mathbin{\tikz [x=1.4ex,y=1.4ex,line width=.2ex, red] \draw (0,0) -- (1,1) (0,1) -- (1,0);}$}%

\newcommand{\lina}[1]{\textcolor{Green}{LS:~#1}}

\newcommand{\maybe}[1]{\textcolor{orange!70!red}{#1}}

\newcommand{\rev}[1]{#1}
\newcommand{\Lina}[1]{{\color{violet} #1}}
\newcommand{\rem}[1]{{\color{Red} #1}}
\newcommand{\todo}[1]{{\color{Orange!80!Red} #1}}

\newcommand{\StateNew}[1]{\algrenewcommand{\alglinenumber}[1]{\footnotesize A##1:}\State #1}

\captionsetup[subfigure]{font=footnotesize}



\RUNAUTHOR{Dahan, Sela, and Amin}

\RUNTITLE{Network Inspection for Detecting Strategic Attacks}

\TITLE{Network Inspection for Detecting Strategic Attacks}

\ARTICLEAUTHORS{%
\AUTHOR{Mathieu Dahan}
\AFF{School of Industrial and Systems Engineering, Georgia Institute of Technology, Atlanta, GA 30332, \EMAIL{mathieu.dahan@isye.gatech.edu}} 
\AUTHOR{Lina Sela}
\AFF{Department of Civil, Architectural and Environmental Engineering, University of Texas, Austin, TX 78712, \EMAIL{linasela@utexas.edu}}
\AUTHOR{Saurabh Amin}
\AFF{Laboratory for Information and Decision Systems, Massachusetts Institute of Technology, Cambridge, MA 02139, \EMAIL{amins@mit.edu}}
} 

\ABSTRACT{\rev{This article studies a problem of strategic network inspection, in which a defender (agency) is tasked with detecting the presence of multiple attacks in the network. An inspection strategy entails monitoring the network components, possibly in a randomized manner, using a given number of detectors. We formulate the network inspection problem $(\mathcal{P})$ as a large-scale bilevel optimization problem, in which the defender seeks to determine an inspection strategy with minimum number of detectors that ensures a target expected detection rate under worst-case attacks. We show that optimal solutions of $(\mathcal{P})$ can be obtained from the equilibria of a large-scale zero-sum game. 
Our equilibrium analysis involves both game-theoretic and combinatorial arguments, and leads to a computationally tractable approach to solve $(\mathcal{P})$. 
Firstly, we construct an approximate solution by utilizing solutions of minimum set cover (MSC) and maximum set packing (MSP) problems, and evaluate its detection performance. In fact, this construction generalizes some of the known results in network security games. Secondly, we leverage properties of the optimal detection rate to iteratively refine our MSC/MSP-based solution through a column generation procedure. Computational results on benchmark water networks demonstrate the scalability, performance, and operational feasibility of our approach. The results indicate that utilities can achieve a high level of protection in large-scale networks by strategically positioning a small number of detectors.}

}

\SUBJECTCLASS{Games/group decisions:  noncooperative; Search and surveillance; Programming: integer: applications.}

\maketitle

%


\eqspace


\ifarXiv

\else
\newpage
\fi

\section{Introduction}\label{sec:intro}


\rev{Ensuring the security of critical infrastructures such as water, oil and gas, and power distribution systems is crucial for the welfare and prosperity of our society. These infrastructure networks span huge geographical areas, and are inherently vulnerable to both intentional and unintentional threats. In most jurisdictions, public utilities and municipalities are the primary entities responsible for ensuring the infrastructure reliability and service quality, and employ various degrees of oversight to manage and respond to emergency situations. In recent years, numerous incidents have been reported that highlight the inherent vulnerability of infrastructure networks to adversarial events \citep{Physical2016,Dancy2017,Cyber2018}. Such events often result in recurrent service interruptions, and in some cases even pose significant danger to human lives \citep{NYTMexico2019}. In response, governments and public utility commissions are developing new policies and regulations that charge the utilities to proactively recognize the security risks to their infrastructure, and develop specific capabilities to reduce them \citep{NIST}. Public sector operations research can provide new solutions to guide and support the utilities in such risk assessment and mitigation activities.}

%
%
%



\rev{  
In this article, we study a network inspection problem that exploits the capabilities of modern sensing and event detection technology to monitor an infrastructure network against strategic attacks. Our objective is to design inspection strategies that can effectively detect adversarial failure events in a large-scale network, and hence limit and reduce the operational losses faced by utilities due to undetected events. 
The underlying technological motivation is the commercial availability of smart detectors that can be easily operated by the utility personnel and flexibly positioned at different locations in the network \citep{PGE2010}. These detectors are integrated systems with advanced capabilities \citep{PHILLIPS20131,XING2019291} such as: (i) On-board sensing to collect state measurements at fine temporal resolution; (ii) Accurate and timely detection of faulty events using data analytics; and (iii) Real-time communication to transmit data as well as event alerts to remote utility personnel. Our approach can help utilities leverage these capabilities in detecting targeted and/or random disruptions by providing useful guidelines for selecting the number of detectors, monitoring locations, and schedule of inspection operations in order to satisfy a target detection performance.}

\rev{
Specifically, we consider a bilevel optimization formulation of the strategic network inspection problem and focus on the question: \emph{How many detectors are required and how to position them in the network to detect multiple adversarial attacks?} In our formulation, we assume that the set of locations that can be utilized for monitoring the network and the set of network components that can be accessed and targeted by the attacker (malicious entity) are predefined. The utility (defender) aims to minimize the number of detectors to achieve a desirable attack-detection performance, while the attacker seeks to avoid detection (i.e., maximize undetected attacks). 
Importantly, we allow the choice of randomized inspection strategies by the defender, which is a departure from the traditional fixed sensing paradigm. Practically, randomized inspection entails shifting and/or mobilizing the available detectors over a subset of locations in the network. In fact, randomized strategies are known to be an effective defense mechanism in various applications, as listed in Table~\ref{List_app}. However, such strategies cannot be adopted in practice unless they are simple and cost-effective to execute, and provide strong performance guarantees. 
In this article, we focus on inspection strategies that have these desirable features.} 
 

\OneAndAHalfSpacedXI
\begin{table}[h]\small
\centering
\caption{Applications of \rev{the} network inspection problem.}
\ifarXiv\begin{tabular}{|c|c|c|c|}
\else\begin{tabular}{|l|l|l|l|}
\fi
\hline Inspection setting & Network & Type of detector & Type of attacks  \\\hline
\hline Urban patrolling & city streets  & police unit & robbery \\
\hline Network security & information network & firewall & cyberattack on server \\
\hline Sensing of gas and water networks & gas and water pipelines & leak/pressure sensor & pipe disruption \\
\hline Interdiction of illegal goods & transportation network & police officer & drug trafficking  \\
\hline Infiltration game & water channel & electric cable & malicious infiltration \\\hline
\end{tabular}
\label{List_app}
\end{table}

\ifarXiv
\OneAndAHalfSpacedXI
\else
\DoubleSpacedXI

\vspace{-0.8cm}
\fi

\subsection{\rev{Main} Contributions}

In Section \ref{sec:form}, we introduce a generic detection model \rev{that} captures the key features of modern inspection systems with respect to sensing technology for event detection and flexibility of positioning. 
We use this detection model to \rev{formulate the bilevel optimization problem, denoted  $(\mathcal{P})$, in which the defender first selects a randomized positioning of detectors, and the attacker responds by targeting one or more network components. The attacker seeks to maximize the expected number of undetected attacks, while the defender aims to minimize the number of detectors required to ensure that the expected detection rate under worst-case attacks is above a pre-specified threshold.}

\rev{Our approach to solve the problem $(\mathcal{P})$ involves analyzing the equilibrium properties of a zero-sum game, denoted $\Gamma$, where the defender (resp. attacker) seeks to minimize (resp. maximize) the expected number of undetected attacks (Proposition~\ref{Eq_P}). However, the sets of players' actions in $\Gamma$ grow combinatorially with the size of the network, thus making the equilibrium computation challenging in itself. In Section \ref{general}, we derive structural properties that are satisfied by all Nash equilibria of $\Gamma$.}
We present these properties for the most conservative case when the attacker has the ability to spread her attacks across the network. In particular, we show that in any equilibrium \rev{of $\Gamma$}, both players must randomize their actions and use all available resources, and every network component must be monitored with positive probability (\rev{Theorem \ref{all_resources} and Proposition \ref{all_edges}}). 
\rev{Additionally, we prove the important, and rather surprising property,} that the expected detection rate and the inspection strategies in equilibrium do not depend on the attacker's number of resources (Theorem \ref{Constant}). This implies that \rev{the defender} does not need to know precisely the amount of attack resources \rev{in order to monitor the network}. 
\rev{The proofs of these game-theoretic results rely on linear programming duality in zero-sum games, submodularity of the detection function, as well as} \rev{the minimum set cover (MSC) and maximum set packing (MSP) problems, which respectively capture the ``coverage'' and ``spread'' of the network.
}

\rev{Our equilibrium analysis leads to a novel approach to solve the inspection problem $(\mathcal{P})$. Firstly, we obtain lower and upper bounds on the \rev{optimal} expected detection rate in terms of the number of available detectors, and the optimal values of the MSC and MSP problems. A preliminary and specialized version of this result was presented in \cite{7852316}.  Furthermore, we construct an inspection strategy that randomizes the positioning of detectors over an MSC, and derive guarantees on the resulting expected detection performance.
%
%
%
%
%
This provides us with an approximate solution to 
$(\mathcal{P})$ and optimality gap that can be computed by solving the MSC and MSP problems.
%
It turns out this solution is optimal for the special case when the MSCs and MSPs are of same size. Secondly, a consequence of the equilibrium properties is that a column generation-based procedure can be used to iteratively improve our MSC/MSP-based solution to optimality.}

Although our approach to solve problem $(\mathcal{P})$ relies on the MSC and MSP problems which are known to be NP-hard, we find that modern integer programming solvers can solve large instances of these problems. \rev{In Section \ref{sec:app}, we demonstrate the benefits of our solution approach in monitoring large-scale urban water networks facing adversarial disruptions. Our computational study shows that the MSC/MSP-based solution is scalable, provides good performance guarantees, and is easily implementable by the defender. On the other hand, we find that implementing the optimal inspection strategy requires a much higher number of monitored locations and a more complex scheduling of operations. Furthermore, it only provides a marginal improvement in comparison to the simpler MSC-based inspection strategy. Thus, our approach can be used for designing inspection strategies that achieve a desired tradeoff between detection performance and operational feasibility.}

The complete proofs of our results\rev{, as well as additional examples,} are provided in the electronic companion of this article.


\subsection{Related Work}

Our detection model is inspired by modern sensing technology used in detecting leaks and other failure events in pipeline networks for distribution of natural gas \citep{PHILLIPS20131}, and water \citep{Ost1415211920040901,sensors:autom}. \rev{The dominant paradigm} in sensing of these infrastructure networks is to optimally place \rev{a limited number of} sensors for maximizing \rev{a metric of detection performance} \citep{berry,Krause:2008:NSP:1390681.1390689,4729806}. \rev{A myriad of models for the sensor placement problem have been proposed in the literature, including robust formulations} \citep{SELA201855}; for example, 
 \cite{krause+al:jmlr08} proposed an efficient approximation algorithm to maximize the worst-case detection performance against a set of possible failure scenarios. More recently, \cite{8263844} and \cite{Orlin2016} designed approximation algorithms to find a sensor placement that is robust against a subset of sensors' failures. The main feature of this line of work is \emph{fixed sensing}, i.e., continuous operation of sensors placed at fixed locations in the network. 
 

 \rev{However, in large-scale networks, a fixed strategy implemented by a resource-constrained utility inevitably leaves some parts of the network unmonitored. In an adversarial situation, a strategic attacker will target these unmonitored parts to avoid or make their detection more difficult. It follows that a fixed inspection strategy can lead to a significant loss of detection performance and, in turn, compromise the overall security of the infrastructure system.}
%
\rev{In contrast}, it \rev{has been shown} that \emph{randomized strategies} can significantly improve the defender's performance  against worst-case \rev{disruptions or adversarial failure events} \citep{doi:10.1287/opre.43.2.243,BERTSIMAS2016114}. \rev{Practically, randomized inspection strategies can be translated into random scheduling of inspections that can be performed on a day-to-day basis by utility personnel.}
\rev{For example,~\cite{Pita:2008:DAP:1402795.1402819} use randomized strategies for the scheduling of checkpoints and for generating patrolling schedules for canine units to assist the police at the Los Angeles International Airport. \cite{Hochbaum2011} investigate the allocation of mobile sensors to detect transported nuclear weapons based on related radiological dispersion devices.} 
\rev{Finally, water utilities routinely sample water quality at random locations in the distribution system to comply with safety standards such as the Safe Drinking Water Act (SDWA) rules~\citep{tiemann2014safe}}.

\rev{In the context of network security, several models have been proposed for the strategic allocation of defense resources \citep{doi:10.1287/opre.1070.0434,BaykalGursoy2014469,Goyal:2014aa}.
For instance, \cite{doi:10.1287/inte.1060.0252,Bier_2011,doi:10.1111/risa.12333,doi:10.1287/trsc.2017.0749} consider bilevel and trilevel optimization problems to model defender-attacker interactions where each player selects a pure strategy. In contrast, our setting involves randomized strategies, and the combinatorial size of the sets of players' actions does not enable us to solve problem $(\mathcal{P})$ using mixed-integer linear programming techniques. 
 Other models that have been studied include search  games \citep{Gal20140062}  and inspection problems \citep{doi:10.1287/opre.43.2.243,doi:10.1287/opre.46.2.184,smith2008algorithms}. In addition, \cite{10.2307/27644485,Bier_2008,ZHUANG2010409} investigate equilibria in security games with an asymmetric information structure. However, the approaches presented in these papers cannot be applied to solve problem $(\mathcal{P})$.}

\rev{The zero-sum game $\Gamma$ we analyze for solving problem $(\mathcal{P})$} is more general than the classical \rev{\emph{hide-and-seek}} game first introduced by \cite{von1953certain}, and further discussed in \rev{Chapter} 3.2 of \cite{karlin2016game}. In this game, a robber hides in one of a set of ``safe-houses'' located at \rev{intersections of vertical and horizontal roads}, and a police unit simultaneously chooses to travel along one road to find the robber. 
Our \rev{equilibrium analysis} can be applied to solve the generalized hide-and-seek game which involves multiple police units patrolling in a complex street network to find multiple robbers. 
%
%
%
Related to our setting is \rev{also} the work by \cite{mavronicolas}, who consider a security game on a \rev{bipartite} information network in which \rev{servers} are vulnerable to multiple attacks and the defender can install a firewall to protect a subnetwork. \rev{In fact, our analysis approach can be used to derive a more sophisticated defense strategy that installs multiple firewalls to secure more complex information networks against multiple simultaneous attacks. Our game similarly generalizes the \emph{patrolling game} studied in \cite{patrolling}, and the \emph{infiltration games} defined in \cite{Garnaev1997} and in Chapter 2.1 of \cite{garnaev2000search}.}

\section{Problem Description}\label{sec:form}

\rev{In this section, we introduce a generic formulation of strategic network inspection problem. Our formulation is a bilevel optimization model of sequential defender-attacker interaction on an infrastructure network, with each player having access to multiple resources.}

\subsection{Defender and Attacker Models} \label{sec:detect}

\rev{We consider the setting where a defender (utility) is tasked with inspecting an infrastructure network that transports a commodity (e.g., water, oil, natural gas). The network faces risk of adversarial disruptions that can compromise the operational functionality of its set of components, denoted $\edges$. All components in the set $\edges$ are critical in the sense that they are prone to targeted attacks by an attacker (malicious entity). 
To inspect the network and monitor its components, the defender positions a set of {\em detectors} on a set of locations (or nodes), denoted $\nodes$. Each detector is an integrated system comprising an on-board sensing unit, detection software, and communication unit \citep{1219475}. The defender (or the utility’s employees) can flexibly mobilize the detectors from one network node to another. For our purposes, the sets $\edges$ and $\nodes$ are predefined. 
}


\rev{For example, in the context of a municipal water network, the set $\edges$ represents the system's components that can be accessed and targeted by an attacker, and the set $\nodes$ represents the access points where detectors can be deployed (e.g., manholes, valves, or fire hydrants).
%
Targeted physical or remote attacks to the network can induce disruptions such as damage to pipelines, backflow at fire hydrants, or sabotage of valves \citep{MONROE201837,cyberphysicalWDS2020}. Such disruption events typically result in local perturbations in the state variables (water flow rate and pressure) that progressively propagate to other parts of the network. If the defender (water utility) has positioned a detector at a node that experiences perturbations from a disruption event, the on-board sensing unit can measure the change in state variables \citep{allen:smart,seshan}. These measurements can then be processed to detect the occurrence of the event. Clearly, the ability of the defender to detect such disruption events depends on her inspection strategy, i.e., how the available detectors are positioned in the network.} 

\rev{Formally,} when a detector is positioned at node $i\in\nodes$ \rev{by the defender, the following steps govern its attack detection capability: 
} Firstly, the \rev{sensing unit} collects relevant state measurements from node $i$. These measurements capture the state of a subset of components $\set{i} \in 2^\edges$\rev{, i.e., the detector at node $i$ monitors the components in $\set{i}$}. Secondly, the detection software processes these measurements and generates a diagnostic signal indicating the number of \rev{disruption events (or attacks)} present within the component set $\set{i}$. Thirdly, the communication unit transmits the diagnostic signal to the defender. \rev{We assume that the} cost of data collection, processing, and transmission is negligible in comparison to the cost of procuring the detector. 
For a detector positioned at node $i\in\nodes$, we refer to the set $\set{i}$ as a \emph{monitoring set}, because under the aforementioned \rev{setting}, an attack to any component $e\in\edges$ can be detected if and only if $e\in\set{i}$. The tuple $\mathcal{G}\coloneqq(\nodes, \edges, \{\set{i}, \ i \in\nodes\})$ represents the \emph{detection model} of the network. Without loss of generality, we assume that each component in $\edges$ can be monitored from at least one node in $\nodes$.  

\rev{Importantly, we consider} that the defender has access to only a limited number of detectors for network inspection. This limitation results from the economic and operational constraints of the defender. For simplicity, we \rev{also suppose} that all detectors are homogeneous in terms of their monitoring and detection capabilities, and cost. Let $b_1\in\mathbb{N}$ be the number of available detectors that can be simultaneously positioned on distinct nodes in $\nodes$. We denote a \emph{detector positioning} by a set $\Def\in 2^\nodes$ of nodes that receive detectors. The set of feasible detector positionings is then defined by $\mathcal{A}_1\coloneqq \{\Def\in 2^\nodes \ | \  |\Def|\leq b_1\}$. For a given detector positioning $\Def\in \mathcal{A}_1$, let $\set{\Def}\coloneqq \cup_{i\in \Def}\, \set{i}$ denote the set of components that are monitored by at least one detector in $\Def$.

To count the number of components in any given subset of components of $\edges$ that can be monitored using an arbitrary detector positioning, we define a \emph{detection function} $\detect: 2^{\nodes} \times 2^{\edges} \longrightarrow \mathbb{N}$. For a detector positioning $\Def \in 2^\nodes$ and a subset of components $\att \in 2^\edges$, the value of $\detect(\Def,\att)$ is the number of components of $\att$ that are monitored by at least one detector positioned in $\Def$, i.e.:
\begin{align}
\forall (\Def,\att) \in 2^{\nodes} \times 2^{\edges},  \quad \valdet{\Def}{\att}  \coloneqq  |\set{\Def} \cap \att|.\label{detect_sets}
\end{align}

\rev{Note that under our detection model, if the components of $\att$ face attack-induced disruptions, the number of attacks detected by the detector positioning $\Def$ is $\valdet{\Def}{\att}$.}
The detection function \rev{satisfies} two natural properties: 


\begin{enumerate}
\item For any subset of components $\att \in 2^\edges$, $\valdet{\cdot}{\att}$ is submodular and monotone:
\begin{align*}
\forall \att \in 2^\edges, \ \forall (\Def,\Def^\prime) \in (2^\nodes)^2, \quad \left\{\begin{array}{l}
\valdet{\Def \cup \Def^\prime}{\att} + \valdet{\Def \cap \Def^\prime}{\att} \leq \valdet{\Def}{\att} + \valdet{\Def^\prime}{\att},\ifarXiv\mcr\else \\ \fi
\Def \subseteq \Def^\prime \ \Longrightarrow\ \valdet{\Def }{\att} \leq \valdet{\Def^\prime}{\att}.\end{array}\right.
\end{align*}
That is, adding a detector to a smaller detector positioning increases the number of monitored components in $\att$ by at least as many as when adding that detector to a larger detector positioning.

\item For any detector positioning $\Def \in 2^\nodes$, $\valdet{\Def}{\cdot}$ is finitely additive (a direct consequence of \eqref{detect_sets}):
\begin{align*}
\forall \Def \in 2^\nodes,\ \forall (\att,\att^\prime) \in (2^\edges)^2 \ | \ \att \cap \att^\prime = \emptyset, \quad \valdet{\Def}{\att \cup \att^\prime} = \valdet{\Def}{\att} + \valdet{\Def}{\att^\prime}.
\end{align*}
\end{enumerate}

\rev{Similar} to the defender, the attacker is also resource-constrained, in that she can attack a subset of components $\att\in 2^\edges$ of the network of size no larger than $b_2\in\mathbb{N}$; we refer to such a subset as an \emph{attack plan}. This \rev{constraint models the attacker's limited ability to gain} access to network components and \rev{disrupt them}. The set of all attack plans is given by $\mathcal{A}_2\coloneqq \{\att\in 2^\edges \ | \ |\att|\leq b_2\}$.

\rev{In fact, our solution approach and results on its guarantees can be extended to the model of imperfect detection, where each detector only detects a disruption in its monitoring set with independent probability $\lambda \in [0,1]$. Given a detector positioning $\Def \in 2^\nodes$, and an attack plan $\att \in 2^\edges$, the average number of detected attacks would be given by $\sum_{e \in \att}\left(1 - (1-\lambda)^{|\{i \in \Def \, | \, e \in \set{i}\}|}\right)$, which is also submodular and monotone with respect to $\Def$ (see Chapter 2 of \cite{Fujishige2005}), and finitely additive with respect to $\att$. For ease of exposition, we henceforth assume the model of perfect detection, given by \eqref{detect_sets}.}



\subsection{Network Inspection Problem}

\rev{We are now in the position to introduce our network inspection problem, which we define as a bilevel optimization model. In this problem, the defender (referred to as player 1 or \defender) first chooses an inspection strategy to monitor network components using minimum number of detectors. After observing the defender's action, the attacker (referred to as player 2 or \attacker) targets one or more components to induce disruption events. A typical assumption in infrastructure defense is that of an informed attacker who knows the defender's capabilities. Thus, we assume that both players know the detection model $\mathcal{G}$. At this stage, we also assume that the defender knows the number of attack resources $b_2$, although we will later show that our solution to the network inspection problem does not depend on it.}





\rev{The detector positionings (resp. attack plans) are} realized from a chosen probability distribution on the set $\mathcal{A}_1$ (resp. $\mathcal{A}_2$). 
Specifically, the defender and attacker respectively choose \rev{an} inspection strategy $\sigma^1 \in 
\Delta(\mathcal{A}_1)$ and \rev{an} attack strategy $\sigma^2 \in  \Delta(\mathcal{A}_2)$, where 
$\Delta(\mathcal{A}_1)\coloneqq\{\sigma^1 \in [0,1]^{|\mathcal{A}_1|}\ | \ \sum_{{\Def} \in \mathcal{A}_1} \sigma^1_\Def = 1 \}$ and $\Delta(\mathcal{A}_2) \coloneqq \{\sigma^2 \in [0,1]^{|\mathcal{A}_2|}\ | \ \sum_{{\att} \in \mathcal{A}_2} \sigma^2_{\att} = 1 \}$ denote the \rev{mixed} strategy sets. Here, $\sigma^1_\Def$ (resp. $\sigma^2_\att$) represents the probability assigned to the detector positioning $\Def$ (resp. attack plan $\att$) by the defender's strategy $\sigma^1$ (resp. the attacker's strategy $\sigma^2$).  


For ease of exposition, we denote $\valdet{i}{e} \coloneqq \valdet{\{i\}}{\{e\}}$  \rev{for all} $(i,e) \in \nodes \times \edges$. \rev{We will also refer to the degenerate mixed-strategies $\mathds{1}_{\{\Def\}} \in \Delta(\mathcal{A}_1)$ and  $\mathds{1}_{\{\att\}} \in \Delta(\mathcal{A}_2)$ as $S$ and $T$, respectively.}  The \emph{support} of $\sigma^1 \in \Delta(\mathcal{A}_1)$ (resp. $\sigma^2 \in \Delta(\mathcal{A}_2)$) is defined as $\supp(\sigma^1) = \{\Def \in \mathcal{A}_1 \ | \ \sigma^1_{\Def} >0 \}$ (resp. $\supp(\sigma^2) = \{\att \in \mathcal{A}_2 \ | \ \sigma^2_{\att} >0 \}$). The \emph{node basis} of a strategy $\sigma^1 \in \Delta(\mathcal{A}_1)$, denoted  $\nbasis{\sigma^1}\coloneqq \{i \in \nodes \ | \ \mathbb{P}_{\sigma^1} (i \in \Def)>0\}$, is the set of nodes that are inspected with non-zero probability \rev{by the defender}. Analogously, the \emph{component basis} of a strategy $\sigma^2 \in \Delta(\mathcal{A}_2)$, denoted $\ebasis{\sigma^2}\coloneqq\{e \in \edges \ | \ \mathbb{P}_{\sigma^2}(e \in \att) >0\}$, is the set of components that are targeted with positive probability \rev{by the attacker}. 



\rev{We now present the inner and outer problem of our bilevel optimization model.

\emph{Inner problem:} In our model, the attacker responds to the defender's inspection strategy by choosing an attack strategy, with the objective of maximizing the expected number of attacks that remain undetected by the defender. Formally, given an inspection strategy $\sigma^1 \in \Delta(\mathcal{A}_1)$, the attacker aims to find an attack strategy $\sigma^2 \in \Delta(\mathcal{A}_2)$ that maximizes the following objective: 
\begin{align}
U(\sigma^1,\sigma^2)\coloneqq \mathbb{E}_{(\sigma^1,\sigma^2)}[|\att| - \valdet{\Def}{\att}].\label{payoff2}
\end{align} 
We denote $B_2(\sigma^1,b_2)\coloneqq \argmax_{\sigma^2 \in \Delta(\mathcal{A}_2)}U(\sigma^1,\sigma^2)$ the set of attack strategies that are best responses to $\sigma^1$. 

\emph{Outer problem:} The defender seeks to minimize the number of detectors and also ensure that her chosen inspection strategy provides a certain level of detection performance against the attacker's best response strategy.
%
We use the following metric of detection performance: 
For a given strategy profile $\sigma \in \Delta(\mathcal{A}_1) \times \Delta(\mathcal{A}_2)$, the \emph{expected detection rate}, denoted $\sdet{}$, is the expectation (under $\sigma$) of the ratio between the number of attacks that are detected and the total number of attacks:
\begin{align}
\sdet{} \coloneqq \Expboth{}{}{\dfrac{\valdet{\Def}{\att}}{|\att|}}. \label{Exp_detection_rate}
\end{align}

Thus, the defender aims to find a minimum-resource inspection strategy while ensuring that the expected detection rate is no less than a pre-specified threshold level $\alpha\in [0,1]$ against worst-case attack plans.  This can be written as the following network inspection problem:\hypertarget{(P)}{}
%
%
%
 \begingroup
 \addtolength{\jot}{\spacingalign}
\begin{align}
(\mathcal{P}): \quad \underset{b_1,\, \sigma^{1}}{\text{minimize}}& \quad b_1\nonumber\\
\text{subject to}& \quad r(\sigma^{1},\sigma^{2}) \geq \alpha,  \quad \forall \sigma^{2} \in B_2(\sigma^{1},b_2) \label{all_NE}\\
& \quad \sigma^1 \in \Delta(\mathcal{A}_1)\nonumber \\
& \quad b_1 \in \mathbb{N}.\nonumber
\end{align}
 \endgroup

Specifically, constraints \eqref{all_NE} ensure that for a given number of attack resources $b_2$, the expected detection rate induced by the chosen number of detectors $b_1$ and their randomized positioning $\sigma^1$ is at least $\alpha$ under the attacker's best response to $\sigma^1$. 
The target detection rate $\alpha$ reflects the performance requirement that the defender's inspection strategy must satisfy (for example, due to a regulatory imposition). We denote $b_1^*$ the optimal value of \ECOP.

}




\rev{More generally, the problem \ECOP} captures some of the key features of network inspection in strategic settings; see Table~\ref{List_app} for a comparison of various applications. Firstly, the detection model $\mathcal{G}$ is generic in that it represents the detection capability of the defender, without making further modeling assumptions about how the monitoring sets $\set{i}$ ($i\in\nodes$) depend on specific aspects such as \rev{the} sensing technology employed by detectors, the different means that the attacker may use in targeting a component, and the network's topological structure. Secondly, it considers multiple resources on the part of both \rev{players}. This is a particularly desirable feature \rev{when} the attacker can simultaneously attack multiple components across the network, and the defender's inspection involves positioning multiple detectors  in order to monitor a large number of critical components. 
\rev{However, \ECOP is a challenging problem to solve. Indeed, bilevel optimization problems are known to be NP-hard \citep{doi:10.1137/0913069}, and in our case the number of possible detector positionings  grows combinatorially with the number of available detectors and the size of $\nodes$. Thus, we must leverage structural properties of the problem to solve it in a scalable manner.}

\section{\rev{Game-Theoretic Analysis}}\label{general}

\rev{In this section, we derive the key properties satisfied by optimal solutions of our network inspection problem \ECOP. We start by studying \attacker's best response function $B_2$, and analyze the corresponding zero-sum game. This will in turn help derive a scalable solution approach to \ECOP. }

\subsection{\rev{Zero-Sum Game}}

\rev{
Given the detection model $\mathcal{G} = (\nodes, \edges, \{\set{i}, \ i \in\nodes\})$, and the players' resources $b_1$ and $b_2$, we consider the zero-sum game in normal form $\Gamma(b_1,b_2) \coloneqq \langle \{1,2\}, (\Delta(\mathcal{A}_1), \Delta(\mathcal{A}_2)), (-U,U) \rangle$. In this game, \defender (resp. \attacker) selects an inspection strategy $\sigma^1 \in \Delta(\mathcal{A}_1)$ (resp. an attack strategy $\sigma^2 \in \Delta(\mathcal{A}_2)$) and seeks to minimize (resp. maximize) the expected number of undetected attacks (see \eqref{payoff2}).

}

A strategy profile $({\sigma^1}^\ast,{\sigma^2}^\ast) \in \Delta(\mathcal{A}_1) \times \Delta(\mathcal{A}_2)$ is a mixed strategy \emph{Nash Equilibrium} (NE) of the game $\Gamma(b_1,b_2)$ if:
\begin{align}
\rev{\forall (\sigma^1,\sigma^2) \in \Delta(\mathcal{A}_1) \times \Delta(\mathcal{A}_2), \quad U({\sigma^1}^*,{\sigma^2}) \leq U({\sigma^1}^*,{\sigma^2}^*) \leq U({\sigma^1},{\sigma^2}^*) \label{best}.}
\end{align}
We denote the set of NE of the game $\Gamma(b_1,b_2)$ as $\nash(b_1,b_2)$. Also, when there is no confusion, we \rev{simply} refer to $\Gamma(b_1,b_2)$, $\Sigma(b_1,b_2)$,  \rev{and $B_2(\sigma^1,b_2)$} as $\Gamma$, $\Sigma$, \rev{and $B_2(\sigma^1)$, respectively}.




%
\rev{Since $\Gamma$ is a zero-sum game, the set of NE $\Sigma$} can be obtained by solving the following \rev{pair of dual linear programming problems}:\hypertarget{LPs}{}
\begin{align*}
(\text{LP}_1)  \quad \displaystyle \rev{\min_{\sigma^1 \in \Delta(\mathcal{A}_1)} \max_{\att \in \mathcal{A}_2} {U}(\sigma^1,\att)} \quad\quad \quad  \vrule{} \quad \quad \quad (\text{LP}_2) \quad \displaystyle \max_{\sigma^2 \in \Delta(\mathcal{A}_2)} \min_{\Def \in \mathcal{A}_1} \rev{{U}(\Def,\sigma^2)}.
\end{align*}
\rev{We refer to the optimal value of \lpo and \lpt as the value of the game $\Gamma(b_1,b_2)$, denoted by $U^*(b_1,b_2)$.} \rev{In principle,} linear programming techniques can be used to compute NE of $\Gamma$.  However,  the computation of \lpo and \lpt quickly becomes intractable as the size of the network increases. In particular, due to the size of the players' sets of actions ($|\mathcal{A}_1|= \sum_{k=0}^{b_1}{|\nodes| \choose k}$ and $|\mathcal{A}_2|= \sum_{l=0}^{b_2}{|\edges| \choose l}$), the number of variables and constraints in both linear programs can be huge. For example, for a network consisting of $200$ nodes and components, and $b_1 = b_2 = 10$, computing the equilibria of game $\Gamma(b_1,b_2)$ entails solving linear programs containing $2.37\cdot 10^{16}$ variables and constraints. For large bimatrix games, \cite{Lipton:2003:PLG:779928.779933} provide an algorithm to compute an $\epsilon-$NE in $n^{O(\ln n/\epsilon^2)}$ time, \rev{where $n$ is the number of pure strategies available to each player}. However, for realistic instances of the game $\Gamma$, \rev{this number} can easily reach values for which their algorithm is practically inapplicable.


\rev{Next,} we develop new results to study the equilibrium characteristics of the game $\Gamma(b_1,b_2)$, given any parameters $b_1$ and $b_2$. Our equilibrium characterization utilizes two combinatorial optimization problems, formulated as minimum set cover and maximum set packing problems. \rev{This characterization enables us to analyze} the detection performance \rev{in equilibrium}\rev{, which in turn reveals properties satisfied by optimal solutions of our problem \ECOP}.




\subsection{Set Cover and Set Packing Problems}\label{2 Comb Pbs}

We say that a set of nodes $S \in 2^{\nodes}$ is a \emph{set cover} if and only if every component in $\edges$ can be monitored by at least one detector positioned in $\Def$, i.e., $\valdet{S}{e}=1$, \rev{for all} $e \in \edges$. A set of nodes $\Def \in 2^\nodes$ is a \emph{minimal set cover} if $\Def$ is a set cover that is minimum with respect to inclusion, i.e., if any node of $\Def$ is removed, the resulting set is not a set cover anymore. A set of nodes $S \in 2^{\nodes}$ is a \emph{minimum set cover} (MSC) if and only if it is an optimal solution of the following problem:\hypertarget{(MSC)}{}
\begin{align}
(\mathcal{I}_{\text{MSC}}):  \ \underset{\Def \in 2^\nodes}{\text{minimize}} \ |\Def| \quad \text{subject to } \ \valdet{S}{e}=1, \quad \forall e \in \edges.
\label{def_minimum_set_cover}
\end{align}

Solving  (\hyperlink{(MSC)}{$\mathcal{I}_{\text{MSC}}$}) amounts to determining the minimum number of detectors and their positioning to monitor all network components.
We denote the set (resp. the size) of MSCs by $\msc$ (resp. $\smsc$). Since we assumed that each component can be monitored from at least one node in the network (Section \ref{sec:detect}),  (\hyperlink{(MSC)}{$\mathcal{I}_{\text{MSC}}$}) is feasible and $\smsc$ exists.



We say that a set of components $T \in 2^{\edges}$ is a \emph{set packing} if and only if a detector positioned at any node $i$ can monitor at most one component in $\att$, i.e., $\valdet{i}{\att}\leq1,$ \rev{for all} $i \in \nodes$. A set of components $T \in 2^{\edges}$ is a \emph{maximum set packing} (MSP) if and only if it optimally solves the following problem:\hypertarget{(MSP)}{}
\begin{align}
(\mathcal{I}_{\text{MSP}}): \underset{\att \in 2^\edges}{\text{maximize}} \ |\att| \quad \text{subject to } \ \valdet{i}{\att}\leq1, \quad \forall i \in \nodes.
\label{def_extended_matching}
\end{align}

Solving  (\hyperlink{(MSP)}{$\mathcal{I}_{\text{MSP}}$}) amounts to finding the maximum number of ``independent'' components, i.e., a set of components of maximum size such that monitoring each component requires a unique detector.
We denote the set (resp. the size) of MSPs by $\emm$ (resp. $\semm$). 

Although, (\hyperlink{(MSC)}{$\mathcal{I}_{\text{MSC}}$}) and (\hyperlink{(MSP)}{$\mathcal{I}_{\text{MSP}}$}) are known to be NP-hard problems, modern mixed-integer optimization solvers can be used to optimally solve them for realistic problem instances; see Section~\ref{sec:app}. Furthermore, their integer programming formulations have linear programming relaxations that are dual of each other (see \rev{Chapter} 13.1 of \cite{Vazirani:2001}). This implies that $\semm \leq \smsc$. 

MSCs and MSPs represent the network's \emph{coverage} and \emph{spread}, respectively: $\smsc$ represents the minimum number of detectors required by \defender to completely monitor the network, and $\semm$ represents the maximum number of attack resources for which \attacker can spread her attacks across the network. In fact, solving $\Gamma(b_1,b_2)$ is trivial when $b_1 \geq \smsc$, because \defender can monitor all network components by deterministically positioning the detectors on an MSC. \rev{Such a detector positioning satisfies constraints~\eqref{all_NE} for any target detection rate $\alpha$.} A direct consequence is that the optimal number of detectors in \ECOP, \rev{$b_1^*$}, is at most $\smsc$.

\rev{On the other hand, a practically  relevant (and interesting) case is} when \attacker's number of attack resources is less than the size of MSPs, i.e., $b_2 < \semm$.  This case captures the situations in which the network is ``large enough'' in that \attacker can exhaust her ability to spread attacks, thereby making it most challenging for \defender to detect the attacks using her inspection strategy. Furthermore, when $b_2 \geq \semm$, a larger number of attack resources improves \defender's ability to detect some of the attacks. Thus, an inspection strategy that ensures the target detection performance for the case $b_2 < \semm$ can also be applied when $b_2 \geq \semm$. Henceforth, our analysis primarily focuses on the case when $b_1 < \smsc$ and $b_2 < \semm$ (see \rev{Figure}~\ref{Regimes}). We discuss the other cases whenever relevant.



\begin{figure}[htbp]
\centering
\begin{tikzpicture}[x=0.8cm,y=0.58cm]

\def \FontFig {\small}
\draw[thick][->] (0,0) -- (12,0) node[anchor=north] {\FontFig$b_1$};
\draw	(0,-0.1) node[anchor=north] {\FontFig$0$}
		(6,-0.1) node[anchor=north] {\FontFig$n^*$};

\draw[thick] (6,-0.07) -- (6,0.07);	
		
\draw[thick][->] (0,0) -- (0,5) node[anchor=east] {\FontFig$b_2$};

\draw	(-0.1,2.5) node[anchor=east] {\FontFig$m^*$};		
\draw[thick] (-0.07,2.5) -- (0.07,2.5);	

\draw[thick] (6,0) -- (6,5);
\draw[thick] (6,2.5) -- (0,2.5);

%

\draw[draw=black,fill=gray,opacity=0.2]  (0.02,0.02) rectangle (5.98,2.48);

\draw	(9,2.85) node{\FontFig\textcolor{black}{Complete monitoring}};
\draw	(9,2.15) node{\FontFig\textcolor{black}{(trivial game)}};

\draw	(3,4.1) node{\FontFig\textcolor{black}{Redundant attacks}};
\draw	(3,3.4) node{\FontFig\textcolor{black}{(easy detections)}};

\draw	(3,1.6) node{\FontFig\textcolor{black}{Case of interest}};
\draw	(3,0.9) node{\FontFig\textcolor{black}{(large network)}};


\end{tikzpicture}

\vspace{-0.2cm}

\caption{\rev{Three cases based on the magnitude of $b_1$ and $b_2$ relative to $\smsc$ and $\semm$.}}
\label{Regimes}
\end{figure}

\subsection{Equilibrium \rev{Analysis of Game $\Gamma(b_1,b_2)$}}\label{Main_case}

\rev{We proceed in three steps:}
\rev{Firstly, we derive bounds on the value of the game $\Gamma$ based on exact or approximate solutions to the MSC and MSP problems (Proposition~\ref{best_set_cover})}. \rev{Secondly, we show that every NE satisfies certain structural properties (Theorem \ref{all_resources} and Proposition \ref{all_edges}); these properties establish a connection between the zero-sum game $\Gamma$ and problem \ECOP (Proposition~\ref{Eq_P}). Finally, we derive properties satisfied by the expected detection rate in equilibrium of $\Gamma$ (Theorem~\ref{Constant}).}
 

\rev{\textbf{Step 1: MSC/MSP-based bounds on the value of the game $\Gamma$.}}
\rev{Recall that NE and the value of the game $\Gamma$ are respectively given by the optimal solutions and optimal value of the linear programs \lpo and \lpt.}
%
%
\rev{To} derive bounds on the optimal value of \lpo and \lpt, along with mixed strategies that achieve these bounds, \rev{we utilize} the following construction:

\begin{lemma}\label{algebra1}
Consider a set of nodes $\Def \in 2^\nodes$ of size $n\geq b_1$, and a set of components $\att \in 2^\edges$ of size $m \geq b_2$. Then, there exists a strategy profile, denoted $(\sold{\Def}{b_1}{},\sola{\att}{b_2}{}) \in \Delta(\mathcal{A}_1)\times\Delta(\mathcal{A}_2)$, whose node basis and component basis are $\Def$ and $\att$, respectively, and such that:
\begin{align}
&\forall i \in \Def, \ \mathbb{P}_{\sold{\Def}{b_1}{}}(i \text{ is inspected by \defender}) = \frac{b_1}{n}, \label{set_sensing_prob}\\
&\forall e \in \att, \ \mathbb{P}_{\sola{\att}{b_2}{}}(e \text{ is targeted by \attacker}) = \frac{b_2}{m}.\label{set_disruption_prob}
\end{align}

\end{lemma}
%

For  details on the construction of $(\sold{\Def}{b_1}{},\sola{\att}{b_2}{})$, we refer to Lemma~\ref{algebra2}. The main idea behind the construction of the inspection strategy $\sold{\Def}{b_1}{}$ is to ``cycle'' over size-$b_1$ subsets of $\Def$, such that every node of $\Def$ is inspected with an identical probability given by \eqref{set_sensing_prob}; similarly for the attack strategy $\sola{\att}{b_2}{}$. 
\rev{We can use Lemma~\ref{algebra1} to derive bounds on the value of the game $\Gamma(b_1,b_2)$ using set covers and set packings:}

\rev{
\begin{proposition}\label{best_set_cover}
The value of the game  $\Gamma(b_1,b_2)$ is upper-bounded by $b_2\left(1 - \frac{b_1}{|\Def^\prime|}\right)$ for every minimal set cover $\Def^{\prime} \in 2^\nodes$, and is lower-bounded by $\max\left\{0,b_2\left(1 - \frac{b_1}{|\att^\prime|}\right)\right\}$ for every set packing $\att^{\prime} \in 2^\edges$ of size at least $b_2$. Furthermore, these bounds are achieved by $\sold{\Def^\prime}{b_1}{}$ and $\sola{\att^{\prime}}{b_2}{}$, respectively:
\begin{align*}
\max\left\{0,b_2\left(1 - \frac{b_1}{|\att^\prime|}\right)\right\} = \min_{\Def \in \mathcal{A}_1} U(S,\sola{\att^{\prime}}{b_2}{}) \leq U^*(b_1,b_2) \leq \max_{\att \in \mathcal{A}_2} U(\sold{\Def^\prime}{b_1}{},\att) = b_2 \left(1 - \dfrac{b_1}{|\Def^\prime|} \right).
\end{align*}

%
%
%
%
\end{proposition}

}

Recall \rev{from Section~\ref{2 Comb Pbs}} that if \defender had at least $\smsc$ detectors (i.e., $b_1 \geq \smsc$), an equilibrium inspection strategy would be to position $\smsc$ detectors on an MSC. \rev{Proposition~\ref{best_set_cover} shows that,} even for the case when \defender has strictly less than $\smsc$  detectors, a set cover is a good candidate for node basis. Analogously, a good candidate for component basis is a set packing. Indeed, if \attacker targets components that are spread apart, then it will be difficult for \defender to detect many of these attacks using the available detectors. Thus, by targeting a set packing, \attacker can ensure that a single detector can detect at most one attack.
\rev{We observe that decreasing the size of the minimal set cover and increasing the size of the set packing tighten the bounds on the value of the game $\Gamma$. Thus, the best lower (resp. upper) bound on $U^*(b_1,b_2)$ is $\max\left\{0,b_2\left(1-\frac{b_1}{\semm}\right)\right\}$ (resp. $b_2\left(1 - \frac{b_1}{\smsc}\right)$).}

\rev{\textbf{Step 2: Equilibrium Properties.}} The second step consists of deriving structural properties satisfied by every NE \rev{of $\Gamma$}. 
An important property is that when $b_1 < \smsc$ and $b_2 < \semm$, \emph{any} equilibrium strategy for each player necessarily randomizes over actions that use all available resources.

\begin{theorem}\label{all_resources}
\rev{In any equilibrium of $\Gamma(b_1,b_2)$, where
$b_1<\smsc$ and $b_2<\semm$}, \defender must choose an inspection strategy that randomizes over detector positionings of size exactly $b_1$, and \attacker must randomize her attacks over sets of $b_2$ components.
%
%
%
\begin{align}
&\forall ({\sigma^1}^*,{\sigma^2}^*) \in \nash, \ \forall \Def \in \supp({\sigma^1}^*), \ |\Def| =   b_1, \label{res1}\\
&\forall ({\sigma^1}^*,{\sigma^2}^*) \in \nash, \ \forall \att \in \supp({\sigma^2}^*), \ |\att| =   b_2. \label{res2}
\end{align}

Then, the NE of $\Gamma$ can be obtained by solving the following two linear programs:\hypertarget{LPsb}{}
\begin{align*}
\rev{(\overline{\text{LP}_1}) \quad \min_{\sigma^1 \in  \Delta(\overline{\mathcal{A}_1})} \max_{\att \in \overline{\mathcal{A}_2}} {U}(\sigma^1,\att)
\quad \quad \quad \vrule \quad \quad \quad
(\overline{\text{LP}_2}) \quad \max_{\sigma^2 \in  \Delta(\overline{\mathcal{A}_2})} \min_{\Def \in \overline{\mathcal{A}_1}} {U}(\Def,\sigma^2)}
\end{align*}
where $\overline{\mathcal{A}_1} \coloneqq \{\Def \in 2^\nodes \ | \ |\Def| = b_1\}$ and $\overline{\mathcal{A}_2} \coloneqq \{\att \in 2^\edges \ | \ |\att| = b_2\}$.
\end{theorem}

Although it is intuitive that both players \emph{should} use all available resources, this result \rev{shows} that both players \emph{must necessarily} do so. Property~\eqref{res1} is proven by showing that any additional detector can be utilized by \defender to strictly improve her payoff, which holds because the network is ``large'' (captured by the inequality $b_1 < \smsc$). Similarly, property~\eqref{res2} is proven by showing that any additional attack resource can be used by \attacker to strictly improve her payoff. This argument combines the fact that  \defender cannot monitor all network components with a single detector positioning, and that \attacker can spread her attacks \rev{across the network} (since $b_2 < \semm$). In addition, showing \eqref{res2} involves using  the features of the detection function $\detect$, Proposition~\ref{best_set_cover}, and the properties of (\hyperlink{(MSC)}{$\mathcal{I}_{\text{MSC}}$}) and (\hyperlink{(MSP)}{$\mathcal{I}_{\text{MSP}}$}). \rev{Theorem}~\ref{all_resources} also holds when $b_1 < \smsc$ and $b_2 = \semm$. However, counterexamples can be found when $b_1 \geq \smsc$ or $b_2 > \semm$; see Section~\ref{Other Case}.

From \eqref{res1} and \eqref{res2}, we conclude that the NE of the game $\Gamma$ can be obtained by solving smaller linear programs. \rev{Particularly}, the number of variables and constraints can be reduced from 1 + $\sum_{k=0}^{b_1}{|\nodes| \choose k}$ and  $1 +\sum_{l=0}^{b_2}{|\edges| \choose l}$  for \lpo, to $1 +{|\nodes| \choose b_1}$ and $1 +{|\edges| \choose b_2}$ for \lpob; similar reduction applies between \lpt and \lptb.
Although \lpob and \lptb can be used to compute NE for small-sized networks, this approach is still not scalable to large-sized networks. 


\rev{Importantly, when the network is ``large enough'', i.e., when $b_1 < \smsc$ and $b_2 < \semm$,  we can build on Proposition~\ref{best_set_cover} and Theorem~\ref{all_resources} to establish the following result, which connects \ECOP and $\Gamma$:}




\rev{
\begin{proposition}\label{Eq_P}

\rev{For an inspection strategy $\sigma^{1^*} \in \Delta(\mathcal{A}_1)$ that maximizes $\min_{\sigma^2 \in B_2(\sigma^1,b_2)} r(\sigma^1,\sigma^2)$, any best response to $\sigma^{1^*}$ randomizes over attack plans of size $b_2$:}
\begin{align}
    \rev{\forall \sigma^{1^*} \in \argmax_{\sigma^1 \in \Delta(\mathcal{A}_1)}\min_{\sigma^2 \in B_2(\sigma^1,b_2)} r(\sigma^1,\sigma^2), \ \forall \sigma^2 \in B_2(\sigma^{1^*},b_2), \ \forall \att \in \supp(\sigma^2), \quad |\att| = b_2.}\label{Att_BR}
\end{align}

Then, the optimal value of \ECOP is the smallest number of detectors for which the expected detection rate in equilibrium of $\Gamma$ is at least $\alpha$. 
\begin{align*}
b_1^* = \argmin\{b_1 \in \mathbb{N} \ | \ r(\sigma^*) \geq \alpha, \ \forall \sigma^* \in \Sigma(b_1,b_2)\}.
\end{align*}
Furthermore, an equilibrium inspection strategy of $\Gamma(b_1^*,b_2)$ is an optimal inspection strategy of \ECOP.



\end{proposition}}

\rev{
Thus, an optimal solution of \ECOP can be obtained by solving the game $\Gamma$ and determining the expected detection rate in equilibrium.
 Next, we focus on the node bases of inspection strategies in equilibrium, which represent the sets of nodes that are inspected with positive probability by \defender.}  



\begin{proposition}\label{all_edges}
In any NE $(\sigma^{1^*},\sigma^{2^*}) \in \nash$, the node basis $\nbasis{\sigma^{1^*}}$ is a set cover.

Furthermore, both players must necessarily randomize their actions in equilibrium.
\end{proposition}


The proof of this result is based on a best-response argument, and uses the fact that from any inspection strategy that leaves one or more components completely unmonitored, we can construct another  strategy that strictly improves \defender's payoff. This argument is completed by repositioning some detectors and evaluating the resulting change in \defender's payoff, which involves exploiting the submodularity of the detection function $\detect$, \rev{the lower bound on the value of the game ${\Gamma}$ (Proposition~\ref{best_set_cover})}, and the fact that the players must use all resources in equilibrium (\rev{Theorem}~\ref{all_resources}). Interestingly, this result may not hold when $b_2 \geq \semm$: In that case, \attacker may target components that are ``close'' to each other, which can result in \defender leaving some components completely unmonitored to focus on the ones for which targeted attacks are easier to detect; see Section~\ref{Other Case} for  an example.


\rev{
Proposition~\ref{all_edges} provides an important insight for planning a network inspection operation. To position and operate detectors on the network, the defender typically needs to ``prepare'' a subset of locations (nodes).
%
 For example, such locations need a secure connection between the detectors' sensing unit and the infrastructure network, as well as reliable power supply for sensing and transmitting the measurements. The number of distinct locations that need to be prepared can be minimized by finding an equilibrium inspection strategy that has a node basis of minimum size. From Proposition~\ref{all_edges}, we deduce that this number is \emph{at least} $\smsc$.
%

}

\rev{\textbf{Step 3: Properties of the expected detection rate in equilibrium.} 
We now conclude our game-theoretic analysis of $\Gamma$ by focusing on the equilibrium expected  detection rate. In particular, we combine Proposition~\ref{best_set_cover} and Theorem~\ref{all_resources} to obtain the following parametric bounds on the expected detection rate in equilibrium:}
\begin{align}
\rev{\forall \sigma^* \in \Sigma(b_1,b_2), \quad \dfrac{b_1}{\smsc} \leq\sdet{*} \leq \min\left\{\dfrac{b_1}{\semm},1\right\}.}\label{Eq:Bounds_r}
\end{align}

\rev{Firstly, we note that the lower and upper bounds on the expected detection rate in equilibrium are nondecreasing with respect to $b_1$.} The intuition is that the more detectors \defender has, the more attacks she will be able to detect. Secondly, these bounds are also nonincreasing with respect to $\smsc$ and $\semm$. Indeed, as the network size becomes larger, both $\smsc$ and $\semm$ increase because each monitoring set covers a smaller fraction of the network. Thus, it is more difficult for \defender to detect attacks (with the same number of detectors) in larger-sized networks, reducing her detection performance.

\rev{Next, \eqref{Eq:Bounds_r} and Proposition~\ref{best_set_cover} imply that given an MSC $\Def^{min} \in \msc$, the expected detection rate by positioning $b_1$ detectors according to $\sold{\Def^{min}}{b_1}{}$ provides the following detection guarantee:} 
\begin{align}
\rev{\displaystyle\min_{\sigma^2 \in \Delta(\mathcal{A}_2)} r(\sold{\Def^{min}}{b_1}{},\sigma^2) \,= \, \frac{b_1}{\smsc} \,\geq\, \frac{\max\{b_1,\semm\}}{\smsc} r(\sigma^*), \quad \forall \sigma^* \in \nash(b_1,b_2).}\label{Eq:detect_guarantee}
\end{align}
\rev{Property~\eqref{Eq:detect_guarantee} shows that} by using $\sold{\Def^{min}}{b_1}{}$ as inspection strategy, \defender is guaranteed an expected detection rate of at least $\frac{b_1}{\smsc}$, regardless of the attack strategy chosen by \attacker. \rev{Thus, this guarantee applies even if the disruptions are caused by random failures or by an attacker who does not always select a best response strategy.}
%
%
In fact, the relative difference between the expected detection rate in equilibrium and when \defender chooses $\sold{\Def^{min}}{b_1}{}$  is upper bounded by $1 - \frac{\max\{b_1,\semm\}}{\smsc}$; we refer to this bound as the \emph{relative loss of performance}.



We note that when $\smsc$ and $\semm$ become closer to each other (or equivalently, as the duality gap between (\hyperlink{(MSC)}{$\mathcal{I}_{\text{MSC}}$}) and (\hyperlink{(MSP)}{$\mathcal{I}_{\text{MSP}}$}) decreases), \rev{the gaps between the upper and lower bounds in Proposition~\ref{best_set_cover} and~\eqref{Eq:Bounds_r} also become narrower.} When $\smsc = \semm$, the results in  \rev{Proposition~\ref{best_set_cover} and~\eqref{Eq:Bounds_r}} can be tightened as follows: If $\smsc = \semm$ (in addition to $b_1 < \smsc$ and $b_2 < \semm$), \rev{then} $(\sold{\Def^{min}}{b_1}{},\sola{\att^{max}}{b_2}{})$ is a NE \rev{of $\Gamma(b_1,b_2)$}, and the \rev{value of the game $\Gamma$} and the equilibrium expected detection rate are given by:
\begin{align}
\rev{\forall (\sigma^{1^*},\sigma^{2^*}) \in \nash(b_1,b_2),}& \quad
\rev{U ({\sigma^1}^*,{\sigma^2}^*) =   b_2  \left(1 - \dfrac{  b_1  }{\smsc}\right),}\label{eq_payoff}\\
\forall \sigma^* \in \nash(b_1,b_2),& \quad \sdet{*} = \dfrac{b_1}{\smsc}.\label{eq_srate}
\end{align}

\rev{In fact, this} result generalizes the equilibrium characterization of prior results on \rev{a class of} security games. Indeed, our MSC/MSP-based characterization of NE applies to any detection model for which $\smsc = \semm$ holds. In Table~\ref{List_games}, we list some of the classical models \rev{reported in the literature} that fall in this category, and compare their features with those of the game \rev{$\Gamma$}. The table \rev{also} compares the combinatorial objects underlying our equilibrium characterization with their settings. \rev{Thus, our analysis generalizes the equilibrium analysis of these security games when both players have multiple resources.}



\OneAndAHalfSpacedXI

\begin{table}[htbp]\footnotesize
\centering
\caption{\rev{Comparison of security games.}}
\ifarXiv
\begin{tabular}{C{5.9em}|C{6em}|C{6.5em}|C{8.5em}|C{4.50em}|C{5.5em}|C{7.4em}|}
\else
\begin{tabular}{L{5.9em}|L{6em}|L{6.5em}|L{8.5em}|L{4.50em}|L{5.5em}|L{7.4em}|}
\fi
\cline{2-7}  & \multicolumn{3}{c|}{Detection model} & Resources  &\multicolumn{2}{c|}{Combinatorial objects used } \\
 & \multicolumn{3}{c|}{(monitoring locations, components, monitoring sets)}&def./att.&\multicolumn{2}{c|}{for equilibrium characterization}\\
\hline \multicolumn{1}{|C{6em}|}{\rev{$\Gamma(b_1,b_2)$}}  & $\nodes$ & $\edges$ &$\set{i}, \ i \in \nodes$ & $b_1 \geq 1$ $b_2 \geq 1$ & MSC & MSP\\
\hline \multicolumn{1}{|C{5.9em}|}{\cite{karlin2016game}} & street roads & safe-houses &safe-houses located on road $i$ & $b_1 = 1$ $b_2 = 1$ & minimum line cover & maximum matching\\
\hline \multicolumn{1}{|C{5.9em}|}{\cite{patrolling}} & \rev{walk} & \rev{node and start time} & \rev{(node, time) tuples belonging to walk $i$} & \rev{$b_1 = 1$ $b_2 = 1$} & \rev{minimum covering set} & \rev{maximum independent set}\\
\hline \multicolumn{1}{|C{5.9em}|}{\cite{garnaev2000search}} & cable location& channel sections& sections covered from the cable's location $i$ & $b_1 = 1$ $b_2 = 1$ & minimum interval cover & maximum independent infiltration set \\
\hline \multicolumn{1}{|C{5.9em}|}{\cite{mavronicolas}} & network edge & network nodes & end nodes of edge $i$  & $b_1 = 1$ $b_2 \geq 1$ & minimum edge cover & maximum independent set\\
\hline \multicolumn{1}{|C{5.9em}|}{\cite{Garnaev1997}} & \rev{node and time step} & \rev{walk} & \rev{walks containing the (node, time) tuple $i$} & \rev{$b_1 \geq 1$ $b_2 = 1$} & \rev{set of covering calendars} & \rev{set of wait-and-run walks}\\\hline
\end{tabular}
\label{List_games}
\end{table}


%
\ifarXiv
\OneAndAHalfSpacedXI
\else
\DoubleSpacedXI
\fi

\rev{Finally, we observe in \eqref{eq_srate} that when $\smsc = \semm$, the expected detection rate in equilibrium \emph{does not} depend on the attack resources $b_2$. We are able to generalize this result and show that the expected detection rate in equilibrium satisfies an important and rather surprising property:}



\begin{theorem}\label{Constant}
Given \defender's resources $b_1 \in \mathbb{N}$, the expected detection rate in equilibrium is identical in any game $\Gamma(b_1,b_2)$, with $b_2 < \semm$; we denote it as $r^*_{b_1}$.
\begin{align}
&\forall b_1 \in \mathbb{N}, \ \exists \, r^*_{b_1} \in [0,1] \ | \ \forall b_2 < \semm, \ \forall \sigma^* \in \nash(b_1,b_2), \quad  \sdet{*} = r^*_{b_1}. \label{Most Important}
\end{align}

\rev{Furthermore, inspection strategies in equilibrium of the game $\Gamma(b_1,1)$ are also inspection strategies in equilibrium of any game $\Gamma(b_1,b_2)$ with $b_2 < \semm$.}

\end{theorem}

\rev{Property~\eqref{Most Important}} is the result of both game-theoretic and combinatorial aspects of our problem. The proof starts by upper bounding the attack probabilities of each component in equilibrium. This is done by accounting for \attacker's ability to spread her attacks in the network -- which is evaluated by Proposition~\ref{best_set_cover} and MSPs -- and by 
exploiting the submodularity of the detection function $\detect$ with respect to the first variable. These bounds enable us to further characterize \attacker's equilibrium strategies for any game $\Gamma(b_1,b_2$), with $b_2 < \semm$. Specifically, consider an attack strategy $\sigma^{2^*}$ in equilibrium of $\Gamma(b_1,1)$, and let $\rho_{\sigma^{2^*}}(e)$ denote the resulting probability with which each component $e \in \edges$ is targeted. Then, for any $b_2 < \semm$, by applying Farkas' lemma, we show the existence of an attack strategy in equilibrium of $\Gamma(b_1,b_2)$ such that the probability \rev{that}  $e \in \edges$ is targeted is given by $b_2\rho_{\sigma^{2^*}}(e)$.
From the additivity of the detection function $\detect$ with respect to the second variable, we deduce that \rev{the value of the game $\Gamma$}  is linear with respect to $b_2$\rev{, and can be expressed as follows:}  
\begin{align}
\forall b_1 < \smsc, \ \forall b_2 < \semm, \quad \rev{U^*(b_1,b_2) =  (1 - r^*_{b_1})b_2.}
\label{General_Formulation}
\end{align}
Finally, \rev{Theorem}~\ref{all_resources} \rev{implies} that the \rev{equilibrium} expected detection rate is independent of $b_2$.
This whole argument \rev{holds} because the network is large in comparison to \attacker's resources, i.e., $b_2 < \semm$. While Theorem~\ref{Constant} also holds when $b_2 = \semm$, \rev{Section~\ref{Other Case} illustrates a counterexample} when $b_2 > \semm$.

\rev{Note that} for the special case when $\smsc = \semm$, $r^*_{b_1} = \frac{b_1}{\smsc}$ (from \eqref{eq_srate}) and we find again \eqref{eq_payoff} from \eqref{General_Formulation}.  \rev{Other implications} of Theorem~\ref{Constant} \rev{and Proposition~\ref{Eq_P}} are that the optimal value of \ECOP does not depend on $b_2$, and that equilibrium inspection strategies in the game \rev{$\Gamma(b_1^*,1)$ are optimal inspection strategies of \ECOP.}
Therefore, we can solve the problem \ECOP by considering that $b_2 =1$. \rev{Thus, the defender \emph{does not} need to know the actual number of attack resources, \emph{so long as} $b_2 < \semm$.}

This conclusion provides a significant advantage from a computational viewpoint. Recall from \rev{Theorem}~\ref{all_resources} that \rev{equilibrium} inspection strategies of $\Gamma(b_1,b_2)$ are the optimal solutions of \lpob. Now, given $b_1 <\smsc$, and by considering that $b_2 = 1$, the optimal value of \lpob is the \rev{expected \emph{undetection} rate in equilibrium $1-r^*_{b_1}$} (see \eqref{General_Formulation}), and its optimal solutions are inspection strategies in equilibrium of \emph{any} game $\Gamma(b_1,b_2)$ with $b_2 < \semm$.  Thus, \lpob can now be reformulated with ${|\nodes| \choose b_1} + 1$ variables and only $|\edges| + 1$ constraints, and \rev{can be solved using} column generation \citep{doi:10.1287/opre.8.1.101}. \rev{In fact, the additivity of the detection function $\operatorname{F}$ implies that equilibrium attack strategies can also be computed with a second column generation algorithm. In Appendix, we present a procedure for computing NE of the game $\Gamma(b_1,b_2)$ with $b_1< \smsc$ and $b_2 < \semm$ in the general case $\semm \leq \smsc$.} 
%
%
%
%
%
%
%
\rev{Next, we leverage Theorem~\ref{Constant} and the MSC/MSP-based bounds on the expected detection rate in equilibrium (property~\eqref{Eq:Bounds_r}) to derive a scalable solution approach to \ECOP.}






\section{\rev{Solution of} the Network Inspection Problem}\label{sec:Answer}

\rev{In this section, we utilize the equilibrium properties of the game $\Gamma$ to solve the problem \ECOP. Our approach provides an approximate solution to \ECOP based on MSCs and MSPs that can be further improved with a refinement procedure.}

%
%
%
%
%
%

\subsection{\rev{MSC/MSP-Based Solution}}\label{Sec:MSC/MSP}





To motivate our approach, let us \rev{again} consider the special case of $\smsc = \semm$.  From \eqref{eq_srate}, we conclude that the minimum number of detectors that are needed for the expected detection rate to be at least $\alpha$ in equilibrium is $b_1^* = \lceil \alpha \smsc\rceil$. Besides, \rev{for an MSC $\Def^{min}$, Proposition~\ref{best_set_cover} implies that $\sold{\Def^{min}}{b_1^*}{}$ is an equilibrium inspection strategy of $\Gamma(b_1^*,b_2)$. Thus, when $\smsc = \semm$, we know from Proposition~\ref{Eq_P}  that an optimal solution of the network inspection problem \ECOP is  given by $(\lceil \alpha \smsc \rceil,\sold{\Def^{min}}{\lceil \alpha \smsc \rceil}{})$.}

\rev{For the general case $\semm \leq \smsc$, we make the following observations: Firstly, the lower bound on the equilibrium expected detection rate, given in \eqref{Eq:Bounds_r}, ensures the target detection rate $\alpha$ is satisfied with $\lceil \alpha \smsc \rceil$ detectors. Secondly, the upper bound in \eqref{Eq:Bounds_r} implies that \defender needs at least $\lceil \alpha \semm\rceil$ detectors to \rev{meet the target detection performance}. Consequently, the optimal value of \ECOP satisfies $ \lceil \alpha \semm\rceil \leq b_1^* \leq  \lceil \alpha \smsc\rceil$. \rev{Finally, \eqref{Eq:detect_guarantee} ensures that our inspection strategy constructed over an MSC (according to Lemma~\ref{algebra1}) satisfies constraints~\eqref{all_NE}.} 
These observations lead to the following \textbf{MSC/MSP-based solution:} 

\emph{For any MSC $\Def^{min} \in \msc$ and any number of attack resources $b_2 < \semm$, $(\lceil \alpha \smsc \rceil,\sold{\Def^{\text{min}}}{\lceil \alpha \smsc \rceil}{})$ is an approximate solution of \ECOP, with optimality gap given by $\lceil \alpha \smsc \rceil - \lceil \alpha \semm \rceil$.}}

We now summarize the main advantages of \rev{this MSC/MSP-based solution}. Firstly, it reduces to a significant extent the size of the optimization problems that are involved in computing a solution. 
Indeed, \rev{although \lpob can be used to solve \ECOP with $b_2 = 1$,} the number of variables and constraints required is equal to \rev{${|\nodes| \choose b_1} +1$ and $|\edges| +1$}, respectively. On the other hand, the number of variables and constraints of (\hyperlink{(MSC)}{$\mathcal{I}_{\text{MSC}}$}) is only $|\nodes|$ and $|\edges|$, respectively. \rev{Similarly, the number of variables and constraints of (\hyperlink{(MSP)}{$\mathcal{I}_{\text{MSP}}$}) is only $|\edges|$ and $|\nodes|$, respectively.}

Secondly, solving a single instance of (\hyperlink{(MSC)}{$\mathcal{I}_{\text{MSC}}$}) and (\hyperlink{(MSP)}{$\mathcal{I}_{\text{MSP}}$}) enables us to derive a solution to problem \ECOP for any target detection rate $\alpha$. \rev{Furthermore, for any $b_2 < \semm$, the loss in detection performance and the optimality gap associated with our solution can directly be computed from $\smsc$ and $\semm$.}
In contrast, \rev{for a given target detection rate $\alpha$,} computing an optimal solution of \ECOP using \rev{\lpob requires solving it for each value of $b_1$.}

\rev{Thirdly, the MSC-based inspection strategy derived from our approach is desirable from a practical viewpoint. Since $\sigma^1(\Def^{min},b_1)$ is a uniform distribution, it can easily be translated into a schedule that determines how the detectors are positioned in the network and mobilized between the locations (nodes) that have been prepared for the purpose of monitoring. 
Furthermore, the number of distinct locations being monitored by $\sigma^1(\Def^{min},b_1)$, which is represented in our model by the node basis size, is only $\smsc$. From Proposition~\ref{all_edges}, we recall that the node basis size in equilibrium is at least $\smsc$, and this number is optimal when $\smsc = \semm$.
%
%
%
%
%
This suggests that our MSC-based inspection strategy is simple to implement.}

Finally, we note that while the above-mentioned results require computing an MSC and an MSP (both NP-hard problems),  modern mixed-integer optimization solvers can be used to optimally solve them (see Section \ref{sec:app}). For extremely large-sized problems, these solvers may not be able to solve (\hyperlink{(MSC)}{$\mathcal{I}_{\text{MSC}}$}) and (\hyperlink{(MSP)}{$\mathcal{I}_{\text{MSP}}$}) to optimality. Still, we can extend our results based only on the computation of a set cover and a set packing. Given a set cover $\Def^\prime$ and a set packing $\att^\prime$ obtained from a heuristic or greedy algorithm \citep{Chvatal-79,Hifi-97}, we can conclude that \rev{$(\lceil \alpha |\Def^{\prime}|\rceil, \sold{\Def^{\prime}}{\lceil\alpha |\Def^\prime|\rceil}{})$} is an approximate solution of \ECOP.
The associated optimality gap is given by $\lceil\alpha |\Def^\prime|\rceil - \lceil\alpha |\att^\prime|\rceil$,
%
%
which decreases as the size of the set cover decreases and the size of the set packing increases.

\subsection{\rev{Refinement Procedure}}{\label{conclusions}}

\rev{Despite the above-listed advantages of our MSC/MSP-based solution, finding an optimal solution to \ECOP (i.e., an inspection strategy utilizing less number of detectors than $\lceil \alpha \smsc \rceil$) might be desirable. Next, we develop a procedure that iteratively refines the MSC/MSP-based solution proposed in Section~\ref{Sec:MSC/MSP} to provide a stronger performance guarantee, until it reaches optimality of \ECOP. This procedure relies on a column generation algorithm to optimally solve \lpob for $b_2 = 1$, and obtains an inspection strategy in equilibrium of any game $\Gamma(b_1,b_2)$ with $b_2 < \semm$ (Theorem~\ref{Constant}).

%
%
%
}


\rev{Each iteration of the column generation algorithm involves solving a master problem and a subproblem. Essentially, the master problem is a restricted version of \lpob, where only a subset of variables are considered. Once the master problem is solved, the optimal dual variables are used to construct the subproblem, which involves finding the variable in the unrestricted \lpob with lowest reduced cost. Specifically, given a subset $\mathcal{I} \subseteq \overline{\mathcal{A}_1}$ of indices, the master problem is given by:}\hypertarget{Pcg}{}
\begin{align*}
\rev{\begin{array}{cclc}
(M_{\text{CG}}(\mathcal{I})): \quad & \text{minimize} & \displaystyle z&\ifarXiv\mcr\else\\\fi
&\text{subject to} & z + \displaystyle \sum_{\Def \in\mathcal{I}}\valdet{\Def}{e}\sigma^{1}_{\Def}\geq 1, & \forall e \in \edges\\
&&  \displaystyle\sum_{\Def \in \mathcal{I}} \sigma^{1}_\Def = 1&\\
&& \sigma^1_{\Def} \geq 0, & \forall \Def \in \mathcal{I}.
\end{array}}
\end{align*}

Let \rev{$(\sigma^{1^*}, z^*) \in \mathbb{R}_+^{|\overline{\mathcal{A}_1}|} \times \mathbb{R}$ (resp. $(\rho^*, z^{\prime^*}) \in \mathbb{R}_+^{|\edges|} \times \mathbb{R}$)} denote the optimal primal (resp. dual) solution of \pcg. The reduced cost associated with each $\Def \in \overline{\mathcal{A}_1}$ is given by \rev{$-\sum_{e \in \edges}\valdet{\Def}{e}\rho_e^* - z^{\prime^*}$}. Therefore, the detector positioning with the \rev{lowest} reduced cost can be obtained by solving a maximum weighted covering set problem, \rev{where the component weights are the optimal dual variables obtained from the master problem.} \rev{The subproblem can be formulated as the following integer program:}\hypertarget{Dcg}{}
\begin{align*}
\begin{array}{ccll}
(S_{\text{CG}}(\rho^*)): \quad &\text{maximize} & \displaystyle \sum_{e \in \edges} \rho_e^* y_e&\\
&\text{subject to} & y_e \leq \displaystyle \sum_{\{i \in \nodes \, | \, e \in \set{i}\}} x_i, & \ \forall e \in \edges\\
&&  \displaystyle\sum_{i \in \nodes} x_i = b_1&\\
&& x_i, y_e \in \{0,1\}, & \ \forall i \in \nodes, \ \forall e \in \edges.
\end{array}
\end{align*}

If the optimal value of \dcg is no more than \rev{$-z^{\prime^*}$}, then this proves that the optimal primal solution of \pcg, $(\sigma^{1^*}, z^*)$,  is also an optimal solution of \lpob. However, if the optimal value of \dcg is more than \rev{$-z^{\prime^*}$}, then we add the detector positioning corresponding to the optimal solution of  \dcg to the set of indices $\mathcal{I}$.  \rev{The master problem} \pcg is then solved with the new set of indices $\mathcal{I}$. \rev{In fact, this algorithm can be warm-started by considering $\mathcal{I} = \supp(\sold{\Def^{min}}{b_1}{})$, and repeated until an optimal solution of \pcg is found.}

Thus, we finally arrive at the following computational procedure to \emph{exactly} solve  problem \ECOP:

\OneAndAHalfSpacedXI

\begin{algorithm}[H]

\caption*{\rev{\textbf{Refinement Procedure: Optimal Solution of \ECOP}}}
\label{ALG5}
\hypertarget{ALG:Ref}{}
\rev{
\vspace{0.2cm}
\hspace*{\algorithmicindent} \textbf{Input}: Detection model $\mathcal{G} = (\nodes,\edges,\{\set{i}, \ i \in \nodes\})$, target detection rate $\alpha \in [0,1]$, MSC $\Def^{min} \in \msc$ of size $\smsc$, and MSP $\att^{max} \in \emm$ of size $\semm$.

\hspace*{\algorithmicindent} \textbf{Output}: Number of detectors $b_1^*$ and inspection strategy $\sigma^{1^*}$.

\begin{algorithmic}[1]

\StateNew{Run a binary search method in the discrete interval $\llbracket \lceil \alpha \semm \rceil, \lceil \alpha \smsc \rceil\rrbracket$:}

\State{\hspace*{\algorithmicindent} Select $b_1$}
\State{\hspace*{\algorithmicindent} $\mathcal{I} \gets \supp(\sold{\Def^{min}}{b_1}{})$}
\State{\hspace*{\algorithmicindent} Solve \lpob by considering $b_2=1$ using column generation:}\label{alg:CG}
\State{\hspace*{\algorithmicindent} \hspace*{\algorithmicindent} $(\sigma^{1^*},z^*), (\rho^*,z^{\prime^*}) \gets$ optimal primal and dual solutions of \pcg}\label{primal_dual_alg}
\State{\hspace*{\algorithmicindent} \hspace*{\algorithmicindent} $(x^*,y^*) \gets$ optimal solution of \dcg}

\State{\hspace*{\algorithmicindent} \hspace*{\algorithmicindent} \textbf{If $- \sum_{e \in \edges} \rho^*_e y^*_e - z^{\prime^*} < 0$, then}}

\State{\hspace*{\algorithmicindent} \hspace*{\algorithmicindent}  \hspace*{\algorithmicindent}  $\mathcal{I} \gets \mathcal{I} \cup \supp(x^*)$ and go to \aref{primal_dual_alg}}


\State{\hspace*{\algorithmicindent} \hspace*{\algorithmicindent} \textbf{else}}

\State{\hspace*{\algorithmicindent} \hspace*{\algorithmicindent}  \hspace*{\algorithmicindent}  Output $\sigma^{1^*}$ and $r_{b_1}^* = 1-z^*$}

\State{\hspace*{\algorithmicindent} \hspace*{\algorithmicindent} \textbf{end if}}\label{end_CG}

%
%
%

\State{\hspace*{\algorithmicindent} Terminate the binary search with $b_1^* = \argmin \{b_1 \in \llbracket \lceil \alpha \semm \rceil, \lceil \alpha \smsc \rceil\rrbracket\ | \ r^*_{b_1} \geq \alpha\}$}
\end{algorithmic}}

%
%
%
%

\end{algorithm}

\ifarXiv
\OneAndAHalfSpacedXI
\else
\DoubleSpacedXI
\fi

After each iteration of the column generation algorithm \aref{primal_dual_alg}-\aref{end_CG} on \lpob~\rev{for a given $b_1 \in \llbracket \lceil \alpha \semm \rceil, \lceil \alpha \smsc \rceil\rrbracket$}, let $\sigma^{1^\prime}$ and \rev{$1-r^\prime$} respectively denote the current inspection strategy and value of the objective function; note that $r^\prime = \min_{e \in \edges}r(\sigma^{1^\prime},e)$. Then, one can derive performance guarantees for $\sigma^{1^\prime}$ by solving (\hyperlink{(MSP)}{$\mathcal{I}_{\text{MSP}}$}), similarly to~\eqref{Eq:detect_guarantee}. Indeed, given $\semm$, an upper bound on the relative loss in detection performance is given by $\ell^\prime = 1 - \frac{ \max\{b_1,\semm\}}{b_1} r^\prime$. 
\rev{When} the \rev{support of the} MSC-based inspection strategy $\sold{\Def^{min}}{b_1}{}$ is used \rev{to warm-start} the column generation algorithm, the first iteration of the master problem will give $r^\prime = \frac{b_1}{\smsc}$, for which we find again the expression of the loss in detection performance in~\eqref{Eq:detect_guarantee}. Then, \rev{$\ell^\prime$ decreases as the number of iterations of the column generation algorithm increases}. \rev{Note that if $r^\prime \geq \alpha$, then $(b_1,\sigma^{1^\prime})$ is a feasible solution of \ECOP, with optimality gap given by $b_1 - \lceil \alpha \semm\rceil$.} 

\rev{When} \lpob is solved to optimality for a given $b_1$, the optimal dual variables of \lpob  represent the probabilities with which the network components are targeted by \attacker in an equilibrium of the game $\Gamma(b_1,1)$. In \rev{Appendix}, we \rev{derive a procedure that uses} these probabilities to construct an attack strategy in equilibrium of $\Gamma(b_1,b_2)$ for $b_2 < \semm$.

A downside of this refinement procedure is that \rev{it can output a significantly complex inspection strategy; for example, one that  randomizes over $|\edges|$ detector positionings on a node basis of size $b_1 |\edges|$, as opposed to our MSC-based inspection strategy $\sold{\Def^{min}}{b_1}{}$ that uniformly randomizes over $\smsc$ detector positionings on a node basis of size $\smsc$.}
%
%
%
 Thus, scheduling \rev{a network inspection} according to this new strategy \rev{would likely require} a larger level of \rev{preparation and operational capability on the part of the defender. Our approach enables the defender to compute and choose an inspection strategy with a tradeoff between detection performance and ease of implementation.}

\section{Computational Results}\label{sec:app}

%
%
%

In this section, we demonstrate the scalability and performance guarantees of our approach for large-scale networks. We consider a batch of benchmark water distribution networks \rev{varying in their size and complexity} that are \rev{typically} used to test network monitoring algorithms. Table \ref{tab:4} lists the characteristics of the 13 networks considered in our study. The data for these networks can be found in \cite{doi:10.1002/j.1551-8833.2008.tb09659.x,exeter,jolly}. \rev{The water networks in our study range from medium-sized to very large-sized networks serving populations from 3,000 to 250,000 consumers \citep{EPA}.} All network simulations were implemented in Matlab, and all optimization problems were solved using the Gurobi solver on a computer with a 2.3 GHz 8-Core Intel Core i9 processor and 32 GB of RAM.

\rev{We consider an application of problem \ECOP, in which pipelines are subject to attack-induced disruptions}. 
To detect these attacks, we consider that the water utility has access to the relevant \rev{sensing} technology, \rev{such as pressure loggers that can easily be mounted at various nodes (e.g., access points such as valves and hydrants), and shifted from one node to another~\citep{allen:smart,infrasense,XING2019291}}.  
\rev{For this application,} the set of monitoring locations $\nodes$ is given by the set of network nodes, and the set $\edges$ \rev{of critical components} is the set of pipes. Then, for each possible monitoring location $i \in \nodes$, we compute the monitoring set $\set{i}$ (defined in Section~\ref{sec:detect}). In our study, monitoring sets are computed through simulations using a threshold-based detection model,  as proposed in \cite{deshpande2013optimal,sensors:autom}.

\rev{We then apply our solution approach for each network (\rev{see Section~\ref{Sec:Additional_Ex} for an illustrative example}):} We solve (\hyperlink{(MSC)}{$\mathcal{I}_{\text{MSC}}$}) to compute the number of detectors \rev{$\lceil \alpha \smsc \rceil$ that are sufficient to achieve the target detection rate $\alpha$}, and \rev{determine an MSC $\Def^{min}$ that should be prepared by the water utility for inspection. Then, the utility's schedule of inspections can be generated from the inspection strategy $\sold{\Def^{min}}{\lceil \alpha \smsc \rceil}{}$.}
%
%
%
%
%
%
Next, we solve (\hyperlink{(MSP)}{$\mathcal{I}_{\text{MSP}}$}), which enables us to evaluate the performance of our solution, i.e., we compute the optimality gap \rev{$\lceil \alpha \smsc \rceil - \lceil \alpha \semm \rceil$} given in \rev{Section}~\ref{Sec:MSC/MSP} and the relative loss of performance \rev{$1 - \frac{\max\{\lceil\alpha \smsc \rceil,\semm\}}{\smsc}$} derived from \eqref{Eq:detect_guarantee}. The computational results are summarized in Table~\ref{tab:4}.

\begin{table}[h]\footnotesize
  \centering
  \caption{Network data and computational results \rev{of the MSC/MSP-based solution}, $\alpha = 0.75$.}
\tabcolsep=0.09cm
	\renewcommand{\arraystretch}{\spacingcomput} 
	\ifarXiv
	\begin{tabular}{|c|clc|clclc|c|c|c|}
	\else
	\begin{tabular}{|l|lll|ll|ll|l|l|l|}
	\fi
\hline
   	 \multirow{ 2}{*} {\textbf{Network}} &  \textbf{Total length} &  \textbf{No. of}  &  \textbf{No. of} &  
 \textbf{Running}& \textbf{time [s]}&\multirow{ 2}{*} {$\mathbf{m^*}$}& \multirow{ 2}{*} {$\mathbf{n^*}$}& \multirow{ 2}{*} {$\mathbf{\boldsymbol{\lceil} \boldsymbol{\alpha} \smsc \boldsymbol{\rceil}}$} & \textbf{Optimality} & \textbf{Relative loss}\\
    &  \textbf{[km]} & \textbf{pipes}   & \textbf{nodes}&\textbf{(}\hyperlink{(MSP)}{$\boldsymbol{\mathcal{I}}_{\textbf{MSP}}$}\textbf{)}& \textbf{(}\hyperlink{(MSC)}{$\boldsymbol{\mathcal{I}}_{\textbf{MSC}}$}\textbf{)}&& && \textbf{gap}& \textbf{of performance}\\
\hline
   \textbf{bwsn1}   & 37.56 & 168   & 126 &  0.05 & 0.11 & 7 &7 & 6 & 0\,\% & 0\,\% \\
    \textbf{ky3}  & 91.29 &  366   & 269 &  0.01 & 0.03 & 15 & 15 & 12 & 0\,\% & 0\,\% \\
    \textbf{ky5}  & 96.58 &  496   & 420 &  0.02 & 0.05 & 18 & 19 & 15& 1 (7.14\,\%)& 5.3\,\% \\
    \textbf{ky7}  & 137.05 &  603   & 481 &  0.09 & 0.08 & 28 & 28  & 21 & 0\,\% & 0\,\%\\
    \textbf{ky6}  & 123.20 &644   & 543 &   0.08 & 0.06 & 24 & 24 & 18 & 0\,\% & 0\,\%\\
    \textbf{ky1}  & 166.60 & 907   & 791 & 0.03 & 0.08 & 31 & 31 & 24 & 0\,\% &  0\,\%\\
    \textbf{ky13}  & 153.30 & 940   & 778 &  0.06 & 0.08 & 28 & 30 & 23 & 2 (9.52\,\%)& 6.7\,\%\\
    \textbf{ky2}  & 152.25 & 1124  & 811 &  0.39 & 0.41 & 18 & 19 & 15& 1 (7.14\,\%)& 5.3\,\%\\
    \textbf{ky4}  & 260.24 & 1156  & 959 & 0.03 & 0.05 & 62 & 64 &  48 & 1 (2.13\,\%) &3.1\,\%\\
    \textbf{ky8} & 247.34 & 1614  & 1325 &  0.14 & 0.22 & 45 & 45 & 34 & 0\,\% & 0\,\%\\
    \textbf{dover} & 779.86 &  16000 & 14965  &  4.34 & 8.36 & 119 & 121 & 91 & 1 (1.11\,\%)& 1.7\,\%\\
    \textbf{bswn2} & 1,844.04 &  14822 & 12523  & 0.77 & 4.06 &  352 & 361 & 271 & 7 (2.65\,\%)& 2.5\,\%\\
    \textbf{mnsr} & 476.67 &  25484 & 24681  & 58.89 & 68.67 & 50 & 52 & 39 & 1 (2.63\,\%) & 3.8\,\%\\
\hline
    \end{tabular}
  \label{tab:4}%
\end{table}%

We note that the sizes of MSCs and MSPs are equal for 6 out of the 13 networks. Thus, for these 6 networks, our \rev{MSC/MSP-based solution is optimal for \ECOP}.
For the remaining 7 networks, we note that the relative difference between $\smsc$ and $\semm$ is small, which implies that our estimate of the optimal value of (\hyperlink{(P)}{$\mathcal{P}$}), $\lceil \alpha \smsc \rceil$, is close to the optimal value $b_1^*$. Additionally, we can see from Table~\ref{tab:4} that the loss in detection performance by choosing $\sold{\Def^{min}}{\lceil \alpha \smsc \rceil}{}$ in comparison to the \rev{optimal} performance  is \rev{also} small ($2.7\,\%$ on average over all networks). 


\rev{Furthermore}, the time to solve  (\hyperlink{(MSC)}{$\mathcal{I}_{\text{MSC}}$}) and (\hyperlink{(MSP)}{$\mathcal{I}_{\text{MSP}}$}) is fairly small. For networks with less than 1500 nodes and components, Gurobi computes an optimal solution in less than half a second, which directly enables us to construct an \rev{approximate solution to \ECOP}. 
For larger networks, we can obtain $\smsc$ and $\semm$ in about a minute. Thus, our \rev{MSC/MSP-based solution} is scalable to large-scale networks.

\rev{Next, we run the refinement procedure presented in Section~\ref{conclusions} to improve our solution for the 7 networks for which $\semm < \smsc$. Table~\ref{tab:CG} summarizes the computational results for the 4 networks for which the procedure terminated in a reasonable time.}

\rev{
\begin{table}[h]\small
  \centering
  \caption{\rev{Computational results of the refinement procedure, $\alpha = 0.75$.}}
\tabcolsep=0.09cm
	\renewcommand{\arraystretch}{\spacingcomput} 
	\rev{
    \ifarXiv
     \begin{tabular}{|c|C{5em}|C{2em}|C{3em}|C{6em}|C{7em}|c|c|}
    \else
    \begin{tabular}{|l|L{5em}|L{2em}|L{3em}|L{6em}|L{7em}|l|l|}
    \fi
\hline
   	 \multirow{ 4}{*} {\textbf{Network}} & \multirow{ 2}{*} {\textbf{Running}} &\multirow{ 4}{*} {$\mathbf{b_1^*}$}  &  \multirow{ 4}{*} {$\mathbf{r_{b_1^*}^*}$}  &  
\multicolumn{2}{c|}{\textbf{Improvement}} &  \multirow{ 4}{*} {$\mathbf{\boldsymbol{|}\boldsymbol{\supp}\boldsymbol{(}\boldsymbol{\sigma}^{1^*}\boldsymbol{)}\boldsymbol{|}}$} &  \multirow{ 4}{*} {$\mathbf{\boldsymbol{|}\boldsymbol{\nodes}_{\boldsymbol{\sigma}^{1^*}}\boldsymbol{|}}$} \\
   \cline{5-6}  & &   & &  \multirow{ 2}{*} {\textbf{\# detectors}}  & \textbf{detection}  & &\\
&  \multirow{ 2}{*} {\textbf{time [s]}}  &&&&\textbf{performance} &&
\\ 
&&&& $\mathbf{\boldsymbol{\lceil} \boldsymbol{\alpha} \smsc \boldsymbol{\rceil} - b_1^*}$ &  $\mathbf{r^*_{b_1^*}- b_1^*/\smsc}$&&
\\[0.05cm] \hline
    \textbf{ky5}  & 22.74 &  14   & 0.75 &  1 & 0.0132 (1.79\,\%) & 27	& 94 \\
    \textbf{ky13}  & 643.86 & 22   & 0.7582 &  1 & 0.0248 (3.39\,\%) & 47	& 205 \\
    \textbf{ky2}  & 153.11 & 14  & 0.75 &  1 & 0.0132 (1.79\,\%) & 39	 & 133\\
    \textbf{ky4}  & 2901.90 & 47  & 0.7510 & 1 & 0.0165 (2.26\,\%) & 73 & 311\\
\hline
    \end{tabular}}
  \label{tab:CG}%
\end{table}%
}






\rev{For instance, we find that for network ky13, an optimal solution of \ECOP requires 1 fewer detector than the MSC/MSP-based solution does to satisfy the target detection performance $\alpha$.
%
Furthermore the equilibrium expected detection rate $r^*_{b_1^*}$ improves by 3.39\,\% the detection performance of the MSC-based inspection strategy $\sold{\Def^{min}}{b_1^*}{}$. Finally, we find that the optimal inspection strategy $\sigma^{1^*}$ has a support of size 47, and randomizes over 205 distinct locations.}

\rev{Table~\ref{tab:CG} shows that for 4 out the remaining 7 networks, the refinement procedure optimally solves \ECOP. First, we observe that the optimal solutions require only one fewer detector than the MSC-based solution. Secondly, given $b_1^*$ detectors, the improvement between the optimal and MSC-based inspection strategies is between $1.79\,\%$ and $3.79\,\%$, and is achieved under 50 minutes. We note that solving (\hyperlink{(MSP)}{$\mathcal{I}_{\text{MSP}}$}) significantly reduces the runtime of the refinement procedure by limiting the binary search to the interval $\llbracket \lceil \alpha \semm \rceil,\lceil \alpha \smsc \rceil \rrbracket$: \lpob is solved for only one value of $b_1$ for networks ky5, ky2, and ky4, and is solved for two values of $b_1$ for network ky13.


%
%

However, for the 3 larger networks (dover, bswn2, mnsr), the refinement procedure did not terminate after 72 hours of runtime. One of the main reasons is that the restricted master problem \pcg faces degeneracy issues when the subset of variables $\mathcal{I}$ is small: Even if the subproblem \dcg finds a variable with negative reduced cost, adding that variable to $\mathcal{I}$ does not change the new optimal solution of \pcg. On the other hand, when the subset of variables $\mathcal{I}$ is large, the runtime of one iteration of the column generation algorithm \aref{primal_dual_alg}-\aref{end_CG} is large.


In Figure~\ref{fig:net9_ky4}, we compare the column generation algorithm applied to \lpob for the ky4 network with and without the MSC-based warm-start. Additional figures are presented in Section~\ref{Add:Figures}. 
}

\begin{figure}[htbp]
        \centering
        \begin{subfigure}[t]{0.45\textwidth} \centering
              \begin{tikzpicture}[scale=0.93]
\begin{semilogxaxis}[
  grid=major, 
  grid style={dashed,gray!30}, 
  xlabel=Runtime {[s]},
  ylabel= Worst-case detection rate: \ $\min r(\sigma^1{,}\cdot)$,
   font=\small,
  xmin=10,
  xmax=10000,
  ymin=0,
  ymax=1,
 ytick={0,0.25, 0.5,0.75,1},
  legend pos=south east,
	legend cell align={left},
  legend style={draw=none, font = \footnotesize},
]

  \addplot[color = red, very thick] table[x=Time2, y=Objective2, col sep=comma, comment chars={\%}] {Net9_ky4_b1_47.csv};

 \addplot[color = blue, very thick] table[x=Time, y=Objective, col sep=comma, comment chars={\%}] {Net9_ky4_b1_47.csv};

\legend{No warm-start, MSC warm-start}
 
\end{semilogxaxis}

\end{tikzpicture}

        \end{subfigure}
        \quad \
         \begin{subfigure}[t]{0.45\textwidth} \centering
              \begin{tikzpicture}[scale=0.93]
\begin{semilogxaxis}[
  grid=major, 
  grid style={dashed,gray!30}, 
  xlabel= Runtime {[s]},
  ylabel= Node basis size: \ $|\nodes_{\sigma^1}|$,
   font=\small,
  xmin=10,
  xmax=10000,
  ymin=0,
  ymax=750,
 ytick={0,150,300,450,600,750},
  legend pos=south east,
	legend cell align={left},
  legend style={draw=none, font = \footnotesize},
]

  \addplot[color = red, very thick] table[x=Time2, y=Node_Basis_Size2, col sep=comma, comment chars={\%}] {Net9_ky4_b1_47.csv};

 \addplot[color = blue, very thick] table[x=Time, y=Node_Basis_Size, col sep=comma, comment chars={\%}] {Net9_ky4_b1_47.csv};

\legend{No warm-start, MSC warm-start}
 
\end{semilogxaxis}

\end{tikzpicture}

        \end{subfigure}

       \caption{\rev{Results of column generation applied to  ($\overline{\text{LP}_1}$)  for the ky4 network ($b_1 = 47$).}}   \label{fig:net9_ky4}
\end{figure}

\rev{Figure~\ref{fig:net9_ky4} shows that for this network, initiating the column generation algorithm \aref{alg:CG} with the variables corresponding to the detector positionings in the support of the MSC-based strategy reduces the runtime by half. Interestingly, we observe a peak in the size of the node basis, i.e., the number of distinct monitored locations, of the solution $\sigma^1$ to the restricted master problem \pcg. In the first iterations, \attacker's best response (given by the dual variables of \pcg) targets parts of the network that are not monitored by the inspection strategy. Thus, the algorithm first ``explores'' the network and positions the detectors in a greedy-like manner on locations that are more spread out. As the inspection strategy improves, \attacker selects attack strategies that are more evenly spread in the network. This in turn forces the algorithm to consolidate the support of the inspection strategy and position detectors on more strategic locations.}



\rev{
Finally, we note that MSC-based strategies are significantly simpler than the optimal inspection strategies. For instance, for the ky4 network, the MSC-based strategy randomizes over $\smsc = 64$ different locations, while the optimal strategy randomizes over 311 different locations. Similarly, for the ky2 network, the MSC-based strategy has a support of size  $\smsc = 19$, while the support of the optimal strategy is of size 39.}

\rev{In conclusion, our computational results show a tradeoff between the optimal and MSC-based strategies. Specifically, the optimal strategies only provide a marginal improvement in terms of number of utilized detectors and detection performance. On the other hand, implementing the optimal strategies would require a much higher level of effort in preparing the detectors' locations and in scheduling the inspections.
Thus, depending on her operational capabilities, the defender can decide to implement a simple MSC-based strategy with good performance guarantees, or a more complex optimal strategy. %
}

%
%

\section{\rev{Final} Remarks}{\label{Real_conclusions}}


\rev{In this article, we studied a generic yet practically relevant} formulation of \rev{a large-scale bilevel optimization problem for} strategic network inspection. In this problem, the defender seeks \rev{a randomized inspection strategy that utilizes minimum number of detectors}, while ensuring that the expected detection performance against \rev{worst-case} attack plans is above a \rev{desirable} threshold. We developed a novel approach \rev{that analyzes the equilibria of a zero-sum game, which} enables us to solve the inspection problem for large-scale networks along with performance guarantees.

Our \rev{equilibrium analysis} involves: $(i)$ deriving useful qualitative properties satisfied by all NE of the zero-sum game; \rev{$(ii)$ obtaining bounds on the expected detection rate in equilibrium based on solutions of the MSC and MSP problems; and $(iii)$ showing that, in equilibrium, the expected detection rate and inspection strategies are independent of the attack resources.} 

\rev{Our equilibrium analysis leads to a tractable approach to solve the inspection problem: First, the MSC and MSP problems are solved to obtain an approximate solution that estimates the required number of detectors (with optimality gap), and provides an inspection strategy with guarantees on the expected detection performance. Then, a column generation-based procedure further improves the guarantees of our solution. We demonstrated the scalability and performance of our approach for the allocation of sensing resources in large-scale urban water networks facing security attacks. Our results highlight an important tradeoff between the optimal and MSC/MSP-based solutions in terms of performance guarantees and ease of implementation.}

A future research question is to solve the inspection problem under a more refined detection model that accounts for imperfect detection of attacks (and other types of compromises). Typically, the diagnostic ability of sensing technology can be represented by a probabilistic detection rate for any given false alarm rate. In fact, \rev{as mentioned in Section~\ref{sec:detect},} the guarantees provided by our approach can be extended (via simple scaling) to the case when the detection probability is a priori known and homogeneous \rev{for} all detectors. The general case of heterogeneous detection rates can be addressed by extending our detection model; in particular, by adding a weight to each inspected node to represent the probability of detecting an attack within the node's monitoring set. 

\rev{Another research question is to extend our solution approach} to account for the heterogeneity of network components in terms of their criticality to the overall network functionality. In principle, this case can be addressed by adding weights to the \rev{detection function}. However, in many practical situations, the defender can only qualitatively distinguish the criticality of various components (high versus low). In such cases, our approach for strategic network inspection can be applied to each group of components with homogeneous criticality levels, and the inspection strategies for individual groups can be then integrated based on the defender's operational constraints.

\rev{Overall, the outcomes of the proposed approach can be used to inform and guide public utilities to design inspection strategies for protecting critical infrastructure against intentional threats. The results indicate that a small number of defense resources, if allocated in a strategic manner, can be sufficient to achieve a high level of protection in large-scale networks, which is especially appealing for budget-constrained utilities. With advances in sensing and detection technologies,  randomized inspection strategies, such as the ones proposed in this work, are expected to be critical for reducing risks and for building greater resilience in critical infrastructure systems.}



%
%
%

%

%


 \begin{APPENDIX}{\rev{Column Generation and Nash Equilibria of $\Gamma$}}\label{Sec:Important}

%

\ifarXiv
\OneAndAHalfSpacedXI
\else
\DoubleSpacedXI
\fi

\rev{In Section~\ref{conclusions}, we presented a column generation algorithm to solve the linear program \lpob for $b_2 = 1$ and obtain an inspection strategy in equilibrium of any game $\Gamma(b_1,b_2)$ with $b_2 < \semm$. Next, we discuss how a second column generation algorithm can be applied to derive attack strategies in equilibrium of any game $\Gamma(b_1,b_2)$ with $b_2 < \semm$.}

\rev{Consider \lpob for $b_2 = 1$, and assume that the column generation algorithm \aref{alg:CG} finds an optimal solution. Then, the optimal dual variables $(\rho_e^*)_{e \in \edges}$ of \pcg represent} the probabilities with which each component can be targeted in equilibrium of the game $\Gamma(b_1,1)$. In the proof of Theorem~\ref{Constant}, we show how to reallocate these probabilities to create an attack strategy in equilibrium of $\Gamma(b_1,1)$ with the additional property that each component is not targeted with probability more than $\frac{1}{\semm}$. Then, given $b_2 < \semm$, Lemma~\ref{Big Support} \rev{and the proof of Theorem~\ref{Constant}} show that an attack strategy in equilibrium of $\Gamma(b_1,b_2)$ can be computed by solving the following feasibility problem: \rev{Find $\sigma^2 \in \mathbb{R}_+^{|\overline{\mathcal{A}_2}|}$ such that $\sum_{\{\att \in \overline{\mathcal{A}_2} \, | \, e \in \att\}} \sigma^2_\att = b_2 \rho^*_e$ for every $e \in \edges$.}
%
%
This can be done by considering the following auxiliary \rev{linear} problem:\hypertarget{(F)}{}
\begin{align*}
\begin{array}{rll}
(\mathcal{F}(b_2\rho^*)): \quad  \underset{\sigma^2,s}{\text{minimize}} & \displaystyle \sum_{e \in \edges} s_e &\\
\text{subject to} & \displaystyle\sum_{\{\att \in \overline{\mathcal{A}_2} \, | \, e \in \att\}} \sigma^2_\att + s_e = b_2 \rho^*_e,  & \ \forall e \in \edges\\
& \sigma^2 \geq \bs{0}_{|\overline{\mathcal{A}_2}|}, \ s \geq \bs{0}_{|\edges|}. & \end{array}
\end{align*}

Problem \Fea can also be solved using column generation, with $(\sigma^2,s) = (\bs{0}_{|\overline{\mathcal{A}_2}|},b_2\rho^*)$ as initial feasible solution. Given the current restricted master problem generated by the column generation algorithm, let $\beta^* \in \mathbb{R}^{|\edges|}$ denote its optimal dual variables. Then, the index ${\att^*} \in \overline{\mathcal{A}_2}$ with lowest reduced cost is given by $\att^* \in \argmax_{\att \in \overline{\mathcal{A}_2}}\sum_{e \in \att} \beta^*_e$. \rev{Therefore, $\att^*$ can be efficiently computed by simply finding the $b_2$ highest values of $\beta_{e}^*$. }
%
%
\rev{Lemma~\ref{Big Support} guarantees that the optimal value of \Fea is 0, and an optimal solution is an equilibrium attack strategy of $\Gamma(b_1,b_2)$.} 

\rev{In conclusion, we obtain the following procedure for computing NE of the game $\Gamma(b_1,b_2)$ in the general case $\semm \leq \smsc$:}

\OneAndAHalfSpacedXI

\begin{algorithm}[H]

\caption*{\rev{\textbf{Procedure: NE of $\Gamma(b_1,b_2)$}}}
\label{ALG6}

\rev{
\vspace{0.2cm}
\quad \textbf{Input}: Detection model $\mathcal{G} = (\nodes,\edges,\{\set{i}, \ i \in \nodes\})$, and players' resources $b_1 < \smsc$ and $b_2 < \semm$.

\quad \textbf{Output}: Strategy profile $(\sigma^{1^*},\sigma^{2^*}) \in \Delta(\overline{\mathcal{A}_1})\times \Delta(\overline{\mathcal{A}_2})$.

\begin{algorithmic}


\State{Solve \lpob by considering $b_2=1$ using column generation:}
\State{\hspace*{\algorithmicindent} \hspace*{\algorithmicindent} $(\sigma^{1^*},z^*), (\rho^*,z^{\prime^*}) \gets$ optimal primal and dual solutions of \lpob}

\State{Reallocate probabilities in $\rho^*$ so that $\rho^*_{e} \leq \frac{1}{\semm}$ for every $e \in \edges$}

\State{Solve \Fea using column generation:}
\State{\hspace*{\algorithmicindent} \hspace*{\algorithmicindent} $(\sigma^{2^*},\mathbf{0}_{|\overline{\mathcal{A}_2}|})\gets$ optimal primal solution of \Fea}

%
%
%
%
%

%
%
%

\end{algorithmic}}

%
%
%
%

\end{algorithm}

\ifarXiv
\OneAndAHalfSpacedXI
\else
\DoubleSpacedXI
\fi

 \end{APPENDIX}


%
%

\ifarXiv

\else
\vspace{-0.75cm}
\fi

\ACKNOWLEDGMENT{This work was supported by NSF grant CNS 1239054, NSF CAREER award CNS 1453126, the UT Austin New Faculty Start Up Grant, and MIT Schoettler Fellowship. We are  grateful to \"Ozlem Ergun\rev{, the Associate Editor,} and the reviewers who handled the first round of submission, for useful suggestions. We are also grateful to Ali Jadbabaie, Patrick Jaillet, Asuman E. Ozdaglar,  Georgia Perakis, and Zuo-Jun Max Shen for their feedback.}



\OneAndAHalfSpacedXI

\rev{
\bibliographystyle{informs2014} 
\bibliography{References.bib} 

\begin{thebibliography}{61}
\providecommand{\natexlab}[1]{#1}
\providecommand{\url}[1]{\texttt{#1}}
\providecommand{\urlprefix}{URL }

\bibitem[{Alderson et~al.(2015)Alderson, Brown, \protect\BIBand{}
  Carlyle}]{doi:10.1111/risa.12333}
Alderson DL, Brown GG, Carlyle WM (2015) Operational models of infrastructure
  resilience. \emph{Risk Anal.} 35(4):562--586.

\bibitem[{Alderson et~al.(2018)Alderson, Brown, Carlyle, \protect\BIBand{}
  Wood}]{doi:10.1287/trsc.2017.0749}
Alderson DL, Brown GG, Carlyle WM, Wood RK (2018) Assessing and improving the
  operational resilience of a large highway infrastructure system to worst-case
  losses. \emph{Transportation Sci.} 52(4):1012--1034.

\bibitem[{Allen et~al.(2011)Allen, Preis, Iqbal, Stitangarajan, Lim, Girod,
  \protect\BIBand{} Whittle}]{allen:smart}
Allen M, Preis A, Iqbal M, Stitangarajan S, Lim HN, Girod L, Whittle AJ (2011)
  Real time in-network monitoring to improve operational efficiently. \emph{J.
  Amer. Water Works Assoc.} 103(7):63--75.

\bibitem[{Alpern et~al.(2011)Alpern, Morton, \protect\BIBand{}
  Papadaki}]{patrolling}
Alpern S, Morton A, Papadaki K (2011) Patrolling games. \emph{Oper. Res.}
  59(5):1246--1257.

\bibitem[{Barrett(2018)}]{NIST}
Barrett MP (2018) \emph{Framework for Improving Critical Infrastructure
  Cybersecurity} (National Institute of Standards and Technology, Gaithersburg,
  MD).

\bibitem[{Baykal-G{\"{u}}rsoy et~al.(2014)Baykal-G{\"{u}}rsoy, Duan, Poor,
  \protect\BIBand{} Garnaev}]{BaykalGursoy2014469}
Baykal-G{\"{u}}rsoy M, Duan Z, Poor HV, Garnaev A (2014) Infrastructure
  security games. \emph{Eur. J. Oper. Res.} 239(2):469--478.

\bibitem[{Berry et~al.(2006)Berry, Hart, Phillips, Uber, \protect\BIBand{}
  Watson}]{berry}
Berry J, Hart W, Phillips C, Uber J, Watson J (2006) Sensor placement in
  municipal water networks with temporal integer programming models. \emph{J.
  Water Res. Plan. Man.} 132(4):218--224.

\bibitem[{Bertsimas et~al.(2016)Bertsimas, Nasrabadi, \protect\BIBand{}
  Orlin}]{BERTSIMAS2016114}
Bertsimas D, Nasrabadi E, Orlin JB (2016) On the power of randomization in
  network interdiction. \emph{Oper. Res. Lett.} 44(1):114--120.

\bibitem[{Bier \protect\BIBand{} Haphuriwat(2011)}]{Bier_2011}
Bier VM, Haphuriwat N (2011) Analytical method to identify the number of
  containers to inspect at {U.S.} ports to deter terrorist attacks. \emph{Ann.
  Oper. Res.} 187(1):137--158.

\bibitem[{Bier et~al.(2008)Bier, Haphuriwat, Menoyo, Zimmerman,
  \protect\BIBand{} Culpen}]{Bier_2008}
Bier VM, Haphuriwat N, Menoyo J, Zimmerman R, Culpen AM (2008) Optimal resource
  allocation for defense of targets based on differing measures of
  attractiveness. \emph{Risk Anal.} 28(3):763--770.

\bibitem[{Brown et~al.(2006)Brown, Carlyle, Salmerón, \protect\BIBand{}
  Wood}]{doi:10.1287/inte.1060.0252}
Brown G, Carlyle M, Salmerón J, Wood K (2006) Defending critical
  infrastructure. \emph{INFORMS J. Applied Analytics} 36(6):530--544.

\bibitem[{Chakrabarti et~al.(2009)Chakrabarti, Kyriakides, \protect\BIBand{}
  Eliades}]{4729806}
Chakrabarti S, Kyriakides E, Eliades D (2009) Placement of synchronized
  measurements for power system observability. \emph{IEEE Trans. Power
  Delivery} 24(1):12--19.

\bibitem[{Chong \protect\BIBand{} Kumar(2003)}]{1219475}
Chong CY, Kumar SP (2003) Sensor networks: evolution, opportunities, and
  challenges. \emph{Proc. IEEE} 91(8):1247--1256.

\bibitem[{Chvatal(1979)}]{Chvatal-79}
Chvatal V (1979) A greedy heuristic for the set-covering problem. \emph{Math.
  Oper. Res.} 4(3):233--235.

\bibitem[{Cormican et~al.(1998)Cormican, Morton, \protect\BIBand{}
  Wood}]{doi:10.1287/opre.46.2.184}
Cormican KJ, Morton DP, Wood RK (1998) Stochastic network interdiction.
  \emph{Oper. Res.} 46(2):184--197.

\bibitem[{Dahan et~al.(2016)Dahan, Perelman, \protect\BIBand{} Amin}]{7852316}
Dahan M, Perelman LS, Amin S (2016) Network sensing for security against link
  disruption attacks. \emph{Proc. 54th Allerton Conf. Comm., Control, Comput.
  \emph{(IEEE, Piscataway, NJ)}}, 808--815.

\bibitem[{Dancy \protect\BIBand{} Dancy(2017)}]{Dancy2017}
Dancy JR, Dancy VA (2017) Terrorism and oil \& gas pipeline infrastructure:
  Vulnerability and potential liability for cybersecurity attacks. \emph{Oil
  and Gas, Nat. Resources \& Energy J.} 2(6):579--619.

\bibitem[{Dantzig \protect\BIBand{} Wolfe(1960)}]{doi:10.1287/opre.8.1.101}
Dantzig GB, Wolfe P (1960) Decomposition principle for linear programs.
  \emph{Oper. Res.} 8(1):101--111.

\bibitem[{Deshpande et~al.(2013)Deshpande, Sarma, Youcef-Toumi,
  \protect\BIBand{} Mekid}]{deshpande2013optimal}
Deshpande A, Sarma SE, Youcef-Toumi K, Mekid S (2013) Optimal coverage of an
  infrastructure network using sensors with distance-decaying sensing quality.
  \emph{Automatica} 49(11):3351--3358.

\bibitem[{Fujishige(2005)}]{Fujishige2005}
Fujishige S (2005) \emph{Submodular Functions and Optimization, \emph{volume 58
  - 2nd ed.}} (Elsevier, Amsterdam).

\bibitem[{Gal \protect\BIBand{} Casas(2014)}]{Gal20140062}
Gal S, Casas J (2014) Succession of hide{\textendash}seek and
  pursuit{\textendash}evasion at heterogeneous locations. \emph{J. Roy. Soc.
  Interface} 11(94):20140062.

\bibitem[{Garnaev(2000)}]{garnaev2000search}
Garnaev A (2000) \emph{Search Games and Other Applications of Game Theory}.
  Lecture Notes in Economics and Mathematical Systems (Springer, Berlin
  Heidelberg).

\bibitem[{Garnaev et~al.(1997)Garnaev, Garnaeva, \protect\BIBand{}
  Goutal}]{Garnaev1997}
Garnaev A, Garnaeva G, Goutal P (1997) On the infiltration game. \emph{Int. J.
  Game Theory} 26(2):215--221.

\bibitem[{Giustolisi et~al.(2008)Giustolisi, Savic, \protect\BIBand{}
  Kapelan}]{apulian}
Giustolisi O, Savic D, Kapelan Z (2008) Pressure-driven demand and leakage
  simulation for water distribution networks. \emph{J. Hydraul. Engrg.}
  134(5):626--635.

\bibitem[{Goyal \protect\BIBand{} Vigier(2014)}]{Goyal:2014aa}
Goyal S, Vigier A (2014) Attack, defense, and contagion in networks. \emph{Rev.
  Econ. Stud.} 81(4):1518--1542.

\bibitem[{Hansen et~al.(1992)Hansen, Jaumard, \protect\BIBand{}
  Savard}]{doi:10.1137/0913069}
Hansen P, Jaumard B, Savard G (1992) New branch-and-bound rules for linear
  bilevel programming. \emph{SIAM J. Sci. Stat. Comput.} 13(5):1194--1217.

\bibitem[{Hassanzadeh et~al.(2020)Hassanzadeh, Rasekh, Galelli, Aghashahi,
  Taormina, Ostfeld, \protect\BIBand{} Banks}]{cyberphysicalWDS2020}
Hassanzadeh A, Rasekh A, Galelli S, Aghashahi M, Taormina R, Ostfeld A, Banks
  MK (2020) A review of cybersecurity incidents in the water sector. \emph{J.
  Environ. Engrg.} 146(5):03120003.

\bibitem[{Hifi(1997)}]{Hifi-97}
Hifi M (1997) A genetic algorithm-based heuristic for solving the weighted
  maximum independent set and some equivalent problems. \emph{J. Oper. Res.
  Soc.} 48(6):612--622.

\bibitem[{Hochbaum \protect\BIBand{} Fishbain(2011)}]{Hochbaum2011}
Hochbaum DS, Fishbain B (2011) Nuclear threat detection with mobile distributed
  sensor networks. \emph{Ann. Oper. Res.} 187(1):45--63.

\bibitem[{Jolly et~al.(2014)Jolly, Lothes, Bryson, \protect\BIBand{}
  Ormsbee}]{jolly}
Jolly MD, Lothes AD, Bryson S, Ormsbee L (2014) Research database of water
  distribution system models. \emph{J. Water Res. Plan. Man.} 140(4):410--416.

\bibitem[{Karlin \protect\BIBand{} Peres(2016)}]{karlin2016game}
Karlin A, Peres Y (2016) \emph{Game Theory, Alive} (AMS, Providence, RI).

\bibitem[{Krause et~al.(2008{\natexlab{a}})Krause, McMahan, Guestrin,
  \protect\BIBand{} Gupta}]{krause+al:jmlr08}
Krause A, McMahan B, Guestrin C, Gupta A (2008{\natexlab{a}}) Robust submodular
  observation selection. \emph{J. Mach. Lear. Res.} 9:2761--2801.

\bibitem[{Krause et~al.(2008{\natexlab{b}})Krause, Singh, \protect\BIBand{}
  Guestrin}]{Krause:2008:NSP:1390681.1390689}
Krause A, Singh A, Guestrin C (2008{\natexlab{b}}) Near-optimal sensor
  placements in gaussian processes: Theory, efficient algorithms and empirical
  studies. \emph{J. Mach. Learn. Res.} 9:235--284.

\bibitem[{Lipton et~al.(2003)Lipton, Markakis, \protect\BIBand{}
  Mehta}]{Lipton:2003:PLG:779928.779933}
Lipton RJ, Markakis E, Mehta A (2003) Playing large games using simple
  strategies. \emph{Proc. 4th ACM Conf. Elec. Commerce \emph{(ACM, New York,
  NY)}}, 36--41.

\bibitem[{Mavronicolas et~al.(2008)Mavronicolas, Papadopoulou, Philippou,
  \protect\BIBand{} Spirakis}]{mavronicolas}
Mavronicolas M, Papadopoulou V, Philippou A, Spirakis P (2008) A network game
  with attackers and a defender. \emph{Algorithmica} 51(3):315--341.

\bibitem[{Monroe et~al.(2018)Monroe, Ramsey, \protect\BIBand{}
  Berglund}]{MONROE201837}
Monroe J, Ramsey E, Berglund E (2018) Allocating countermeasures to defend
  water distribution systems against terrorist attack. \emph{Reliab. Engrg.
  Sys. Safe.} 179:37--51.

\bibitem[{Naureen et~al.(2018)Naureen, Collins, \protect\BIBand{}
  Vamburkar}]{Cyber2018}
Naureen MS, Collins R, Vamburkar M (2018) Cyberattack pings data systems of at
  least four gas networks. \emph{Bloomberg \emph{(April 3)},}
  \url{https://www.bloomberg.com/news/articles/2018-04-03/day-after-cyber-attack-a-third-gas-pipeline-data-system-shuts}.

\bibitem[{Orlin et~al.(2018)Orlin, Schulz, \protect\BIBand{}
  Udwani}]{Orlin2016}
Orlin JB, Schulz AS, Udwani R (2018) Robust monotone submodular function
  maximization. \emph{Math. Programming} 172(1):505--537.

\bibitem[{Ostfeld \protect\BIBand{} Salomons(2004)}]{Ost1415211920040901}
Ostfeld A, Salomons E (2004) Optimal layout of early warning detection stations
  for water distribution systems security. \emph{J. Water Res. Plan. Man.}
  130(5):377--385.

\bibitem[{Owolabi(2016)}]{Physical2016}
Owolabi T (2016) Nigerian militant group claims attack on oil pipeline in
  {N}iger {D}elta. \emph{Reuters \emph{(September 29)},}
  \url{https://www.reuters.com/article/us-nigeria-oil-idUSKCN11Z0XE}.

\bibitem[{Perelman et~al.(2008)Perelman, Maslia, Ostfeld, \protect\BIBand{}
  Sautner}]{doi:10.1002/j.1551-8833.2008.tb09659.x}
Perelman L, Maslia ML, Ostfeld A, Sautner JB (2008) Using
  aggregation/skeletonization network models for water quality simulations in
  epidemiologic studies. \emph{J. Amer. Water Works Assoc.} 100(6):122--133.

\bibitem[{PG\&E(2010)}]{PGE2010}
PG\&E (2010) Pipeline accident report: Pacific gas and electric company natural
  gas transmission pipeline rupture and fire. Technical report, National
  Transportation Safety Board, Washington, DC.

\bibitem[{Phillips et~al.(2013)Phillips, Ackley, Crosson, Down, Hutyra,
  Brondfield, Karr, Zhao, \protect\BIBand{} Jackson}]{PHILLIPS20131}
Phillips NG, Ackley R, Crosson ER, Down A, Hutyra LR, Brondfield M, Karr JD,
  Zhao K, Jackson RB (2013) Mapping urban pipeline leaks: Methane leaks across
  {B}oston. \emph{Environ. Pollut.} 173:1--4.

\bibitem[{Pita et~al.(2008)Pita, Jain, Marecki, Ord\'{o}\~{n}ez, Portway,
  Tambe, Western, Paruchuri, \protect\BIBand{}
  Kraus}]{Pita:2008:DAP:1402795.1402819}
Pita J, Jain M, Marecki J, Ord\'{o}\~{n}ez F, Portway C, Tambe M, Western C,
  Paruchuri P, Kraus S (2008) Deployed {ARMOR} protection: The application of a
  game theoretic model for security at the {L}os {A}ngeles {I}nternational
  {A}irport. \emph{Proc. 7th Internat. Conf. Autonomous Agents Multiagent Sys.
  \emph{(IFAAMAS, Richland, SC)}}, 125--132.

\bibitem[{Powell(2007)}]{10.2307/27644485}
Powell R (2007) Allocating defensive resources with private information about
  vulnerability. \emph{Amer. Polit. Sci. Rev.} 101(4):799--809.

\bibitem[{Sela \protect\BIBand{} Amin(2018)}]{SELA201855}
Sela L, Amin S (2018) Robust sensor placement for pipeline monitoring: Mixed
  integer and greedy optimization. \emph{Adv. Engrg. Inform.} 36:55--63.

\bibitem[{Sela~Perelman et~al.(2016)Sela~Perelman, Abbas, Koutsoukos,
  \protect\BIBand{} Amin}]{sensors:autom}
Sela~Perelman L, Abbas W, Koutsoukos X, Amin S (2016) Sensor placement for
  fault location identification in water networks: A minimum test cover
  approach. \emph{Automatica} 72:166--176.

\bibitem[{Smith \protect\BIBand{} Lim(2008)}]{smith2008algorithms}
Smith JC, Lim C (2008) Algorithms for network interdiction and fortification
  games. Chinchuluun A, Pardalos PM, Migdalas A, Pitsoulis L, eds.,
  \emph{Pareto Optimality, Game Theory And Equilibria \emph{(Springer, New
  York, NY)}}, 609--644.

\bibitem[{Srirangarajan et~al.(2013)Srirangarajan, Allen, Preis, Iqbal, Lim,
  \protect\BIBand{} Whittle}]{seshan}
Srirangarajan S, Allen M, Preis A, Iqbal M, Lim H, Whittle A (2013)
  Wavelet-based burst event detection and localization in water distribution
  systems. \emph{J. Signal Process. Sys.} 72(1):1--16.

\bibitem[{Tiemann(2017)}]{tiemann2014safe}
Tiemann M (2017) Safe drinking water act ({SDWA}): a summary of the act and its
  major requirements. Technical report, Congressional Research Service,
  Washington, DC.

\bibitem[{Tzoumas et~al.(2017)Tzoumas, Gatsis, Jadbabaie, \protect\BIBand{}
  Pappas}]{8263844}
Tzoumas V, Gatsis K, Jadbabaie A, Pappas GJ (2017) Resilient monotone
  submodular function maximization. \emph{Proc. 56th Conf. Dec. Control
  \emph{(IEEE, Piscataway, NJ)}}, 1362--1367.

\bibitem[{{University of Exeter}(2014)}]{exeter}
{University of Exeter} (2014) {Centre for Water Systems}. Accessed October 24,
  2014,
  \url{http://emps.exeter.ac.uk/engineering/research/cws/resources/benchmarks/design-resiliance-pareto-fronts/data-files/}.

\bibitem[{{US Environmental Protection Agency}(2007)}]{EPA}
{US Environmental Protection Agency} (2007) Factoids: Drinking water and ground
  water statistics for 2007. Technical report, Office of Water, US EPA,
  Washington, DC.

\bibitem[{Vazirani(2001)}]{Vazirani:2001}
Vazirani VV (2001) \emph{Approximation Algorithms} (Springer, Berlin
  Heidelberg).

\bibitem[{Von~Neumann(1953)}]{von1953certain}
Von~Neumann J (1953) A certain zero-sum two-person game equivalent to the
  optimal assignment problem. \emph{Contr. Theory Games} 2:5--12.

\bibitem[{Washburn \protect\BIBand{} Wood(1995)}]{doi:10.1287/opre.43.2.243}
Washburn A, Wood K (1995) Two-person zero-sum games for network interdiction.
  \emph{Oper. Res.} 43(2):243--251.

\bibitem[{Wright et~al.(2015)Wright, Abraham, Parpas, \protect\BIBand{}
  Stoianov}]{infrasense}
Wright R, Abraham E, Parpas P, Stoianov I (2015) Control of water distribution
  networks with dynamic {DMA} topology using strictly feasible sequential
  convex programming. \emph{Water Resour. Res.} 51(12):9925--9941.

\bibitem[{Xing \protect\BIBand{} Sela(2019)}]{XING2019291}
Xing L, Sela L (2019) Unsteady pressure patterns discovery from high-frequency
  sensing in water distribution systems. \emph{Water Res.} 158:291--300.

\bibitem[{Yuhas(2016)}]{NYTMexico2019}
Yuhas A (2016) Pipeline erupts in fiery explosion in {M}exico, killing many.
  \emph{The New York Times \emph{(June 18)},}
  \url{https://www.nytimes.com/2019/01/18/world/americas/mexico-gas-pipeline-explosion.html}.

\bibitem[{Zhuang \protect\BIBand{} Bier(2007)}]{doi:10.1287/opre.1070.0434}
Zhuang J, Bier VM (2007) Balancing terrorism and natural disasters—defensive
  strategy with endogenous attacker effort. \emph{Oper. Res.} 55(5):976--991.

\bibitem[{Zhuang et~al.(2010)Zhuang, Bier, \protect\BIBand{}
  Alagoz}]{ZHUANG2010409}
Zhuang J, Bier VM, Alagoz O (2010) Modeling secrecy and deception in a
  multiple-period attacker–defender signaling game. \emph{Eur. J. Oper. Res.}
  203(2):409--418.

\end{thebibliography}
}


\ifarXiv
\OneAndAHalfSpacedXI
\else
\DoubleSpacedXI
\fi

\ECSwitch


\ECHead{\rev{Supplementary Material}}


\section{Preliminary Results}

First, we define the following quantities: 
For a strategy $\sigma^1\in \Delta(\mathcal{A}_1)$ of \defender, the \emph{inspection probability} of node $i \in \nodes$, denoted $\nprob$, is the probability with which $i$ is inspected, i.e.:
\begin{align}
&\forall \sigma^1 \in \Delta(\mathcal{A}_1), \ \forall i \in \nodes,  \quad \nprob \coloneqq \mathbb{E}_{\sigma^1}\left[\mathds{1}_{\{i \in \Def\}}\right] = \sum_{\{\Def \in \mathcal{A}_1 \, | \, i \in \Def\}}\sigma^1_{\Def} .\label{rho_sigma1}
\end{align}




Given a strategy $\sigma^2 \in \Delta(\mathcal{A}_2)$, the \emph{attack probability} of component $e \in \edges$, denoted $\eprob$, is the probability with which $e$ is targeted by $\sigma^2$, i.e.:
\begin{align}
&\forall \sigma^2 \in \Delta(\mathcal{A}_2), \ \forall e \in \edges, \quad \eprob \coloneqq \mathbb{E}_{\sigma^2}\left[\mathds{1}_{\{e \in \att\}}\right]  = \sum_{\{\att \in \mathcal{A}_2 \, | \, e \in \att\}}\sigma^2_{\att} .\label{rho_sigma2}
\end{align}

\ifadditional
Note that the reversed inclusion $\set{\Def\cap\Def^\prime} \supseteq \set{\Def} \cap \set{\Def^\prime}$ does not hold in general (which is the main reason why $\valdet{\cdot}{\att}$ is submodular and not modular). Next, we use the properties of the monitoring sets to derive the properties of the sensing function $\detect$.

\else
\fi

\begin{lemma}\label{Submodularity}
The detection function defined in \eqref{detect_sets} satisfies the following properties:
\begin{enumerate}
\item For any subset of components $\att \in 2^\edges$, $\valdet{\cdot}{\att}$ is submodular and monotone:
\vspace{-0.1cm}
\begin{align}
\forall \att \in 2^\edges, \ \forall (\Def,\Def^\prime) \in (2^\nodes)^2, \quad
&\valdet{\Def \cup \Def^\prime}{\att} + \valdet{\Def \cap \Def^\prime}{\att} \leq \valdet{\Def}{\att} + \valdet{\Def^\prime}{\att},\label{submodular_inequality}\\
&\Def \subseteq \Def^\prime \ \Longrightarrow\ \valdet{\Def }{\att} \leq \valdet{\Def^\prime}{\att}.\label{monotonicity_detection}
\end{align}

\item For any detector positioning $\Def \in 2^\nodes$, $\valdet{\Def}{\cdot}$ is finitely additive:
\vspace{-0.1cm}
\begin{align}
\forall \Def \in 2^\nodes,\ \forall (\att,\att^\prime) \in (2^\edges)^2 \ | \ \att \cap \att^\prime = \emptyset, \quad \valdet{\Def}{\att \cup \att^\prime} = \valdet{\Def}{\att} + \valdet{\Def}{\att^\prime}.\label{additive_equality}
\end{align}
\end{enumerate}
\end{lemma}
\proof{Proof of Lemma~\ref{Submodularity}.}\
\begin{enumerate}
\item 
Consider a subset of components $\att \in 2^\edges$, and a pair of detector positionings $(\Def,\Def^\prime) \in (2^\nodes)^2$. Then, $\set{\Def\cup\Def^\prime} =\set{\Def} \cup \set{\Def^\prime}$ and $\set{\Def\cap\Def^\prime} \subseteq \set{\Def}\cap\set{\Def^\prime}$, and we obtain:
%
%
%
%
%
%
\vspace{-0.1cm}
\begin{align*}
&\valdet{\Def \cup \Def^\prime}{\att} + \valdet{\Def \cap \Def^\prime}{\att} \overset{\eqref{detect_sets}}{=} |\set{\Def \cup \Def^\prime} \cap \att| + |\set{\Def \cap \Def^\prime} \cap \att| = |(\set{\Def} \cap \att) \cup (\set{\Def^\prime} \cap \att)| + |\set{\Def \cap \Def^\prime} \cap \att|\nonumber\\
& = |\set{\Def} \cap \att| + |\set{\Def^\prime} \cap \att| - |\set{\Def} \cap \set{\Def^\prime}\cap \att| + |\set{\Def \cap \Def^\prime} \cap \att| \leq|\set{\Def} \cap \att| + |\set{\Def^\prime} \cap \att| \overset{\eqref{detect_sets}}{=} \valdet{\Def}{\att} + \valdet{\Def^\prime}{\att}.
\end{align*}

Furthermore, if $\Def \subseteq \Def^\prime$, then: $\valdet{\Def}{\att} \overset{\eqref{detect_sets}}{=} |\set{\Def} \cap \att| \leq  |\set{\Def^\prime} \cap \att| \overset{\eqref{detect_sets}}{=}\valdet{\Def^{\prime}}{\att}.$

\item 
Consider a detector positioning $\Def \in 2^\nodes$. Then, \rev{for every} $(\att,\att^\prime) \in (2^\edges)^2$ \rev{such that} $\att \cap \att^\prime = \emptyset, \quad \valdet{\Def}{\att \cup \att^\prime} \overset{\eqref{detect_sets}}{=} |\set{\Def}\cap (\att \cup \att^\prime)| = |\set{\Def} \cap \att| +|\set{\Def} \cap \att^\prime| - |\set{\Def} \cap \att \cap \att^\prime|  \overset{\eqref{detect_sets}}{=} \valdet{\Def}{\att} + \valdet{\Def}{\att^\prime}.$\hfill \Halmos
\end{enumerate}
\endproof


\begin{corollary}\label{useful_(in)equalities}
The detection function defined in \eqref{detect_sets} satisfies the following properties:
\begin{align}
\forall (\Def,\att) \in 2^{\nodes} \times 2^\edges, \ \valdet{\Def}{\att} \leq \sum_{i\in \Def}\valdet{i}{\att}, \label{redundant_nodes}\\
\forall (\Def,\att) \in 2^\nodes \times 2^\edges, \ \valdet{\Def}{\att} = \sum_{e \in \att} \valdet{\Def}{e}. \label{detection_disruptions}
\end{align}
\end{corollary}
\vspace{-0.0cm}
\proof{Proof of Corollary~\ref{useful_(in)equalities}.}\
\begin{enumerate}
\item  
Consider $\att \in 2^\edges$. Since $\valdet{\cdot}{\att}$ is a submodular and nonnegative function, then $\valdet{\cdot}{\att}$ is subadditive, i.e., \rev{for all} $(\Def,\Def^\prime) \in (2^\nodes)^2, \ \valdet{\Def\cup\Def^\prime}{\att} \leq \valdet{\Def}{\att} + \valdet{\Def^{\prime}}{\att}$. 

%

Therefore, by induction, we obtain \rev{that for all} $ \Def \in 2^\nodes, \ \valdet{\Def}{\att} = \valdet{\cup_{i \in \Def}\{i\}}{\att} \leq \sum_{i \in \Def}\valdet{i}{\att}.$

\item Consider $\Def \in 2^\nodes$. Since $\valdet{\Def}{\cdot}$ is \rev{finitely} additive (Lemma~\ref{Submodularity}), we obtain by induction that \rev{for every} $\att \in 2^\edges, \ \valdet{\Def}{\att} = \valdet{\Def}{\cup_{e \in \att}\{e\}} \overset{\eqref{additive_equality}}{=} \sum_{e \in \att}\valdet{\Def}{e}$. \hfill \Halmos
\end{enumerate}
\endproof

\begin{lemma}\label{MM<MSC}
The size of MSPs is no greater than the size of MSCs, i.e., $\semm \leq \smsc$.
\end{lemma}

\proof{Proof of Lemma~\ref{MM<MSC}.}
Consider an MSP $T^{max} = \{e_1,\dots,e_{\semm}\} \in \emm$ and an MSC $S^{min} = \{i_1,\dots,i_{\smsc}\} \in \msc$. Then, we have the desired inequality:
\begin{align*}
\semm &\overset{\eqref{def_minimum_set_cover}}{=} \sum_{l=1}^{\semm}\valdet{S^{min}}{e_l} \overset{\eqref{redundant_nodes}}{\leq} \sum_{l=1}^{\semm}\sum_{k=1}^{\smsc} \valdet{i_k}{e_l} =\sum_{k=1}^{\smsc} \sum_{l=1}^{\semm} \valdet{i_k}{e_l} \overset{\eqref{detection_disruptions}}{=}  \sum_{k=1}^{\smsc}\valdet{i_k}{T^{max}} \overset{\eqref{def_extended_matching}}{\leq} \smsc.
\end{align*}

\vspace{-0.8cm}
\hfill\Halmos
\endproof

\ifadditional
\begin{example}

Consider the network shown in Fig.~\ref{Example1}. Assume a sensing model in which a sensor placed at any node can monitor edges at one-hop vertically, and at two hops horizontally. Thus, $\set{i_1} = \{e_1,e_2,e_3\}$, $\set{i_2} = \{e_1,e_2,e_4\}$, $\set{i_3} = \{e_1,e_2,e_5\}$, $\set{i_4} = \{e_4,e_6,e_7\}$, $\set{i_5} = \{e_5,e_6,e_8\}$, $\set{i_6} = \{e_3,e_9,e_{10}\}$, $\set{i_7} = \{e_7,e_9,e_{10}\}$, $\set{i_8} = \{e_8,e_9,e_{10}\}$.

\def \sizefig {1.4cm}
\def \scalefig {\small}
\def \innersepfig {0.07cm}

\begin{figure}[ht]
\centering
{\scalefig
\begin{tikzpicture}[-,auto,x=\sizefig, y=\sizefig,
  thick,main node/.style={circle,draw,inner sep = \innersepfig,black},main node2/.style={circle,black,draw,fill=blue!60,inner sep = \innersepfig},flow_a/.style ={gray!100}]
\tikzstyle{edge} = [draw,thick,-,black]
\tikzstyle{cut} = [draw,very thick,-]
\tikzstyle{match} = [draw,very thick,black,-, dashed]
\tikzstyle{flow} = [draw,line width = 1.5pt,->,gray!100]

	\node[main node] (1) at (0,2) {$i_1$};
	\node[main node] (2) at (1,2) {$i_2$};
	\node[main node2] (3) at (2,2) {$i_3$};
	\node[main node] (4) at (1,1) {$i_4$};
	\node[main node] (5) at (2,1) {$i_5$};
	\node[main node] (6) at (0,0) {$i_6$};
	\node[main node] (7) at (1,0) {$i_7$};
	\node[main node] (8) at (2,0) {$i_8$};
	
	\path[edge]	
	(1) edge node[left]{$e_3$} (6)
	(4) edge node {$e_6$} (5)
	(2) edge[double,blue!80]  node {$e_2$} (3)
	(2) edge node[left] {$e_4$} (4)
	(6) edge node[below] {$e_9$} (7)
	(7) edge node[below] {$e_{10}$} (8)
	(5) edge node {$e_8$} (8);

	\path[edge]
	(1) edge[double,blue!80] node {$e_1$} (2)
	(3) edge[double,blue!80] node {$e_5$} (5)
	(4) edge node[left] {$e_7$} (7);

\end{tikzpicture}}

\caption{Example network. The monitoring set $\set{i_3} = \{e_1,e_2,e_5\}$ is represented by the double blue edges.} \label{Example1}
\end{figure}

\tr{Consider the example network in Section~\ref{example_detection_sensing}. Then, the corresponding influence matrix is given by:}
\begin{align*}
\bs{F} = \begin{array}{cc} & {\arraycolsep=3pt\begin{array}{cccccccccc} e_1 & e_2 & e_3 &e_4&e_5&e_6 & e_7 & e_8 &e_9 &e_{10}\end{array}}\vspace{0.1cm}\\
{\def\arraystretch{\spacinginfluence}\begin{array}{c}
i_1\\i_2\\i_3\\i_4\\i_5\\i_6\\i_7\\i_8\end{array}}&{\arraycolsep=5pt\def\arraystretch{\spacinginfluence}
  \left({\begin{array}{cccccccccc}
	1 & 1 & 1 & 0 & 0 & 0 &0 &0 &0 &0 \\
	1 & 1 & 0 & 1 & 0 & 0&0 &0  &0 &0\\
	 1 &  1 &  0 &  0 &  1 &  0 & 0 & 0&0 &0\\
	 0 &  0 &  0 &  1 &  0 &  1 & 1 & 0&0 &0\\
	 0 &  0 &  0 &  0 &  1 &  1 & 0 & 1&0 &0 \\
	0 & 0 & 1 & 0 & 0 & 0&0 &0&1 &1\\
	0 & 0 & 0 & 0 & 0 & 0&1 &0&1 &1\\
	0 & 0 & 0 & 0 & 0 & 0&0 &1&1 &1
  \end{array}}\right)}\end{array}
\end{align*}

As mentioned in Section~\ref{example_detection_sensing}, a solution of (\hyperlink{(MSC)}{$\mathcal{I}_{\text{MSC}}$}) is given by $\{i_3,i_4,i_6,i_8\}$ and a solution of (\hyperlink{(MSP)}{$\mathcal{I}_{\text{MSP}}$}) is given by $\{e_3,e_4,e_8\}$. They are illustrated in Fig.~\ref{influence_matrices}.
\begin{figure}[ht]
\vspace{\spacingfigure}
\centering
\begin{align*}
\begin{array}{cc} & {\arraycolsep=3pt\begin{array}{cccccccccc} e_1 & e_2 & e_3 &e_4&e_5&e_6 & e_7 & e_8 &e_9 &e_{10}\end{array}}\vspace{0.1cm}\\
{\def\arraystretch{\spacinginfluence}\begin{array}{c}
i_1\\i_2\\i_3\\i_4\\i_5\\i_6\\i_7\\i_8\end{array}}&{\arraycolsep=5pt\def\arraystretch{\spacinginfluence}
  \left({\begin{array}{cccccccccc}
	1 & 1 & 1 & 0 & 0 & 0 &0 &0 &0 &0 \\
	1 & 1 & 0 & 1 & 0 & 0&0 &0  &0 &0\\
	 \cellcolor{Green}1 &  \cellcolor{Green}1 &  \cellcolor{Green}0 &\cellcolor{Green}  0 &\cellcolor{Green}  1 &\cellcolor{Green}  0 &\cellcolor{Green} 0 &\cellcolor{Green} 0&\cellcolor{Green}0 &\cellcolor{Green}0\\
	 \cellcolor{Green}0 &\cellcolor{Green}  0 &\cellcolor{Green}  0 &\cellcolor{Green}  1 &\cellcolor{Green}  0 &\cellcolor{Green}  1 &\cellcolor{Green} 1 &\cellcolor{Green} 0&\cellcolor{Green}0 &\cellcolor{Green}0\\
	 0 &  0 &  0 &  0 &  1 &  1 & 0 & 1&0 &0 \\
	\cellcolor{Green}0 &\cellcolor{Green} 0 &\cellcolor{Green} 1 &\cellcolor{Green} 0 &\cellcolor{Green} 0 &\cellcolor{Green} 0&\cellcolor{Green}0 &\cellcolor{Green}0&\cellcolor{Green}1 &\cellcolor{Green}1\\
	0 & 0 & 0 & 0 & 0 & 0&1 &0&1 &1\\
	\cellcolor{Green}0 &\cellcolor{Green} 0 &\cellcolor{Green} 0 &\cellcolor{Green} 0 &\cellcolor{Green} 0 &\cellcolor{Green} 0&\cellcolor{Green}0 &\cellcolor{Green}1&\cellcolor{Green}1 &\cellcolor{Green}1
  \end{array}}\right)}\end{array}
  \quad \quad \quad
\begin{array}{cc} & {\arraycolsep=3pt\begin{array}{cccccccccc} e_1 & e_2 & e_3 &e_4&e_5&e_6 & e_7 & e_8 &e_9 &e_{10}\end{array}}\vspace{0.1cm}\\
{\def\arraystretch{\spacinginfluence}\begin{array}{c}
i_1\\i_2\\i_3\\i_4\\i_5\\i_6\\i_7\\i_8\end{array}}&{\arraycolsep=5pt\def\arraystretch{\spacinginfluence}
  \left({\begin{array}{cc>{\columncolor{red!50}}c>{\columncolor{red!50}}cccc>{\columncolor{red!50}}ccc}
	1 & 1 & 1 & 0 & 0 & 0 &0 &0 &0 &0 \\
	1 & 1 & 0 & 1 & 0 & 0&0 &0  &0 &0\\
	 1 &  1 &  0 &  0 &  1 &  0 & 0 & 0&0 &0\\
	 0 &  0 &  0 &  1 &  0 &  1 & 1 & 0&0 &0\\
	 0 &  0 &  0 &  0 &  1 &  1 & 0 & 1&0 &0 \\
	0 & 0 & 1 & 0 & 0 & 0&0 &0&1 &1\\
	0 & 0 & 0 & 0 & 0 & 0&1 &0&1 &1\\
	0 & 0 & 0 & 0 & 0 & 0&0 &1&1 &1
  \end{array}}\right)}\end{array}
\end{align*}
\caption{Illustration of an MSC (left) and an MSP (right) for the influence matrix $\bs{F}$ corresponding to the example network in Section~\ref{example_detection_sensing}.}
\label{influence_matrices}
\end{figure}

\end{example}
\else
\fi

\section{Proofs of Section \ref{general}}\label{Main Proofs}


\begin{lemma}\label{algebra2}
Consider a set of nodes $\Def = \{i_1,\dots,i_{n}\} \in 2^{\nodes}$ of size $n \geq b_1$, and a set of  components $\att = \{e_1,\dots,e_{m}\} \in 2^{\edges}$ of size $m \geq b_2$. We define the following pure actions:
\vspace{-0.0cm}
\begin{align}
\forall k \in \llbracket 1,n\rrbracket, \ &\Def^k= \left\{\begin{array}{ll}
      \{i_{k},\dots,i_{k+b_1 -1}\}& \quad \text{ if }\ k\leq n-b_1+1, \\
    \{i_k,\dots,i_n,i_1,\dots,i_{k+ b_1 -n- 1}\}  & \quad \text{ if }\ k \geq n-b_1 +2,
\end{array}\right. 
\label{S^k}\\
\forall l \in \llbracket 1,m \rrbracket, \ &\att^l = \left\{\begin{array}{ll}
      \{e_{l},\dots,e_{l+b_2 -1}\}& \quad \text{ if }\ l\leq m-b_2+1, \\
    \{e_l,\dots,e_m,e_1,\dots,e_{l+ b_2 -m- 1}\}  & \quad \text{ if }\ l \geq m-b_2 +2,
\end{array}\right.\label{att^k}
\end{align}
and a strategy profile  $(\sold{\Def}{b_1}{},\sola{\att}{b_2}{}) \in \Delta(\mathcal{A}_1)\times\Delta(\mathcal{A}_2)$ supported over $\{\Def^1,\dots,\Def^{n}\}$ and $\{\att^1,\dots,\att^{m}\}$, where:
\vspace{-0.6cm}
\begin{align}
\forall k \in \llbracket 1,n \rrbracket, \ &\sold{\Def}{b_1}{\Def^k} \coloneqq \frac{1}{n},\label{sigma^S}\\
\forall l \in \llbracket 1,m \rrbracket, \ &\sola{\att}{b_2}{\att^l} \coloneqq \frac{1}{m}.\label{sigma^T}
\end{align}

Then,  the strategy profile $(\sold{\Def}{b_1}{},\sola{\att}{b_2}{})$ \rev{satisfies} the following properties:
\begin{enumerate}
\item Each node in $\Def$ (resp. each component in $\att$) is inspected (resp. targeted) with an identical probability given by:
\vspace{-0.4cm}
\begin{align}
&\forall i \in \Def, \ \rho_{\sold{\Def}{b_1}{}}(i) = \frac{b_1}{n}, \label{set_sensing_prob_comp}\\
&\forall e \in \att, \ \rho_{\sola{\att}{b_2}{}}(e) = \frac{b_2}{m}.\label{set_disruption_prob_comp}
\end{align}

\item Each node in $\Def$ (resp. each component in $\att$) belongs to $\antiorder{n}{b_1}$ actions (resp. $\antiorder{m}{b_2}$ actions) in the support of $\sold{\Def}{b_1}{}$ (resp. $\sola{\att}{b_2}{}$):
\begin{align}
&\forall i \in \Def, \ |\{k \in \llbracket 1,\order{n}{b_1}\rrbracket \ | \ i \in \Def^k\}| = \antiorder{n}{b_1}, \label{q1}\\
&\forall e \in \att, \ |\{l \in \llbracket 1,\order{m}{b_2}\rrbracket \ | \ e \in \att^l\}| = \antiorder{m}{b_2}.\label{q2}
\end{align}

\item The following inequality is satisfied:
\begin{align}
\forall e \in \edges, \ \valdet{\Def}{e} \leq \frac{1}{\antiorder{n}{b_1}}\sum_{k=1}^{\order{n}{b_1}}\valdet{\Def^k}{e}.\label{nice_detection_inequality}
\end{align}
\end{enumerate}
\end{lemma}

\proof{Proof of Lemma~\ref{algebra2}.}
We show the result for a set of nodes $\Def \in 2^\nodes$ of size $n \geq b_1$.
First we note that, by construction, each node $i \in S$ belongs to the same number of detector positionings $S^k$, $k \in \llbracket 1,\order{n}{b_1}\rrbracket$. Thus, \eqref{q1} follows from the following calculations:
\begin{align*}
\order{n}{b_1}b_1 & \overset{\eqref{S^k}}{=}   \sum_{k=1}^{\order{n}{b_1}}\sum_{ i^\prime \in \nodes} \mathds{1}_{\{i^\prime \in \Def^k\}}  = \sum_{ i^\prime \in \nodes} |\{k \in \llbracket 1,\order{n}{b_1}\rrbracket \ | \ i^\prime \in \Def^k\}|  = n \times |\{k \in \llbracket 1,\order{n}{b_1}\rrbracket \ | \ i \in \Def^k\}|, \quad \forall i \in \Def.
\end{align*}

Then, we can show \eqref{set_sensing_prob_comp}. For every node $i \in \Def$, we have:
\begin{align*}
\rho_{\sold{\Def}{b_1}{}}(i) &\overset{\eqref{rho_sigma1}}{=}  \sum_{k=1}^{\order{n}{b_1}}  \sold{\Def}{b_1}{\Def^{k}}\mathds{1}_{\{i \in \Def^{k}\}} \overset{\eqref{sigma^S}}{=} \frac{1}{\order{n}{b_1}}\sum_{k=1}^{\order{n}{b_1}}  \mathds{1}_{\{i \in \Def^{k}\}}  = \frac{1}{\order{n}{b_1}}|\{k \in \llbracket 1,\order{n}{b_1}\rrbracket \ | \ i \in \Def^k\}| \overset{\eqref{q1}}{=}  \frac{b_1}{n}.
\end{align*}

An analogous proof can be applied to $\att  \in 2^\edges$ of size  $m \geq b_2$ to show \eqref{set_disruption_prob_comp} and \eqref{q2}.

Finally, let us show \eqref{nice_detection_inequality}. Consider $e \in \edges$. If $\valdet{\Def}{e} = 1$, then \rev{there exists} $i_0 \in \Def$ \rev{such that} $\valdet{i_0}{e} = 1$. Since there are $\antiorder{n}{b_1}$ detector positionings in $\{\Def^k, \ k \in \llbracket 1,\order{n}{b_1}\rrbracket\}$ that contain $i_0$, then $\frac{1}{\antiorder{n}{b_1}}\sum_{k=1}^{\order{n}{b_1}}\valdet{\Def^k}{e} \geq 1 = \valdet{\Def}{e}$. \rev{If $\valdet{\Def}{e}=0$, then $\frac{1}{\antiorder{n}{b_1}}\sum_{k=1}^{\order{n}{b_1}}\valdet{\Def^k}{e} = 0 = \valdet{\Def}{e}$.}
%
%
\hfill \Halmos
\endproof

We illustrate this construction with an example:
\begin{example}\label{algebra_for_game_theory}
Consider a set of three nodes $\Def = \{i_1,i_2,i_3\}$ and suppose that \defender has two detectors ($b_1 = 2$). First, we define three pure actions $\Def^1 = \{i_1,i_2\}$, $\Def^2 = \{i_2,i_3\}$, and $\Def^3 = \{i_3,i_1\}$; see \rev{Figure}~\ref{example_S^k}. The strategy $\sold{\Def}{b_1}{}$ is then obtained by assigning uniform probability (i.e., $\frac{1}{3}$) to each pure action. One can check that each node in $\Def$  is inspected with probability $\frac{2}{3} = \frac{b_1}{|\Def|}$. \hfill $\triangle$

\begin{figure}[htbp]
\centering
\small
\begin{tikzpicture}[-,auto,x=1.3cm, y=0.9cm,
  thick,main node/.style={circle,black,draw,inner sep = 0.07cm},main node2/.style={circle,black,draw,fill=blue!50,inner sep = 0.07cm},flow_a/.style ={gray!100}]
\tikzstyle{edge} = [draw,thick,-,black]
\tikzstyle{detect} = [draw,very thick,Green,-]
\tikzstyle{match} = [draw,very thick,black,-, dashed]
\tikzstyle{flow} = [draw,line width = 1.5pt,->,gray!100]

	\node[main node] (1) at (0,2) {$i_1$};
	\node[main node] (2) at (0,1) {$i_2$};
	\node[main node] (3) at (0,0) {$i_3$};
	
	\node[main node2] (1) at (1,2) {$i_1$};
	\node[main node2] (2) at (1,1) {$i_2$};
	\node[main node] (3) at (1,0) {$i_3$};

	\node[main node] (1) at (2,2) {$i_1$};
	\node[main node2] (2) at (2,1) {$i_2$};
	\node[main node2] (3) at (2,0) {$i_3$};
	
	\node[main node2] (1) at (3,2) {$i_1$};
	\node[main node] (2) at (3,1) {$i_2$};
	\node[main node2] (3) at (3,0) {$i_3$};
\draw[black] (0.5,-0.5) -- (0.5,2.5);

\node[black] (10) at (0,2.9) {$\Def$ };
\node[black] (10) at (1,2.9) {$\Def^1$ };
\node[black] (10) at (2,2.9) {$\Def^2$ };
\node[black] (10) at (3,2.9) {$\Def^3$ };
	

\normalsize
\end{tikzpicture}
\caption{Support of $\sold{\Def}{b_1}{}$ when $\Def$ is composed of three nodes and $b_1 = 2$.} \label{example_S^k}
\end{figure}


\end{example}

%
%

\proof{Proof of Proposition~\ref{best_set_cover}.} \

\begin{enumerate}
\item
Consider a minimal set cover $\Def^{\prime} = \{i_1,\dots,i_n\} \in 2^\nodes$ of size $n$. Necessarily, $\Def^{\prime}$ is such that:
\begin{align}
\forall k \in \llbracket 1,n \rrbracket, \ \exists \, e_k \in \edges \ | \ \valdet{i_k}{e_k} = 1 \text{ and } \valdet{i_j}{e_k} = 0, \quad \forall j \neq k. \label{minimal_set_cover}
\end{align}

\begin{itemize}
\item[--]First, let us show that \rev{$U^*(b_1,b_2) \leq b_2\left(1 - \frac{b_1}{n}\right)$}.  Consider $\sold{\Def^{\prime}}{b_1}{} \in \Delta(\mathcal{A}_1)$ defined in \eqref{sigma^S}.
 Recall that we are in the case when $b_1 < \smsc$, implying that $b_1 < \smsc \leq n$.
Since $\Def^{\prime}$ is a set cover, then \rev{for every} $e \in \edges$, \rev{there exists} $k_e \in \llbracket 1,n\rrbracket$ \rev{such that} $\valdet{i_{k_e}}{e} = 1$. Furthermore, $\detect$ is a nonnegative function. Therefore:
\begin{align}
\forall e \in \edges, \ &\sum_{\Def \in \mathcal{A}_1} \sold{\Def^{\prime}}{b_1}{\Def} \valdet{\Def}{e} \overset{\eqref{monotonicity_detection}}{\geq}  \sum_{\{\Def \in \mathcal{A}_1 \, | \, i_{k_e} \in \Def\}}\sold{\Def^{\prime}}{b_1}{\Def}\overset{\eqref{rho_sigma1}}{=} \rho_{\sold{\Def^{\prime}}{b_1}{}}(i_{k_e})\overset{\eqref{set_sensing_prob_comp}}{=} \frac{b_1}{n}.\label{ineq1}
\end{align}

Thus, we obtain:
\begin{align}
\forall \att \in \mathcal{A}_2, \ \rev{U}(\sold{\Def^{\prime}}{b_1}{},\att) &\overset{\eqref{payoff2}}{=} \rev{|\att| - \sum_{\Def \in \mathcal{A}_1} \sold{\Def^{\prime}}{b_1}{\Def} \valdet{\Def}{\att}\overset{\eqref{detection_disruptions}}{=} |\att| - \sum_{e \in \att}\sum_{\Def \in \mathcal{A}_1} \sold{\Def^{\prime}}{b_1}{\Def} \valdet{\Def}{e}} \nonumber\\
& \rev{\overset{\eqref{ineq1}}{\leq} |\att| - \sum_{e \in \att}\frac{b_1}{n} = \underset{>0}{\underbrace{\left(1 - \frac{b_1}{n}\right)}}\underset{\leq b_2}{\underbrace{|\att|}}  \leq b_2 \left(1 - \frac{b_1}{n}\right).} \label{for later}
\end{align}

\rev{Therefore:} 
$$\rev{U^*(b_1,b_2) = \min_{\sigma^1 \in \Delta(\mathcal{A}_1)}\max_{\att \in \mathcal{A}_2} U(\sigma^1,\att) \leq \max_{\att \in \mathcal{A}_2} U(\sold{\Def^{\prime}}{b_1}{},\att) \leq b_2 \left(1 - \frac{b_1}{n}\right).}$$



\item[--] \rev{Now, let us show that the upper bound on $\max_{\att \in \mathcal{A}_2} U(\sold{\Def^{\prime}}{b_1}{},\att)$ is tight.}
%
Consider $\att^\prime = \{e_1,\dots,e_{b_2}\}$ (where the $e_k$'s are defined in \eqref{minimal_set_cover}). Then, for every $k \in \llbracket 1,b_2 \rrbracket$:  
\begin{align}
\rev{\mathbb{E}_{\sold{\Def^{\prime}}{b_1}{}}[\valdet{\Def}{e_k}]} & \rev{\overset{\eqref{minimal_set_cover}}{=} \sum_{\{\Def \in \mathcal{A}_1 \, | \, i_{k} \in \Def\}} \sold{\Def^{\prime}}{b_1}{\Def}\underset{=1}{\underbrace{\valdet{\Def}{e_k}}} + \sum_{\{\Def \in \mathcal{A}_1 \, | \, i_{k} \notin \Def\}} \underset{=0}{\underbrace{\sold{\Def^{\prime}}{b_1}{\Def} \valdet{\Def}{e_k}}}}\nonumber\\
&\rev{\overset{\eqref{rho_sigma1}}{=} \rho_{\sold{\Def^{\prime}}{b_1}{}}(i_{k}),} \label{to be used soon}
\end{align}
where we combined the fact that the node basis of \rev{$\sold{\Def^{\prime}}{b_1}{}$} is $\Def^\prime$ and that $i_k$ is the only node from $\Def^{\prime}$ that monitors component $e_k$ (by construction). This implies that:
\begin{align*}
\rev{\max_{\att \in \mathcal{A}_2} {U}(\sold{\Def^{\prime}}{b_1}{},\att)  \geq {U}(\sold{\Def^{\prime}}{b_1}{},\att^\prime)  \overset{\eqref{payoff2},\eqref{detection_disruptions},\eqref{to be used soon}}{=} b_2 - \sum_{k=1}^{b_2}\rho_{\sold{\Def^{\prime}}{b_1}{}}(i_{k}) \overset{\eqref{set_sensing_prob_comp}}{=} b_2 \left(1 - \frac{b_1}{n}\right).}
\end{align*}
\rev{Therefore:}
\begin{align}
\rev{\max_{\att \in \mathcal{A}_2}U(\sold{\Def^\prime}{b_1}{},\att) = b_2 \left(1 - \frac{b_1}{n}\right).}\label{Tight_Ineq}
\end{align}
\end{itemize}


\item
Consider a set packing $\att^{\prime} = \{e_1,\dots,e_m\} \in 2^\edges$ of size $m \geq b_2$. 

%
\begin{itemize}
\item[--] First, we show that \rev{$U^*(b_1,b_2) \geq \max\left\{0,b_2 \left(1 - \frac{b_1}{m}\right)\right\}$}. Consider $\sola{\att^{\prime}}{b_2}{} \in \Delta(\mathcal{A}_2)$ defined in \eqref{sigma^T}. 
%
Since $\att^\prime$ is a set packing, then \rev{for all} $i \in \nodes, \ \valdet{i}{\att^\prime} \leq 1$. This implies that:
\begin{align*}
\forall \Def \in \mathcal{A}_1, \ & \rev{U}(\Def, \sola{\att^{\prime}}{b_2}{}) \overset{\eqref{payoff2}}{=} \mathbb{E}_{\sola{\att^{\prime}}{b_2}{}}[|\att| - \valdet{\Def}{\att}] \overset{\eqref{redundant_nodes},\eqref{att^k}}{\geq} b_2 - \sum_{i \in \Def}\mathbb{E}_{\sola{\att^{\prime}}{b_2}{}}[\valdet{i}{\att}]\\
&  \overset{\eqref{rho_sigma2},\eqref{detection_disruptions}}{=} b_2 - \sum_{i \in\Def} \sum_{e \in \edges} \valdet{i}{e}\rho_{\sola{\att^{\prime}}{b_2}{}}(e)  \overset{\eqref{detection_disruptions},\eqref{set_disruption_prob_comp}}{=} b_2 -  \frac{b_2}{m}\sum_{i \in\Def} \underset{\leq 1}{\underbrace{\valdet{i}{\att^\prime}}}  \geq b_2\left( 1 - \frac{b_1}{m} \right).
\end{align*}
\rev{Furthermore, for all $\Def \in \mathcal{A}_1$, $U(\Def,\sola{\att^{\prime}}{b_2}{}) \geq 0$.} \rev{Therefore:} 
\begin{align*}
\rev{U^*(b_1,b_2) = \max_{\sigma^2 \in \Delta(\mathcal{A}_2)}\min_{\Def \in \mathcal{A}_1} U(\Def,\sigma^1)  \geq \min_{\Def \in \mathcal{A}_1}U(\Def,\sola{\att^{\prime}}{b_2}{}) \geq \max\left\{0,b_2 \left(1 - \frac{b_1}{m}\right)\right\}.}
\end{align*}


\item[--]  \rev{Now, let us show that the lower bound on $\min_{\Def \in \mathcal{A}_1} U(\Def,\sola{\att^{\prime}}{b_2}{})$ is tight.}
In Section \ref{sec:detect}, we assumed that each component can be monitored from at least one node. Therefore \rev{for all} $l \in \llbracket 1,m \rrbracket,$ \rev{there exists} $i_l \in \nodes$  \rev{such that} $\valdet{i_l}{e_l} = 1$ (note that the $i_l$'s are distinct since $\att^\prime$ is a set packing).  Now, consider the detector positioning \rev{$\Def^\prime = \{i_1,\dots,i_{\min\{b_1,m\}}\}$}. $\Def^\prime$ monitors \rev{$\{e_1,\dots,e_{\min\{b_1,m\}}\}$}, which enables us to show:
\begin{align*}
&\rev{\min_{\Def \in \mathcal{A}_1} U(\Def,\sola{\att^{\prime}}{b_2}{})}  \rev{\leq U(\Def^\prime, \sola{\att^{\prime}}{b_2}{}) \overset{\eqref{payoff2},\eqref{detection_disruptions}}{=} \mathbb{E}_{\sola{\att^{\prime}}{b_2}{}}[|\att|] - \mathbb{E}_{\sola{\att^{\prime}}{b_2}{}}[\sum_{e \in \att}\valdet{\Def^\prime}{e}]}  \\
&\rev{\overset{\eqref{rho_sigma2}}{=}b_2 - \sum_{e \in \edges}\valdet{\Def^\prime}{e}\rho_{\sola{\att^{\prime}}{b_2}{}}(e) = b_2  - \sum_{l=1}^{\min\{b_1,m\}} \rho_{\sola{\att^{\prime}}{b_2}{}}(e_l) \overset{\eqref{set_disruption_prob_comp}}{=}  \max\left\{0,b_2\left( 1 - \frac{b_1}{m}\right)\right\}.}
\end{align*}

\rev{Therefore,
$\min_{\Def \in \mathcal{A}_1} U(\Def,\sola{\att^{\prime}}{b_2}{}) = \max\left\{0,b_2\left( 1 - \frac{b_1}{m}\right)\right\}$.} 
\hfill \Halmos
\end{itemize}
\end{enumerate}
\endproof

\proof{Proof of Theorem \ref{all_resources}.}\ 

\begin{enumerate}
\item[(i.a)] First, let us show by contradiction that \defender uses all her resources in equilibrium. Suppose that \rev{there exist} $({\sigma^1}^*,{\sigma^2}^*) \in \nash$ \rev{and} $\Def^0 \in \supp({\sigma^1}^*)$ \rev{such that} $|\Def^0| <   b_1$. 
\begin{enumitemize}
\item[--] The first step is to show that \attacker's strategy $\sigma^{2^*}$ necessarily targets with positive probability at least one component  that is not monitored by $\Def^0$. On the contrary, assume that \rev{for every} $e \in \edges: \ \rho_{\sigma^{2^*}}(e) > 0\ \Longrightarrow\ \valdet{\Def^0}{e}=1$. Then, \defender can detect all the attacks of $\sigma^{2^*}$ with the detector positioning $\Def^0$. \rev{Since $\Def^0$ is a best response to $\sigma^{2^*}$, the value of the game is 0.}
%
However, since $b_1 < \smsc$, \rev{there exists} $e^\prime \in \edges$ \rev{such that} $\valdet{\Def^0}{e^\prime} = 0$, i.e., $e^\prime$ is not monitored by $\Def^0$. Since $\Def^0 \in \supp(\sigma^{1^*})$, then $e^\prime$ is not monitored with positive probability. Therefore, if \attacker targets $e^\prime$, she will get a positive payoff,
which contradicts the equilibrium condition \eqref{best} for $(\sigma^{1^*},\sigma^{2^*})$. \rev{Thus}, \rev{there exists} $e_0 \in \edges$ \rev{such that} $\rho_{\sigma^{2^*}}(e_0) >0$ and $\valdet{\Def^0}{e_0} = 0$. 

\item[--] Now, we can show that \defender can \rev{improve} her payoff by \rev{positioning} one more detector. Let us denote $i_0 \in \nodes \backslash S^0$ that satisfies $\valdet{i_0}{e_0} = 1$. Then, by considering the detector positioning $\Def^{\prime} = \Def^0 \cup \{i_0\} \in \mathcal{A}_1$, we obtain that \rev{$U(\Def^{\prime},\sigma^{2^*})  \leq U(\Def^{0},\sigma^{2^*}) - \rho_{\sigma^{2^*}}(e_0) < U(\Def^{0},\sigma^{2^*}), $}
 which violates the equilibrium condition \eqref{best} for $({\sigma^1}^*,{\sigma^2}^*)$. Therefore, \rev{for all} $\Def \in \supp(\sigma^{1^*}), \ |\Def| = b_1$.

\end{enumitemize}

\item[(i.b)] 
Now, let us show that \attacker uses all her resources  in equilibrium. By contradiction, suppose that \rev{there exist} $(\sigma^{1^*},\sigma^{2^*}) \in \nash$ \rev{and} $\att^0 \in \supp(\sigma^{2^*})$ \rev{such that} $|\att^0| <b_2$. 
\begin{enumitemize}
\item[--] The first step is to show that there exists a component $e^{\prime}$ not in $\att^0$ that is not monitored by every detector positioning in the support of $\sigma^{1^*}$.
Let us assume the contrary, i.e., that \rev{for every} $\Def \in \supp(\sigma^{1^*})$ \rev{and every} $e \notin \att^0, \ \valdet{\Def}{e} = 1.$
First, let us denote $\att^1 \subseteq \att^0$ the subset of components of $\att^0$ that are unmonitored by at least one detector positioning $\Def \in \supp(\sigma^{1^*})$, i.e.,  $\att^1 \coloneqq  \bigcup_{\Def \in \supp(\sigma^{1^*})} (\edges\backslash\set{\Def}) \subseteq \att^0.$

For all $S \in \supp(\sigma^{1^*})$, let $k_S$ denote the number of components of $T^1$ that are not monitored by $S$. 
Since every component outside of $\att^{1}$ is monitored by every detector positioning in the support of $\sigma^{1^*}$, then \attacker's best response to $\sigma^{1^*}$ is any \rev{attack strategy $\sigma^2 \in \Delta(\mathcal{A}_2)$ such that for all $\att \in \supp(\sigma^2)$,} $\att^1 \subseteq \att$ (note that $|\att^1| \leq  b_2$). \rev{This implies that:} 
\begin{align}
\rev{\forall \sigma^2 \in B_2(\sigma^{1^*}), \ U(\sigma^{1^*},\sigma^2) = \mathbb{E}_{\sigma^{1^*}}[k_S] \eqqcolon k^*.}\label{Eq_k*}
\end{align}


Consider $\Def \in \supp(\sigma^{1^*})$. We know that $\Def$ leaves $k_\Def$ network components unmonitored, that we denote $e_1,\dots,e_{k_\Def}$. For $l \in \llbracket 1,k_{\Def}\rrbracket$, let  $i_l$ be a node from where a detector can monitor component $e_l$. Then, $
\Def \cup\{i_1,\dots,i_{k_{\Def}}\}$ is a set cover (of size at most $b_1+k_\Def$). By definition of $\smsc$, we deduce that $b_1 + k_S \geq \smsc$. Therefore, \rev{for all} $S \in \supp(\sigma^{1^*}), \ k_S \geq \smsc - b_1$.
\rev{Since we are in the case when $b_2 < \semm$, and we have $\semm \leq \smsc$ (Lemma~\ref{MM<MSC}), then we obtain:} 
\begin{align}
\rev{k^* =  \mathbb{E}_{\sigma^{1^*}}[k_S] \geq \smsc - b_1 \geq \frac{\semm}{\smsc}(\smsc - b_1) > \frac{b_2}{\smsc}(\smsc - b_1).} \label{Ineq_k*}
\end{align}
%
\rev{Since $\sigma^{1^*}$ is an equilibrium inspection strategy, the value of the game is $U^*(b_1,b_2) \overset{\eqref{Eq_k*}}{=} k^*$. However, from Proposition~\ref{best_set_cover}, we know that $U^*(b_1,b_2) \leq b_2\left(1 - \frac{b_1}{\smsc} \right)$, which contradicts \eqref{Ineq_k*}.}

Thus, \rev{there exists} $(e^\prime,\Def^\prime) \in \edges\backslash\att^0 \times \supp(\sigma^{1^*})$ \rev{such that} $\valdet{\Def^\prime}{e^\prime} = 0$.
%
%
%
%


\item[--] Now, we can show that \attacker can increase her payoff by targeting component $e^{\prime}$ and the components in $\att^0$.  Let $\att^{\prime} = \att^0 \cup \{e^{\prime}\} \in \mathcal{A}_2$. Then, $\rev{U}(\sigma^{1^*},\att^{\prime}) \geq  \sigma^{1^*}_{\Def^\prime} + \rev{U}(\sigma^{1^*},\att^{0}) > \rev{U}(\sigma^{1^*},\att^{0})$, 
which is a contradiction. Therefore, \rev{for every $\att \in \supp(\sigma^{2^*}), \ |\att|= b_2$.}
\end{enumitemize}
%
%
\item[(ii)] Finally, we show that \defender's strategies in equilibrium \rev{of $\Gamma$} are the optimal solutions of \lpob. First, from (i), we can deduce that the set of optimal solutions of \lpo is a subset of $\Delta(\overline{\mathcal{A}_1})$. 
%
Therefore, the equilibrium inspection strategies are the optimal solutions of \rev{$\min_{\sigma^1 \in \Delta(\overline{\mathcal{A}_1})} \max_{\att \in \mathcal{A}_2} {U}(\sigma^1,\att).$}

Now, consider an inspection strategy $\sigma^1 \in \Delta(\overline{\mathcal{A}_1})$. Since $\overline{\mathcal{A}_2} \subseteq \mathcal{A}_2$, then we trivially have \rev{$\max_{\att \in \mathcal{A}_2} U(\sigma^1,\att) \geq \max_{\att \in \overline{\mathcal{A}_2}} U(\sigma^1,\att)$}. To obtain the reverse inequality, 
let $\att^0 \in \mathcal{A}_2$ be an attack plan that satisfies $
\att^0 \in \rev{\argmax_{\att \in \mathcal{A}_2}  U(\sigma^1,\att)}.$ Then, consider $\att^\prime \in \overline{\mathcal{A}_2}$ \rev{such that} $\att^0 \subseteq \att^\prime$. We can deduce that \rev{$\max_{\att \in \mathcal{A}_2} U(\sigma^1,\att) =  U(\sigma^1,\att^0)  \leq U(\sigma^1,\att^{\prime}) \leq \max_{\att \in \overline{A_2}} U(\sigma^1,\att)$}.

Therefore,  \rev{for all $\sigma^1 \in \Delta(\overline{\mathcal{A}_1}), \ \max_{\att \in \mathcal{A}_2} U(\sigma^1,\att) = \max_{\att \in \overline{\mathcal{A}_2}} U(\sigma^1,\att)$}, which implies that 
\rev{$
\min_{\sigma^1 \in \Delta(\overline{\mathcal{A}_1})} \max_{\att \in \mathcal{A}_2} {U}(\sigma^1,\att) = \min_{\sigma^1 \in \Delta(\overline{\mathcal{A}_1})} \max_{\att \in \overline{\mathcal{A}_2}} {U}(\sigma^1,\att).$} 
Thus, the equilibrium inspection strategies are the optimal solutions of \lpob.
An analogous proof can be applied to show that \rev{the attack strategies in equilibrium of $\Gamma$} are the optimal solutions of \lptb.\hfill \Halmos
\end{enumerate}
\endproof

\rev{
\proof{Proof of Proposition~\ref{Eq_P}.}\ 

\begin{enumerate}
\item
\rev{Let $\sigma^{1^*} $ be an optimal solution of $\max_{\sigma^1 \in \Delta(\mathcal{A}_1)}\min_{\sigma^2 \in B_2(\sigma^1)} r(\sigma^1,\sigma^2)$. We now show that any best response of \attacker to $\sigma^{1^*}$ uses all resources. Instead, suppose that there exist $\sigma^2 \in B_2(\sigma^{1^*})$ and $\att^0 \in \supp(\sigma^{2})$ such that $|\att^0| <b_2$.} \rev{Analogously to the proof of Theorem~\ref{all_resources}, showing that there exists a component not in $\att^0$ that is not monitored by every detector positioning in the support of $\sigma^{1^*}$ leads to a contradiction.}
%
%
%
\rev{By definition of $\sigma^{1^*}$, we have the following lower bound:}
\begin{align}
\rev{\min_{\sigma^2 \in B_2(\sigma^{1^*})} r(\sigma^{1^*},\sigma^2) \geq \min_{\sigma^2 \in B_2( \sold{\Def^{min}}{b_1}{})} r(\sold{\Def^{min}}{b_1}{},\sigma^2)}.\label{contradict0}
\end{align}

\rev{Note that a consequence of Proposition~\ref{best_set_cover} and \eqref{for later} is that for every $\sigma^2 \in B_2( \sold{\Def^{min}}{b_1}{})$ and every $\att \in \supp(\sigma^2)$, $|\att| = b_2$. Therefore, we have:}
\begin{align}
& \rev{\min_{\sigma^2 \in B_2( \sold{\Def^{min}}{b_1}{})} r(\sold{\Def^{min}}{b_1}{},\sigma^2) \overset{\eqref{payoff2},\eqref{Exp_detection_rate}}{=}\min_{\sigma^2 \in B_2( \sold{\Def^{min}}{b_1}{})} \mathbb{E}_{( \sold{\Def^{min}}{b_1}{},\sigma^2)}\left[1 - \frac{U(S,T)}{|\att|}\right]} \nonumber\\
& \rev{= 1 - \frac{1}{b_2}\max_{\sigma^2 \in B_2( \sold{\Def^{min}}{b_1}{})} U(\sold{\Def^{min}}{b_1}{},\sigma^2) \overset{\eqref{Tight_Ineq}}{=} \frac{b_1}{\smsc}.} \label{contradict1}
\end{align}



\rev{Now, if we assume that \rev{for every} $\Def \in \supp(\sigma^{1^*})$ \rev{and every} $e \notin \att^0, \ \valdet{\Def}{e} = 1,$ we obtain the following upper bound from the calculations in the proof of Theorem~\ref{all_resources}:}
\begin{align*}
\rev{\min_{\sigma^2 \in B_2(\sigma^{1^*})} r(\sigma^{1^*},\sigma^2)} & \rev{\overset{\eqref{payoff2},\eqref{Exp_detection_rate}}{\leq}  \min_{\sigma^2 \in B_2(\sigma^{1^*})} \mathbb{E}_{(\sigma^{1^*},\sigma^2)}\left[1 - \frac{1}{b_2}U(S,T)\right]  \overset{\eqref{Eq_k*}}{=}  1 - \frac{1}{b_2}k^* \overset{\eqref{Ineq_k*}}{<}  \frac{b_1}{\smsc},}
\end{align*}
which contradicts the lower bound obtained from \eqref{contradict0} and \eqref{contradict1}. Thus, \rev{there exists} $(e^\prime,\Def^\prime) \in \edges\backslash\att^0 \times \supp(\sigma^{1^*})$ \rev{such that} $\valdet{\Def^\prime}{e^\prime} = 0$, which contradicts $\att^0$ being in the support of a best response to $\sigma^{1^*}$. Therefore, for every $\sigma^2 \in B_2(\sigma^{1^*})$ and every $\att \in \supp(\sigma^2)$, $|\att| = b_2$.

%
%

\item
Consider $b_1 \in \mathbb{N}$. First, we note that:
\begin{align*}
\exists \, \sigma^1 \in \Delta(\mathcal{A}_1) \ | \ r(\sigma^1,\sigma^2) \geq \alpha, \ \forall \sigma^2 \in B_2(\sigma^1) \ \Longleftrightarrow \max_{\sigma^1 \in \Delta(\mathcal{A}_1)} \min_{\sigma^2 \in B_2(\sigma^1)} r(\sigma^1,\sigma^2)\geq \alpha.
\end{align*}

Then,
from Theorem~\ref{all_resources}, and by definition of the best response function $B_2$, we have:
\begin{align*}
\max_{\sigma^1 \in \Delta(\mathcal{A}_1)} \min_{\sigma^2 \in B_2(\sigma^1)} r(\sigma^1,\sigma^2) \geq \alpha &\overset{\eqref{payoff2},\eqref{Exp_detection_rate},\eqref{Att_BR}}{\Longleftrightarrow} 1 - \frac{1}{b_2} \min_{\sigma^1 \in \Delta(\mathcal{A}_1)} \max_{\sigma^2 \in \Delta(\mathcal{A}_2)} U(\sigma^1,\sigma^2) \geq \alpha \\
& \ \ \; \Longleftrightarrow \ 1 - \frac{1}{b_2}U(\sigma^{1^*},\sigma^{2^*}) \geq \alpha, \ \forall (\sigma^{1^*},\sigma^{2^*}) \in \Sigma(b_1,b_2)\\
& \overset{\eqref{payoff2},\eqref{Exp_detection_rate},\eqref{res2}}{\Longleftrightarrow} \ r(\sigma^{1^*},\sigma^{2^*}) \geq \alpha, \ \forall (\sigma^{1^*},\sigma^{2^*}) \in \Sigma(b_1,b_2).
\end{align*}

Thus, $b_1^* = \argmin\{b_1 \in \mathbb{N} \ | \ r(\sigma^*) \geq \alpha, \ \forall \sigma^* \in \Sigma(b_1,b_2)\}$.

\item
Let $\sigma^{1^*}$ be an optimal solution of \lpo. Then, we have:
\begin{align*}
& \max_{\sigma^2 \in B_2(\sigma^{1^*})} U(\sigma^{1^*},\sigma^2) \leq \max_{\sigma^2 \in B_2(\sigma^{1})} U(\sigma^{1},\sigma^2), \ \forall \sigma^1 \in \Delta(\mathcal{A}_1) \\ 
\overset{\eqref{payoff2},\eqref{Exp_detection_rate},\eqref{res2}}{\Longleftrightarrow} & \max_{\sigma^2 \in B_2(\sigma^{1^*})} (1-r(\sigma^{1^*},\sigma^2))b_2 \leq \max_{\sigma^2 \in B_2(\sigma^{1})} \mathbb{E}_{(\sigma^1,\sigma^2)}[(1 - r(\Def,\att)) \underset{\leq b_2}{\underbrace{|\att|}}], \ \forall \sigma^1 \in \Delta(\mathcal{A}_1)\\ 
\Longrightarrow \ \ \;& \min_{\sigma^2 \in B_2(\sigma^{1^*})}r(\sigma^{1^*},\sigma^2) \geq  \min_{\sigma^2 \in B_2(\sigma^{1})}r(\sigma^{1},\sigma^2), \ \forall  \sigma^1 \in \Delta(\mathcal{A}_1).
\end{align*}

Therefore, $\sigma^{1^*} \in \argmax_{\sigma^{1} \in \Delta(\mathcal{A}_1)}\min_{\sigma^2 \in B_2(\sigma^{1})}r(\sigma^{1},\sigma^2)$. Thus, 
%
an equilibrium inspection strategy of $\Gamma(b_1^*,b_2)$ is an optimal inspection strategy of \ECOP. \hfill
\Halmos
\end{enumerate}

\endproof}

\proof{Proof of Proposition~\ref{all_edges}.}\
\begin{enumerate}
\item
We show the result by contradiction, that is, suppose that \rev{there exists} $({\sigma^1}^*,{\sigma^2}^*) \in \nash$ such that $\nbasis{\sigma^{1^*}}$ is not a set cover. For simplicity, we introduce the following notation:
\begin{align}
\forall e \in \edges, \ \eta_{\sigma^{1^*}}(e) \coloneqq \mathbb{E}_{\sigma^{1^*}}[\valdet{\Def}{e}],\label{eta_detection}
\end{align}
which is the probability with which component $e$ is monitored by $\sigma^{1^*}$.
Let us sort the components in nondecreasing order of monitoring probability: $
\eta_{\sigma^{1^*}}(e_1) \leq \eta_{\sigma^{1^*}}(e_2) \leq \dots \leq \eta_{\sigma^{1^*}}(e_{|\edges|}).$ Then,  $\att^{\prime} = \{e_1,\dots,e_{b_2}\}$ is a best response to $\sigma^{1^*}$ for \attacker  (recall that \attacker uses all her resources; see \rev{Theorem}~\ref{all_resources}).
%
%
Let  $\att^0 = \{e_1,\dots,e_{k}\} \in 2^{\edges}$ denote the components that are not monitored by any detector positioning $\Def \in \supp(\sigma^{1^*})$. Then,  $\eta_{\sigma^{1^*}}(e_1) = \dots = \eta_{\sigma^{1^*}}(e_k) = 0$, and \rev{the value of the game $\Gamma$ is}:
\begin{align}
\rev{U^*(b_1,b_2) = U}(\sigma^{1^*},\att^{\prime}) & \overset{\eqref{payoff2},\eqref{detection_disruptions},\eqref{eta_detection}}{=}  b_2 - \sum_{e \in \att^{\prime}}\eta_{\sigma^{1^*}}(e) =   b_2 - \sum_{i=k+1}^{b_2}\eta_{\sigma^{1^*}}(e_i).\label{inequality_useful_later}
\end{align}

Now, to show the contradiction, we construct another strategy $\widehat{\sigma}^1$ that will provide a better payoff than $\sigma^{1^*}$ to \defender.


\begin{itemize}
\item[--] \textbf{Case 1:} $k \geq b_2$. Then \rev{the value of the game is $b_2$} (it corresponds to zero detections). However, \defender has an incentive to switch her strategy, and by randomizing over the nodes that can monitor the $k$ components in $\att^0$, she will \rev{improve} her payoff; which is a contradiction.
\item[--] \textbf{Case 2:} $k < b_2$. Then \attacker will randomize over attack plans that contain $\att^0$. 
\begin{itemize}
\item[-] \textbf{Case 2.1:} $k\geq b_1$. Then, \rev{$U^*(b_1,b_2) \leq b_2 - b_1$} (since \defender can monitor $b_1$ components in $\att^0$ that are always targeted). However, \rev{from Proposition~\ref{best_set_cover}, we have $U^*(b_1,b_2) \geq b_2 - \frac{b_1b_2}{\semm}> b_2 - b_1$} (since $b_2 < m^*$). Therefore there is a contradiction.

\item[-] \textbf{Case 2.2:} $k < b_1$. Then, the idea is to construct another strategy that positions $k$ detectors to monitor the components in $\att^0$ (that were previously unmonitored), and that randomizes the positioning of the remaining $b_1 - k$ detectors over the node basis of $\sigma^{1^*}$.

For now, assume that \defender has $b_1 - k$ detectors. For any detector positioning $\Def \in \supp(\sigma^{1^*})$ (viewed as a set of nodes), we consider $\sold{\Def}{b_1-k}{}$ defined in \eqref{sigma^S} (in this case, we randomize the \rev{positioning} of $b_1 - k$ detectors over the set $\Def$ of size $b_1$). Recall that $\supp(\sold{\Def}{b_1-k}{}) = \{S^1,\dots,S^{\order{b_1}{b_1-k}}\}$, and that \rev{for all} $l \in \llbracket 1,\order{b_1}{b_1-k} \rrbracket, \ \sold{\Def}{b_1-k}{\Def^l} = \frac{1}{\order{b_1}{b_1-k}}$.
Now, let us construct the following inspection strategy:
\begin{align}
\sigma^{1^{\prime}} = \sum_{\Def \in \supp(\sigma^{1^*})}\sigma^{1^*}_{\Def}\sold{\Def}{b_1-k}{}.\label{sigma1'}
\end{align}

One can check that $\sigma^{1^\prime} \in \Delta(\mathcal{A}_1)$, and is such that 
$\supp(\sigma^{1^{\prime}}) = \cup_{\Def \in \supp(\sigma^{1^*})}\{\Def^1,\dots,\Def^{\order{b_1}{b_1-k}}\}$. 
Thus, $\sigma^{1^{\prime}}$ is a probability distribution that randomizes over detector positionings of size $b_1 - k$. Then, \rev{for every} $\Def \in \supp(\sigma^{1^{\prime}})$, we augment $\Def$ by \rev{positioning} $k$ additional detectors to monitor the subset of  components  $\att^0$ that was previously unmonitored and that is always targeted in equilibrium by \attacker. We denote $\{i_1,\dots,i_k\}$ the \rev{positioning} of such additional detectors, and we denote $\widehat{\Def} = \Def \cup \{i_1,\dots,i_k\}$ \rev{for every} $\Def \in \supp(\sigma^{1^\prime})$ the augmented detector positioning. Then, we consider the probability distribution $\widehat{\sigma}^1$ with support equal to $\cup_{\Def \in \supp(\sigma^{1^*})}\{\widehat{\Def}^1,\dots,\widehat{\Def}^{\order{b_1}{b_1-k}}\}$ and such that:
\begin{align}
\forall \Def \in \supp(\sigma^{1^{\prime}}), \ \sigma^{1^{\prime}}_{\Def} = \widehat{\sigma}^1_{\widehat{\Def}},\label{S=Shat}
\end{align}
i.e., $\widehat{\sigma}^1$ is the same probability distribution as $\sigma^{1^\prime}$ except that it randomizes over the augmented detector positionings present in the support of $\sigma^{1^\prime}$.


Then, we can derive the following calculations which combine the previous construction of $\widehat{\sigma}^1$ with a property of the detection function derived in \eqref{nice_detection_inequality}, and which will lead to a contradiction:
\begin{align}
&\forall \att \in \mathcal{A}_2 \ | \ \att^0 \subset \att \text{ and } |\att| = b_2, \ \mathbb{E}_{\widehat{\sigma}^1}[\valdet{\Def}{\att}] \overset{\eqref{additive_equality}}{=} \mathbb{E}_{\widehat{\sigma}^1}[\underset{=k}{\underbrace{\valdet{\Def}{\att^0}}}] + \mathbb{E}_{\widehat{\sigma}^1}[\valdet{\Def}{\att\backslash\att^0}] \nonumber\\
& = k +\sum_{\Def \in \supp(\sigma^{1^*})} \sum_{l=1}^{\order{b_1}{b_1-k}}\widehat{\sigma}^1_{\widehat{\Def}^l} \valdet{\widehat{\Def}^l}{\att\backslash\att^0} \overset{\eqref{monotonicity_detection},\eqref{S=Shat}}{\geq} k +  \sum_{\Def \in \supp(\sigma^{1^*})} \sum_{l=1}^{\order{b_1}{b_1-k}}\sigma^{1^{\prime}}_{\Def^l} \valdet{\Def^l}{\att\backslash\att^0}\nonumber\\
&\overset{\eqref{sigma^S},\eqref{sigma1'}}{=} k +\sum_{\Def \in \supp(\sigma^{1^*})} \sum_{l=1}^{\order{b_1}{b_1-k}}\frac{1}{\order{b_1}{b_1-k}}\sigma^{1^{*}}_{\Def} \valdet{\Def^l}{\att\backslash\att^0} \overset{\eqref{nice_detection_inequality}}{\geq} k + \sum_{\Def \in \supp(\sigma^{1^*})} \frac{\antiorder{b_1}{b_1-k}}{\order{b_1}{b_1-k}}\sigma^{1^{*}}_{\Def}\valdet{\Def}{\att\backslash\att^0} \nonumber\\
&\overset{\eqref{detection_disruptions},\eqref{eta_detection}}{=} k + \underset{>0}{\underbrace{\frac{b_1 - k}{b_1}}}\sum_{e \in \att\backslash\att^0}\eta_{\sigma^{1^*}}(e) \geq  k + \frac{b_1 - k}{b_1}\sum_{i=k+1}^{b_2}\eta_{\sigma^{1^*}}(e_i)\nonumber\\
& = \sum_{i=k+1}^{b_2}\eta_{\sigma^{1^*}}(e_i) + k\left(1 - \frac{1}{b_1}\sum_{i=k+1}^{b_2}\eta_{\sigma^{1^*}}(e_i)  \right).\label{big_calculation!}
\end{align}


From Proposition~\ref{best_set_cover}, we know that \rev{$U^*(b_1,b_2) \geq b_2 - \frac{b_1b_2}{m^*}$}. Therefore:
\begin{align*}
U(\sigma^{1^*},\sigma^{2^*}) \overset{\eqref{inequality_useful_later}}{=}b_2 - \sum_{i=k+1}^{b_2}\eta_{\sigma^{1^*}}(e_i)  \geq b_2 - \frac{b_1b_2}{m^*} \Longleftrightarrow \frac{1}{b_1}\sum_{i=k+1}^{b_2}\eta_{\sigma^{1^*}}(e_i)  \leq \frac{b_2}{m^*}  < 1,
\end{align*}

since $b_2 < m^*$. Then, by combining the previous inequality with \eqref{big_calculation!}, we obtain:
\begin{align*}
\forall \att \in \mathcal{A}_2 \ | \ \att^0 \subset \att \text{ and } |\att| = b_2, \ \rev{U(\widehat{\sigma}^1,\att) \overset{\eqref{payoff2},\eqref{big_calculation!}}{<} b_2 -\sum_{i=k+1}^{b_2}\eta_{\sigma^{1^*}}(e_i) \overset{\eqref{inequality_useful_later}}{=} U(\sigma^{1^*},\sigma^{2^*}).}
\end{align*}

Since each attack plan in the support of an equilibrium strategy uses all the resources (\rev{Theorem}~\ref{all_resources}), and must contain $\att^0$ (beginning of Case 2), then: \rev{$U(\widehat{\sigma}^1,\sigma^{2^*}) = \mathbb{E}_{\sigma^{2^*}}[U(\widehat{\sigma}^1,\att)]< U(\sigma^{1^*},\sigma^{2^*}),$} which violates the equilibrium condition \eqref{best}.\end{itemize}
\end{itemize}

Therefore, \rev{for all} $({\sigma^1}^*,{\sigma^2}^*) \in \nash, \ \nbasis{\sigma^{1^*}}$ is a set cover.

%
\item \rev{Next, we show that both players must randomize their actions in equilibrium.} \rev{Since the node basis in equilibrium is a set cover and $b_1<n^*$, then} \defender must randomize her detector positionings in equilibrium.
Now, assume that there exists a NE $(\sigma^{1^*},\att) \in \nash$ such that \attacker chooses a pure strategy $\att$ (of size $b_2$ from \rev{Theorem}~\ref{all_resources}). 
\begin{enumitemize}
\item[--] If $b_1 \geq b_2$, then \defender can detect all the attacks in $\att$ by \rev{positioning} $b_2$ detectors at the nodes that can monitor the  components of $\att$, and \rev{the value of the game is 0}. 
However, we showed in the proof of \rev{Theorem}~\ref{all_resources} that there exists a component outside of $\att$ that is not monitored with positive probability by $\sigma^{1^*}$. Therefore, \attacker can increase her payoff by targeting that component, thus leading to a contradiction.

\item[--] If $b_1 < b_2$, then \defender can detect at least $b_1$ attacks in $\att$ by \rev{positioning}  detectors on $b_1$ nodes that can collectively monitor $b_1$ components of $\att$. \rev{Thus, $U^*(b_1,b_2) \leq b_2 - b_1$. However, from Proposition~\ref{best_set_cover}, we know that  $U^*(b_1,b_2)\geq b_2(1 - \frac{b_1}{\semm}) > b_2 - b_1$ (since $b_2 < \semm$), thus leading to a contradiction.}
\end{enumitemize}

Therefore, in equilibrium, both players must randomize their actions.\hfill \Halmos
\end{enumerate}
%
\endproof

\rev{
\begin{proposition}\label{Prop:Bounds_r}
For $b_1<\smsc$ and $b_2<\semm$, the expected detection rate in equilibrium of $\Gamma(b_1,b_2)$ is constant and bounded as follows:
%
%
%
\begin{align*}
\forall \sigma^* \in \Sigma(b_1,b_2), \quad \dfrac{b_1}{\smsc} \leq\sdet{*} \leq \min\left\{\dfrac{b_1}{\semm},1\right\}.
\end{align*}
%
%

Furthermore, given an MSC $\Def^{min} \in \msc$, the expected detection rate by positioning $b_1$ detectors according to $\sold{\Def^{min}}{b_1}{}$ provides the following detection guarantee: \vspace{0.15cm}
\begin{align*}
\displaystyle\min_{\sigma^2 \in \Delta(\mathcal{A}_2)} r(\sold{\Def^{min}}{b_1}{},\sigma^2) \,= \, \frac{b_1}{\smsc} \,\geq\, \frac{\max\{b_1,\semm\}}{\smsc} r(\sigma^{1^*},\sigma^{2^*}), \quad \forall (\sigma^{1^*},\sigma^{2^*}) \in \nash(b_1,b_2).
\end{align*}

%
\end{proposition}

}

\proof{Proof of Proposition~\ref{Prop:Bounds_r}.}\
\rev{Theorem~\ref{all_resources} implies that for all $\sigma^* \in \nash, \ r(\sigma^*) \overset{\eqref{payoff2},\eqref{Exp_detection_rate},\eqref{res2}}{=} 1 - \frac{1}{b_2}U^*(b_1,b_2)$. Therefore, the equilibrium expected detection rate is constant. Then, the MSC/MSP-based bounds in Proposition~\ref{best_set_cover} can be translated to bounds on the equilibrium detection rate as follows: For all $\sigma^* \in \nash, \ \frac{b_1}{\smsc} \leq r(\sigma^*) \leq  \min\left\{\frac{b_1}{\semm},1\right\} = \frac{b_1}{\max\{b_1,\semm\}}.$ 
Finally, we can deduce that for all $\sigma^* \in \nash$, $\frac{\max\{b_1,\semm\}}{\smsc} r(\sigma^*) \leq \frac{b_1}{\smsc} \overset{\eqref{payoff2},\eqref{Exp_detection_rate},\eqref{res2},\eqref{Tight_Ineq}}{=} \min_{\sigma^2 \in \Delta(\mathcal{A}_2)} r{(\sold{\Def^{min}}{b_1}{},\sigma^2)}.$ \hfill
\Halmos}

\endproof

\ifadditional
\begin{corollary}[of Theorem~\ref{bounds_game}]\label{Cor1}
Consider a network $\mathcal{G}$, a sensing model $\{\set{i}, \ i \in \nodes\}$, and  the players' resources $b_1 < \smsc$ and $b_2 < \semm$. If $\smsc = \semm$, then the game $\Gamma(b_1,b_2)$ has the following properties:
\begin{enumerate}
\item
The equilibrium payoffs of both players are constant and are given by:
\begin{align}
\forall (\sigma^{1^*},\sigma^{2^*}) \in \nash(b_1,b_2), \quad \begin{cases} U_1 ({\sigma^1}^*,{\sigma^2}^*) = \dfrac{  b_1  b_2}{\smsc},  \\
U_2 ({\sigma^1}^*,{\sigma^2}^*) =   b_2  \left(1 - \dfrac{  b_1  }{\smsc}\right).\end{cases}\label{eq_payoff_EC}
\end{align}
\item In any equilibrium, the expected detection rate is constant and is given by:
\begin{align}
\forall \sigma^* \in \nash(b_1,b_2), \quad \sdet{*} = \dfrac{b_1}{\smsc}.\label{eq_srate_EC}
\end{align}
\end{enumerate}
\end{corollary}

\proof{Proof of Corollary~\ref{Cor1}.}
By rewriting Thm.~\ref{bounds_game} when $\smsc = \semm$, we obtain $\forall (\sigma^{1^*},\sigma^{2^*}) \in \nash(b_1,b_2):$
\begin{align*}
& \frac{b_1b_2}{\smsc }\leq  U_1(\sigma^{1^*},\sigma^{2^*})  \leq \min\left\{\frac{b_1b_2}{\semm },b_2\right\} = \min\left\{\frac{b_1b_2}{\smsc },b_2\right\} =\frac{b_1b_2}{\smsc },\\
& b_2\left(1 - \frac{b_1}{\smsc }\right) = \max\left\{0,b_2\left(1 - \frac{b_1}{\smsc }\right)\right\} = \max\left\{0,b_2\left(1 - \frac{b_1}{\semm }\right)\right\}\leq  U_2(\sigma^{1^*},\sigma^{2^*})  \leq b_2\left(1 - \frac{b_1}{\smsc }\right),
\end{align*}
since $b_1 <\smsc$.

Similarly, for the expected detection rate in equilibrium, we obtain:
\begin{align*}
\forall \sigma^* \in \nash(b_1,b_2), \ \frac{b_1}{\smsc} \leq r(\sigma^*) \leq \min\left\{\frac{b_1}{\semm},1\right\}= \min\left\{\frac{b_1}{\smsc},1\right\} = \frac{b_1}{\smsc}.
\end{align*}
\hfill \Halmos
\endproof
\else
\fi

\ifadditional
\begin{lemma}\label{permutation}
For any MSC $\Def^{\text{min}} \in \mathcal{S}$ and any MSP $\att^{\text{max}} \in \mathcal{M}$ such that $\smsc = \semm$, each component in $\att^{\text{max}}$ is monitored by only one node in $\Def^{\text{min}}$, and each node in $\Def^{\text{min}}$ monitors only one component in $\att^{\text{max}}$.
%
\end{lemma}
\proof{Proof of Lemma~\ref{permutation}.}
Consider an MSC $\Def^{min} \in \msc$ and an MSP $\att^{max} \in \emm$. Since $\att^{max}$ is an MSP, each node in $\Def^{min}$ monitors at most one component in $\att^{max}$. Now, assume that at least one node in $\Def^{min}$ does not monitor any component in $\att^{max}$. Since $\Def^{min}$ is an MSC, then the $\smsc$ components in $\att^{max}$ are monitored by at most $\smsc-1$ nodes. From Dirichlet's principle, there exists a node in $\Def^{min}$ that monitors at least two components in $\att^{max}$, which is a contradiction. Therefore, each node in $\Def^{min}$ monitors exactly one component in $\att^{max}$. 

Thus, we can define a mapping $\psi: \Def^{min} \longrightarrow \att^{max}$ such that $\forall i \in \Def^{min}, \ \psi(i)$ is the component in $\att^{max}$ that is monitored by node $i$. Now, since $\Def^{min}$ is an MSC, for every component $e \in \att^{max}, \ \exists \, i \in \Def^{min}$ such that $e$ is monitored by $i$. Therefore, $\psi$ is surjective which is equivalent to $\psi$ being injective since its domain and codomain have the same number of elements.
Therefore, each component in $\att^{max}$ is monitored by only one node in $\Def^{min}$.
\hfill
\Halmos
\endproof

This lemma tells us that no matter which MSC and which MSP we consider, there is a one-to-one correspondence between a node in the MSC and the component in the MSP in monitors.

In fact, we can obtain a sufficient \emph{and} necessary condition for a strategy profile constructed over an MSC and MSP to be a NE when $\smsc = \semm$.


%

\begin{proposition}\label{MSC/EMM-NE}
%
If $\smsc = \semm$, $b_1 <\smsc$, and $b_2 < \semm$, then for any MSC $\Def^{min} \in \msc$ and any MSP $\att^{max} \in \emm$, a strategy profile $(\sigma^{1^*},\sigma^{2^*}) \in \Delta(\mathcal{A}_1) \times \Delta(\mathcal{A}_2) $ whose node basis is $\Def^{min}$ and whose component basis is $\att^{max}$ is a NE if and only if:
\begin{align}
\forall i \in \Def^{min}, \quad  \mathbb{P}_{\sigma^{1^*}}(i \text{ is monitored}) = \frac{b_1}{\smsc}, \quad \text{and} \quad \forall e \in \att^{max},\quad \mathbb{P}_{\sigma^{2^*}}(e \text{ is targeted}) = \frac{b_2}{\smsc}. \label{if and only if}
\end{align}
\end{proposition}

\proof{Proof of Proposition~\ref{MSC/EMM-NE}.}
For simplicity, given an optimization problem $(\mathcal{Q})$, we denote $\OPT{(\mathcal{Q})}$ its objective value. Let $\Def^{min} \in \msc$ be an MSC and  $\att^{max} \in \emm$ be an MSP, and assume that they are of same size. Recall that from Proposition~\ref{best_set_cover}, we have  $\OPT{\rlpom{\Def^{min}}} \leq  \OPT{\lpom} =- \OPT{\lptm }\leq - \OPT{\rlptm{\att^{max}}}$. Since $\smsc = \semm$ (and $b_1 < \smsc$), then we also have:
\begin{align*}
\OPT{\rlpom{\Def^{min}}} = b_2 \left(\frac{b_1}{\smsc} - 1 \right)  = - b_2 \left(1 - \frac{b_1}{\semm}  \right)  = - \max\left\{0,b_2 \left(1 - \frac{b_1}{\semm}  \right) \right\} =  - \OPT{\rlptm{\att^{max}}}.
\end{align*}

Therefore, $\OPT{\rlpom{\Def^{min}}} = \OPT{\lpom}$ and  $\OPT{\rlptm{\att^{max}}} = \OPT{\lptm }$. Since \rlpo{\Def^{min}} (resp. \rlpt{\att^{max}}) is a restriction of \lpo (resp. \lpt), then any sensing strategy $\sigma^{1^*}$ (resp. attack strategy $\sigma^{2^*}$) that is optimal for  \rlpo{\Def^{min}} (resp. \rlpt{\att^{max}}) is optimal for \lpo (resp. \lpt). Thus, any strategy profile $(\sigma^{1^*},\sigma^{2^*})$ such that $\sigma^{1^*}$ and $\sigma^{2^*}$ are optimal solutions of \rlpo{\Def^{min}}  and \rlpt{\att^{max}} respectively is a NE. 

To show that a sensing strategy $\sigma^{1^*}$ (whose node basis is $\Def^{min}$) is an optimal solution of \rlpo{\Def^{min}}, it is sufficient to show that $\min_{\sigma^2 \in \Delta(\mathcal{A}_2)}-U_2(\sigma^{1^*},\sigma^2) \geq b_2\left(\frac{b_1}{\smsc} - 1\right) = \OPT{\rlpom{\Def^{min}}}$. By applying \eqref{1st_ineq_max-min} for $\Def^{min}$ (which is a minimal set cover), we know that $\min_{\sigma^2 \in \Delta(\mathcal{A}_2)}-U_2(\sold{\Def^{min}}{b_1}{},\sigma^2) \geq b_2\left(\frac{b_1}{\smsc} - 1\right)$. However, recall that to show this inequality, the only property from $\sold{\Def^{min}}{b_1}{}$ that we used was that its node basis is $\Def^{min}$ and that $\forall i \in \Def^{min}$, $\rho_{\sold{\Def^{min}}{b_1}{}}(i) = \frac{b_1}{\smsc}$. Therefore, any sensing strategy that satisfies the same conditions also satisfies the same inequality and is an optimal solution of \rlpo{\Def^{min}}.

Similarly, we can easily deduce from \eqref{1st_inteq_max-min_bis} that any attack strategy $\sigma^{2^*}$ whose component basis is $\att^{max}$ and which satisfies $\forall e \in \att^{max}, \ \rho_{\sigma^{2^*}}(e) = \frac{b_2}{\smsc}$ also satisfies the inequality $\min_{\sigma^1 \in \Delta(\mathcal{A}_1)} U_2(\sigma^1,\sigma^{2^*}) \geq b_2\left(1 - \frac{b_1}{\smsc} \right)$ and is an optimal solution of \rlpt{\att^{max}}.

%


Thus, any strategy profile $(\sigma^{1^*},\sigma^{2^*})$ whose node basis is $\Def^{min}$, whose component basis is $\att^{max}$, and that satisfies $\rho_{\sigma^{1^*}}(i) = \frac{b_1}{\smsc}, \forall i \in \Def^{min}, \ \text{and} \ \rho_{\sigma^{2^*}}(e) = \frac{b_2}{\smsc}, \ \forall e \in \att^{max},$ is a NE (it is a sufficient condition).

Now, let us show (by contradiction) that this is also a necessary condition.
\begin{enumerate}
\item Consider a NE $(\sigma^{1^*},\sigma^{2^*}) \in \nash$ whose node basis is an MSC $\Def^{min} = \{i_1,\dots,i_{\smsc}\} \in \msc$, and assume that the sensing probability is not identical among the nodes in $\Def^{min}$. Without loss of generality (by reordering the indices), assume that $\rho_{\sigma^{1^*}}(i_1) < \rho_{\sigma^{1^*}}(i_{b_2+1})$.

Consider an MSP $\att^{max} = \{e_1,\dots,e_{\smsc}\} \in \emm$. Thanks to Lemma~\ref{permutation}, and without loss of generality, we can rearrange the indices such that $\forall k \in \llbracket 1,\smsc\rrbracket, \ e_k$ is only monitored by node $i_k$  in $\Def^{min}$. Now consider $\sola{\att^{max}}{b_2}{}$ defined in Lemma~\ref{algebra2}. Since $\sola{\att^{max}}{b_2}{}$ satisfies the above-mentioned sufficient conditions, i.e., its component basis is $\att^{max}$ and each component of $\att^{max}$ is targeted with probability $\frac{b_2}{\smsc}$, then $\sola{\att^{max}}{b_2}{}$ is an optimal solution of \rlpt{\att^{max}}. This implies that $(\sigma^{1^*},\sola{\att^{max}}{b_2}{})$ is a NE. Furthermore, the support of $\sola{\att^{max}}{b_2}{}$ contains $\att^1 = \{e_1\}\cup\{e_2,\dots,e_{b_2}\}$ and $\att^{2} = \{e_{b_2+1}\} \cup \{e_2,\dots,e_{b_2}\}$ (which are different since $b_2 < \semm = \smsc$). Therefore, $\att^1$ and $\att^{2}$ should give the same payoff to \attacker. However, we have the following contradiction:
\begin{align}
& U_2(\sigma^{1^*},\att^1)  \overset{\eqref{payoff2},\eqref{additive_equality}}{=}  |\att^1| - \mathbb{E}_{\sigma^{1^*}}[\valdet{\Def}{e_1}] - \mathbb{E}_{\sigma^{1^*}}[\valdet{\Def}{\att^1\backslash\{e_1\}}]\nonumber\\
& = \underset{= |\att^2|}{\underbrace{|\att^{1}|}} - \mathbb{E}_{\sigma^{1^*}}[\mathds{1}_{\{i_1 \in \Def\}}]  - \mathbb{E}_{\sigma^{1^*}}[\valdet{\Def}{\underset{=\att^{2}\backslash\{e_{b_2+1}\}}{\underbrace{\att^1\backslash\{e_1\}}}}]\label{needs some explanation}\\
 &\overset{\eqref{rho_sigma1}}{=} |\att^{2}| - \rho_{\sigma^{1^*}}(i_{1}) - \mathbb{E}_{\sigma^{1^*}}[\valdet{\Def}{\att^{2}\backslash\{e_{b_2+1}\}}] >  |\att^{2}| - \rho_{\sigma^{1^*}}(i_{b_2+1}) - \mathbb{E}_{\sigma^{1^*}}[\valdet{\Def}{\att^{2}\backslash\{e_{b_2+1}\}}]\nonumber\\
&\overset{\eqref{rho_sigma1}}{=}  |\att^{2}| - \mathbb{E}_{\sigma^{1^*}}[\valdet{\Def}{e_{b_2+1}}] - \mathbb{E}_{\sigma^{1^*}}[\valdet{\Def}{\att^{2}\backslash\{e_{b_2+1}\}}] \overset{\eqref{payoff2},\eqref{additive_equality}}{=}  U_2(\sigma^{1^*},\att^{2}).\label{needs some explanation2}
\end{align}
Note that in \eqref{needs some explanation} (resp. \eqref{needs some explanation2}), we used the fact that the node basis of $\sigma^{1^*}$ is $\Def^{min}$ and that $e_1$ (resp. $e_{b_2+1}$) is only monitored by $i_1$ (resp. $i_{b_2+1}$) in $\Def^{min}$. 

Thus, the sensing probability is necessarily identical among the nodes of $\Def^{min}$: $\exists \, \gamma \in \mathbb{R} \ | \ \forall i \in \Def^{min}, \ \rho_{\sigma^{1^*}}(i) \overset{\eqref{rho_sigma1}}{=} \sum_{\{\Def \in \mathcal{A}_1 \, | \, i \in \Def\}} \sigma^{1^*}_{\Def} = \gamma$. By summing over the nodes of $\Def^{min}$, and by combining Proposition~\ref{all_resources} with the fact that the node basis of $\sigma^{1^*}$ is $\Def^{min}$, we obtain:
\begin{align*}
\smsc\gamma &= \sum_{i \in \Def^{min}}\sum_{\{\Def \in \mathcal{A}_1 \, | \, i \in \Def\}} \sigma^{1^*}_{\Def} =   \sum_{i \in \Def^{min}}\sum_{\Def \in \mathcal{A}_1} \sigma^{1^*}_{\Def}\mathds{1}_{\{i \in \Def\}} =  \sum_{\Def \in \mathcal{A}_1}\sigma^{1^*}_{\Def} \sum_{i \in \Def^{min}} \mathds{1}_{\{i \in \Def\}} \overset{\eqref{res1}}{=} b_1\sum_{\Def \in \mathcal{A}_1}\sigma^{1^*}_{\Def} = b_1.
\end{align*}
Therefore, $\gamma = \frac{b_1}{\smsc}$, meaning that in any NE $(\sigma^{1^*},\sigma^{2^*})$ whose node basis is an MSC, the sensing probability must be equal to $\frac{b_1}{\smsc}$ for all the nodes of the MSC, which proves the necessary condition on the sensing strategy.

\item Similarly, consider a NE $(\sigma^{1^*},\sigma^{2^*}) \in \nash$ whose component basis is an MSP $\att^{max} = \{e_1,\dots,e_{\smsc}\} \in \emm$, and assume that the attack probability is not identical among the components of $\att^{max}$. Again, without loss of generality, we assume that $\rho_{\sigma^{2^*}}(e_{1}) < \rho_{\sigma^{2^*}}(e_{b_1+1})$. 

Consider now an MSC $\Def^{min} = \{i_1,\dots,i_{\smsc}\} \in \msc$. Again, we reorder the indices such that $\forall l \in \llbracket 1,\smsc\rrbracket$, node $i_l$ only monitors component $e_l$ in $\att^{max}$. Analogously, consider the sensing strategy $\sold{\Def^{min}}{b_1}{}$. From the sufficient conditions, we know that $\sold{\Def^{min}}{b_1}{}$ is an optimal solution of \rlpo{\Def^{min}} and that $(\sold{\Def^{min}}{b_1}{},\sigma^{2^*})$ is a NE. Since $\sold{\Def^{min}}{b_1}{}$ assigns positive probabilities on the sensor placements $\Def^1:=\{i_1\}\cup\{i_2,\dots,i_{b_1}\}$ and  $\Def^{2} := \{i_{b_1+1}\} \cup\{i_2,\dots,i_{b_1}\}$, $\Def^{1}$ and $\Def^{2}$ should provide \defender with the same payoff. However, we have the following contradiction:
%
%
%
\begin{align}
&U_1(\Def^1,\sigma^{2^*})  \overset{\eqref{payoff1}}{=} \mathbb{E}_{\sigma^{2^*}}[\valdet{\Def^1}{\att}] \overset{\eqref{detection_disruptions}}{=} \sum_{e \in \edges} \mathbb{E}_{\sigma^{2^*}}[\valdet{\Def^1}{e}\mathds{1}_{\{e \in \att\}}]  \overset{\eqref{rho_sigma2}}{=} \sum_{e\in \edges} \valdet{\Def^1}{e} \rho_{\sigma^{2^*}}(e)\nonumber\\
&=   \rho_{\sigma^{2^*}}(e_{1}) +  \sum_{e\in \att^{max} \backslash\{e_{1}\}} \valdet{\Def^1}{e} \rho_{\sigma^{2^*}}(e) <  \rho_{\sigma^{2^*}}(e_{b_1+1})+  \sum_{e\in \att^{max} \backslash\{e_{1}\}} \valdet{\Def^1}{e} \rho_{\sigma^{2^*}}(e)\label{oneee} \\
& =  \rho_{\sigma^{2^*}}(e_{b_1+1})+  \sum_{e\in  \{e_2,\dots,e_{b_1}\}} \valdet{\Def^1}{e} \rho_{\sigma^{2^*}}(e)  =  \rho_{\sigma^{2^*}}(e_{b_1+1})+  \sum_{e\in  \{e_2,\dots,e_{b_1}\}} \valdet{\Def^{2}}{e} \rho_{\sigma^{2^*}}(e)\label{three}\\
& =  \rho_{\sigma^{2^*}}(e_{b_1+1})+  \sum_{e\in \att^{max} \backslash\{e_{b_1+1}\}} \valdet{\Def^{2}}{e} \rho_{\sigma^{2^*}}(e) =  \sum_{e\in \edges} \valdet{\Def^{2}}{e} \rho_{\sigma^{2^*}}(e)  \overset{\eqref{payoff1},\eqref{detection_disruptions}}{=} U_1(\Def^{2},\sigma^{2^*}).\nonumber
\end{align}
In \eqref{oneee}, we used the fact that the component basis of $\sigma^{2^*}$ is $\att^{max}$. In \eqref{three}, we used the fact that $\Def^1$ only monitors components $\{e_1\}\cup\{e_2,\dots,e_{b_1}\}$ of $\att^{max}$, and that $\Def^1\backslash\{i_1\} = \Def^{2}\backslash\{i_{b_1+1}\}$.

Thus, the attack probability is identical among the components of $\att^{max}$: $\exists \, \gamma^{\prime} \in \mathbb{R} \ | \ \forall e \in \att^{max}, \ \rho_{\sigma^{2^*}}(e) \overset{\eqref{rho_sigma2}}{=} \sum_{\{\att \in \mathcal{A}_2 \, | \, e \in \att\}} \sigma^{2^*}_{\att} = \gamma^{\prime}$. By summing over the components of the MSP and by combining Proposition~\ref{all_resources} with the fact the component basis of $\sigma^{2^*}$ is $\att^{max}$, we obtain:
\begin{align*}
\smsc\gamma^{\prime} &= \sum_{e \in \att^{max}}\sum_{\{\att \in \mathcal{A}_2 \, | \, e \in \att\}} \sigma^{2^*}_{\att} =  \sum_{e \in \att^{max}}\sum_{\att \in \mathcal{A}_2} \sigma^{2^*}_{\att}\mathds{1}_{\{e \in \att\}}=  \sum_{\att \in \mathcal{A}_2}\sigma^{2^*}_{\att}\sum_{e \in \att^{max}} \mathds{1}_{\{e \in \att\}} \overset{\eqref{res2}}{=} b_2\sum_{\att \in \mathcal{A}_2}\sigma^{2^*}_{\att} = b_2.
\end{align*}
Thus, $\gamma^{\prime} = \frac{b_2}{\smsc}$, meaning that in any NE $(\sigma^{1^*},\sigma^{2^*})$ whose component basis is an MSP, the attack probability is equal to $\frac{b_2}{\smsc}$ for all the components of the MSP, which proves the necessary condition on the attack strategy.\hfill \Halmos
\end{enumerate}
%
\endproof
\else
\fi

\ifadditional
\tb{We end our analysis of the case $\smsc = \semm$ with a further characterization of the support of the NE of $\Gamma$. In particular, using the MSCs and the MSPs, we derive necessary conditions that are satisfied by every NE.}
\begin{proposition}\label{necessary_conditions}
If $b_1 <\smsc$, $b_2 < \semm$, and $\smsc = \semm$, then the set of NE has the following properties:
\begin{enumerate}
\item  In any NE, a node receives a sensor with positive probability only if it monitors exactly one component of any MSP:
%
%
\begin{align*}
\forall (\sigma^{1^*},\sigma^{2^*}) \in \nash, \ \forall i \in \nbasis{\sigma^{1^*}}, \ \forall \att^{max} \in \emm, \ \valdet{i}{\att^{max}} = 1.
\end{align*}
Furthermore, in any NE, a sensor placement is chosen with positive probability only if each of its sensors monitors a different component of any MSP:
\begin{align*}
\forall (\sigma^{1^*},\sigma^{2^*}) \in \nash, \ \forall \Def \in \supp(\sigma^{1^*}), \ \forall \att^{max} \in \emm, \ \forall e \in \att^{max}, \ \valdet{\Def}{e} = \sum_{i \in \Def}\valdet{i}{e}.
\end{align*}

\item In any NE, a component is targeted with positive probability only if it is monitored by a unique node of any MSC:
%
%
\begin{align*}
\forall (\sigma^{1^*},\sigma^{2^*}) \in \nash, \ \forall e \in \ebasis{\sigma^{2^*}}, \ \forall \Def^{min} \in \msc, \ \exists \, !\, i \in \Def^{min} \ | \ \valdet{i}{e} = 1.
\end{align*}
\end{enumerate}
\end{proposition}

In other words, if there exists at least one MSP, $\att^{max}$, such that a node $i \in \nodes$ does not monitor any component in $\att^{max}$, then $i$ will never receive a sensor in equilibrium. Besides, Proposition~\ref{necessary_conditions} tells us that if there exists a component $e \in \edges$ that is part of an MSP and such that at least two sensors in a sensor placement $\Def$ monitor $e$, then $\Def$ will never be chosen in equilibrium; since \attacker targets components that are spread across the network, \defender must allocate her sensing resources to avoid redundant detections.

Similarly, if there exists at least one MSC, $\Def^{min}$, such that a component $e$ is monitored by at least two nodes in $\Def^{min}$, then $e$ will never be targeted in equilibrium.

\proof{Proof of Proposition~\ref{necessary_conditions}.}
Let $\sigma^* = (\sigma^{1^*},\sigma^{2^*}) \in \nash$. Consider an MSC $\Def^{min} \in \msc$, an MSP $\att^{max} \in \emm$, and $(\sold{\Def^{min}}{b_1}{},\sola{\att^{max}}{b_2}{})$ constructed in Lemma~\ref{algebra2}.

\begin{enumerate}
\item
Thanks to Lemma~\ref{strategic_eq}, we know that the NE of $\Gamma$ are interchangeable. Proposition~\ref{necessary_conditions} implies that $(\sold{\Def^{min}}{b_1}{},\sola{\att^{max}}{b_2}{})$ defined in \eqref{sigma^S}-\eqref{sigma^T} is a NE. Therefore, $(\sigma^{1^*},\sola{\att^{max}}{b_2}{})$ is also a NE, and we obtain:
\begin{align*}
\frac{b_1b_2}{\smsc} &\overset{\eqref{eq_payoff}}{=} U_1(\sigma^{1^*},\sola{\att^{max}}{b_2}{}) \overset{\eqref{payoff1}}{=} \sum_{\Def \in \mathcal{A}_1}\sum_{\att \in \mathcal{A}_2}\sigma^{1^*}_{\Def}\sola{\att^{max}}{b_2}{\att} \valdet{\Def}{\att} \\
&= \sum_{\Def \in \mathcal{A}_1}\sum_{l=1}^{\order{\smsc}{b_2}}\sigma^{1^*}_{\Def}\sola{\att^{max}}{b_2}{\att^l} \valdet{\Def}{\att^l}   \overset{\eqref{sigma^T}}{=}   \sum_{\Def \in \mathcal{A}_1}\sum_{l=1}^{\order{\smsc}{b_2}}\sigma^{1^*}_{\Def}\frac{1}{\order{\smsc}{b_2}} \valdet{\Def}{\att^l}\\
&  \overset{\eqref{detection_disruptions}}{=}\sum_{\Def \in \mathcal{A}_1}\sigma^{1^*}_{\Def} \frac{1}{\order{\smsc}{b_2}}\sum_{l=1}^{\order{\smsc}{b_2}}\sum_{e \in \att^l} \valdet{\Def}{e} = \sum_{\Def \in \mathcal{A}_1}\sigma^{1^*}_{\Def} \frac{1}{\order{\smsc}{b_2}}\sum_{e \in \edges} \valdet{\Def}{e}\sum_{l=1}^{\order{\smsc}{b_2}}\mathds{1}_{\{e \in \att^l\}}\\
& = \sum_{\Def \in \mathcal{A}_1}\sigma^{1^*}_{\Def} \frac{1}{\order{\smsc}{b_2}}\sum_{e \in \att^{max}} \valdet{\Def}{e}\sum_{l=1}^{\order{\smsc}{b_2}}\mathds{1}_{\{e \in \att^l\}}\\
& \overset{\eqref{q1}}{=}   \sum_{\Def \in \mathcal{A}_1}\sigma^{1^*}_{\Def} \frac{1}{\order{\smsc}{b_2}}\sum_{e \in \att^{max}} \valdet{\Def}{e}\antiorder{\smsc}{b_2} \overset{\eqref{detection_disruptions},\eqref{arithmetic equality}}{=}  \frac{b_2}{\smsc}\sum_{\Def \in \mathcal{A}_1}\sigma^{1^*}_{\Def}\valdet{\Def}{\att^{max}}\\
& = \frac{b_2}{\smsc}\Expone[*]{\valdet{\Def}{\att^{max}}}.
\end{align*}

Therefore, we obtain:
\begin{align}
\forall \sigma^* \in \nash, \ \forall \att^{max} \in \emm, \ \Expone[*]{\valdet{\Def}{\att^{max}}} = b_1.\label{I lost count}
\end{align}

We can deduce further results from this equation. Consider  $\sigma^* \in \nash, \ \att^{max} \in \emm$, then:
\begin{align*}
b_1 & \overset{\eqref{I lost count}}{=} \Expone[*]{\valdet{\Def}{\att^{max}}}  \overset{\eqref{detection_disruptions}}{=} \Expone[*]{\sum_{e \in \att^{max}}\valdet{\Def}{e}}  \overset{\eqref{redundant_nodes}}{\leq} \Expone[*]{\sum_{i \in \Def}\sum_{e \in \att^{max}}\valdet{i}{e}} \\
&\overset{\eqref{detection_disruptions}}{=}  \Expone[*]{\sum_{i \in \Def}\valdet{i}{\att^{max}}}  \overset{\eqref{def_extended_matching}}{\leq} \Expone[*]{\sum_{i \in \Def}1} = \Expone[*]{|\Def|} \overset{\eqref{res1}}{=} b_1.
\end{align*}

Therefore, all the previous inequalities become equalities. The first one implies that:
\begin{align*}
\forall \Def \in \supp(\sigma^{1^*}), \ \forall e \in \att^{max}, \ \valdet{\Def}{e} = \sum_{i \in \Def} \valdet{i}{e}.
\end{align*}

The second induced equality implies that:
\begin{align*}
\forall i \in \nbasis{\sigma^{1^*}}, \ \forall \att^{max} \in \emm, \ \valdet{i}{\att^{max}} = 1.
\end{align*}


\item Similarly, by interchangeability $(\sold{\Def^{min}}{b_1}{},\sigma^{2^*}) \in \nash$. Then:
\begin{align*}
\frac{b_1b_2}{\smsc}  &\overset{\eqref{eq_payoff}}{=} U_1(\sold{\Def^{min}}{b_1}{},\sigma^{2^*})  \overset{\eqref{payoff1}}{=}    \sum_{k=1}^{\order{\smsc}{b_1}}\sum_{\att \in \mathcal{A}_2}\sold{\Def^{min}}{b_1}{\Def^k}\sigma^{2^*}_{\att} \valdet{\Def^k}{\att} \\
&\overset{\eqref{sigma^S}}{=}   \sum_{k=1}^{\order{\smsc}{b_1}}\sum_{\att \in \mathcal{A}_2}\frac{1}{\order{\smsc}{b_1}}\sigma^{2^*}_{\att} \valdet{\Def^k}{\att} \overset{\eqref{detection_disruptions}}{=} \frac{1}{\order{\smsc}{b_1}}\sum_{\att \in \mathcal{A}_2}\sigma^{2^*}_{\att}  \sum_{e \in \att} \sum_{k=1}^{\order{\smsc}{b_1}}\valdet{\Def^k}{e}.
\end{align*}

Since $\Def^{min} \in \msc$, then $\forall e \in \edges, \ \exists \,i_e \in \Def^{min} \ | \ \valdet{i_e}{e} = 1$. Therefore:
\begin{align*}
\forall e \in \edges, \  \sum_{k=1}^{\order{\smsc}{b_1}}\valdet{\Def^k}{e} &=  \sum_{\{k\in \llbracket1,\order{\smsc}{b_1} \rrbracket \, | \, i_e \in \Def^k\}}1 + \sum_{\{k\in \llbracket1,\order{\smsc}{b_1} \rrbracket \, | \, i_e \notin \Def^k\}} \underset{\geq 0 }{\underbrace{\valdet{\Def^k}{e}}} \\
&\geq |\{k\in \llbracket1,\order{\smsc}{b_1} \rrbracket \, | \, i_e \in \Def^k\}| \overset{\eqref{q1}}{=} \antiorder{\smsc}{b_1}.
\end{align*}

Thus, by plugging it in the previous equation, we obtain:
\begin{align*}
\frac{b_1b_2}{\smsc}&\geq   \frac{1}{\order{\smsc}{b_1}}\sum_{\att \in \mathcal{A}_2}\sigma^{2^*}_{\att} \sum_{e \in \att} \antiorder{\smsc}{b_1}= \frac{\antiorder{\smsc}{b_1}}{\order{\smsc}{b_1}} \sum_{\att \in \mathcal{A}_2}\sigma^{2^*}_{\att} |\att| \overset{\eqref{arithmetic equality}}{=} \frac{b_1}{\smsc} \mathbb{E}_{\sigma^{2^*}}[|\att|]  \overset{\eqref{res2}}{=} \frac{b_1}{\smsc}b_2.
\end{align*}

Therefore, we obtain the following equality:
\begin{align}
\forall \att \in \supp(\sigma^{2^*}), \ \forall e \in \att, \ \forall  k\in \llbracket1,\order{\smsc}{b_1} \rrbracket \, | \, i_e \notin \Def^k, \ \valdet{\Def^k}{e} = 0.\label{cannot_detect}
\end{align}


Now, consider $\att \in \supp(\sigma^{2^*})$, $e \in \att$, and let $i_e \in \Def^{min}$ be a node that satisfies $\valdet{i_e}{e} = 1$. Note that \eqref{cannot_detect} can be derived starting from any initial ordering of the node indices in $\Def^{min}$. Consider $i^\prime \in \Def^{min} \ | \ i^{\prime} \neq i_e$. Since $b_1 < \smsc$, then we can find an ordering of the node indices in $\Def^{min}$ such that there is a sensor placement $\Def^k$ in the support of $\sold{\Def^{min}}{b_1}{}$  that satisfies $i^{\prime} \in \Def^k$ and $i_e \notin \Def^k$. From \eqref{cannot_detect}, we deduce that $0 \leq \valdet{i^{\prime}}{e} \overset{\eqref{monotonicity_detection}}{\leq} \valdet{\Def^k}{e} \overset{\eqref{cannot_detect}}{=} 0$. Therefore, for any $i^{\prime} \in \Def^{min}$ such that $i^{\prime}\neq i_e$, $i^{\prime}$ does not monitor $e$, which implies that $e$ is only monitored by $i_e$ in $\Def^{min}$. Thus:
\begin{align*}
\forall e \in \ebasis{\sigma^{2^*}}, \ \forall \Def^{min} \in \msc, \ \exists \, !\, i \in \Def^{min} \ | \ \valdet{i}{e} = 1.
\end{align*}
\end{enumerate}
\hfill \Halmos
\endproof

\else
\fi

\ifadditional
\subsection{Proof of Section~\ref{sec:Answer}}


\begin{proposition}\label{Cor2}
Consider a network $\mathcal{G}$ and sensing model $\{\set{i}, \ i \in \nodes \}$ that satisfy $\smsc = \semm$, a target detection rate $\alpha \in [0,1]$, and \attacker's resources $b_2<\semm$. Then, for any MSC $\Def^{\text{min}} \in \mathcal{S}$ and any MSP $\att^{\text{max}} \in \mathcal{M}$, an optimal solution of \ECOP is given by $\lceil \alpha \smsc \rceil$, $(\sold{\Def^{\text{min}}}{\lceil \alpha \smsc \rceil}{},\sola{\att^{\text{max}}}{b_2}{})$.
\end{proposition}

\proof{Proof of Proposition~\ref{Cor2}.}

When $\smsc = \semm$, we know from Cor.~\ref{Cor1} that $\forall \sigma^* \in \nash, \ r(\sigma^*) = \frac{b_1}{\smsc}$. Therefore, \ECOP can be rewritten as follows:
\begingroup
\addtolength{\jot}{\spacingalign}
\begin{align*}
\underset{b_1,\, \sigma^\dag}{\text{minimize}}& \quad b_1\nonumber\\
\text{subject to}& \quad \frac{b_1}{\smsc} \geq \alpha, \\
& \quad \sigma^\dag \in \nash(b_1,b_2),\\
& \quad b_1 \in \mathbb{N}.\nonumber
\end{align*}
\endgroup
Then, the optimal value of \ECOP in that case is $b_1^\dag = \lceil \alpha \smsc \rceil$. 

Now, consider an MSC $\Def^{min} \in \msc$ and an MSP $\att^{max} \in \emm$. From Lemma~\ref{algebra1} and Proposition~\ref{necessary_conditions}, we know that $(\sold{\Def^{min}}{b_1^\dag}{},\sola{\att^{max}}{b_2}{})\in \nash(b_1^\dag,b_2)$, i.e., is a NE of the game induced by  $b_1^\dag$ and $b_2$. Therefore, $\lceil \alpha \smsc \rceil, (\sold{\Def^{min}}{b_1^\dag}{},\sola{\att^{max}}{b_2}{})$ is an optimal solution of \ECOP.
\hfill \Halmos
\endproof

\proof{Proof of Proposition~\ref{Prop_interim}}
Recall that in the general case $\semm \leq \smsc$, we admit a relaxation of \ECOP and consider instead the following mathematical program with equilibrium constraints:
\begingroup
\addtolength{\jot}{\spacingalign}
\begin{align}
(\mathcal{P}_{\epsilon}): \quad \underset{b_1,\, \sigma^\dag}{\text{minimize}}& \quad b_1\nonumber\\
\text{subject to}& \quad \sdet{*} \geq \alpha, \quad \quad \forall \sigma^* \in \nash(b_1,b_2) \label{all_NE_epsilon}\\
& \quad \sigma^\dag \in \nash_\epsilon(b_1,b_2)\nonumber\\
& \quad b_1 \in \mathbb{N},\nonumber
\end{align}
\endgroup
for some $\epsilon\geq 0$.

In this case, we know from the lower bound in Thm.~\ref{bounds_game} that $\forall b_1 < \smsc, \forall \sigma^* \in \nash(b_1,b_2), \quad  \dfrac{b_1}{\smsc} \leq\sdet{*}.$ Therefore, constraint \eqref{all_NE_epsilon} is satisfied if $b_1 \geq b_1^\prime := \lceil \alpha \smsc\rceil$. Now, consider an MSC $\Def^{min} \in \msc$ and an MSP $\att^{max} \in \emm$.  We know from Thm.~\ref{MSC-MSP-eNE} that $(\sold{\Def^{min}}{b_1^\prime}{},\sola{\att^{max}}{b_2}{}) \in \nash_{\epsilon}(b_1^\prime,b_2)$, where $\epsilon = b_1^\prime b_2\left( \frac{1}{\max\{b_1^\prime,\semm\}} - \frac{1}{\smsc}\right)$. Therefore, $b_1^\prime, (\sold{\Def^{min}}{b_1^\prime}{},\sola{\att^{max}}{b_2}{})$ is a feasible solution of $(\mathcal{P}_{\epsilon})$ (with the same $\epsilon$), and the corresponding objective value is $b_1^\prime$.

Finally, from the upper bound in Thm.~\ref{bounds_game}, we know that $\forall b_1 < \smsc, \forall \sigma^* \in \nash(b_1,b_2), \quad  \sdet{*} \leq \min\left\{\frac{b_1}{\semm},1\right\}$. Therefore, constraint \eqref{all_NE_epsilon} cannot be satisfied if $b_1 < \lceil \alpha \semm \rceil$. This implies that an optimality gap associated with $b_1^\prime, (\sold{\Def^{min}}{b_1^\prime}{},\sola{\att^{max}}{b_2}{})$ is given by $\lceil \alpha \smsc \rceil - \lceil \alpha \semm \rceil$.

\hfill \Halmos
\endproof
\else
\fi


\begin{lemma}\label{Big Support}
Given $b_2 < \semm$, let $\rho \in  [0,1]^{|\edges|}$ that satisfies $\sum_{e \in \edges} \rho_e = b_2$. Then, there exists an attack strategy $\sigma^2 \in \Delta(\mathcal{A}_2)$ that satisfies $\rho_{\sigma^2}(e) = \rho_e$ \rev{for all} $e \in \edges$. 
\end{lemma}

\proof{Proof of Lemma~\ref{Big Support}.}

Given $b_2 < \semm$, let $\bs{A}$ be the $|\edges| \times {|\edges| \choose b_2}$ binary matrix whose rows (resp. columns) are indexed by the components (resp. the size-$b_2$ subsets) of $\edges$, and which satisfies \rev{for all} $(e,\att) \in \edges \times \overline{\mathcal{A}_2}, \ a_{e,\att} = \mathds{1}_{\{e \in \att\}}$. Then, given $\rho \in [0,1]^{|\edges|}$ that satisfies $\sum_{e \in \edges}\rho_e = b_2$, we must show that the following system of equations has a feasible solution:
\vspace{-0.3cm}
%
\begin{align*}
&\bs{A}\sigma^2 = \rho\\[-0.2cm]
&\boldsymbol{1}^T_{|\overline{\mathcal{A}_2}|}\sigma^2 = 1\\[-0.2cm]
&\sigma^2 \geq \bs{0}_{|\overline{\mathcal{A}_2}|}.
\end{align*}


Since each $\att \in \overline{\mathcal{A}_2}$ is of size $b_2$, it is easy to see that $\frac{1}{b_2}\bs{1}_{|\edges|}^T \bs{A} = \bs{1}_{|\overline{\mathcal{A}_2}|}^T$. Furthermore, since $\frac{1}{b_2}\bs{1}_{|\edges|}^T \rho = 1$, it implies that if $\sigma^2$ satisfies $\bs{A}\sigma^2 = \rho$, it also satisfies $\bs{1}_{|\overline{\mathcal{A}_2}|}^T \sigma^2 = 1$. Therefore, we only need to show that there exists $\sigma^2 \geq \bs{0}_{|\overline{\mathcal{A}_2}|}$ such that $\bs{A}\sigma^2 = \rho$.
By Farkas' lemma, such a solution exists if and only if there does not exist $w \in \mathbb{R}^{|\edges|}$ such that $w^T\bs{A} \leq \bs{0}_{|\overline{\mathcal{A}_2}|}^T$ and $w^T\rho >0$. 

Let $w \in \mathbb{R}^{|\edges|}$ that satisfies $w^T\bs{A} \leq \bs{0}_{|\overline{\mathcal{A}_2}|}^T$, and let us order the components in $\edges$ so that $w_{e_1} \geq \dots \geq w_{e_{|\edges|}}$. For notational simplicity, let $w_k \coloneqq w_{e_k}$ and $\rho_k \coloneqq \rho_{e_k}$ \rev{for all} $k \in \llbracket 1,|\edges|\rrbracket$. Note that since $\att^1 = \{e_1,\dots,e_{b_2}\} \in \overline{\mathcal{A}_2}$, we have $\sum_{k=1}^{b_2} w_k = (w^T\bs{A})_{\att^1} \leq 0$. Then, we obtain:
\begin{align*}
w^T\rho &= \sum_{k=1}^{|\edges|} w_k \rho_k = \underset{\leq 0}{\underbrace{\sum_{k=1}^{b_2} w_k}} + \sum_{k=b_2+1}^{|\edges|} \underset{\leq w_{b_2}}{\underbrace{w_k}} \underset{\geq 0}{\underbrace{\rho_k}} - \sum_{k=1}^{b_2} \underset{\geq w_{b_2}}{\underbrace{w_k}} \underset{\geq 0}{\underbrace{(1-\rho_k)}} \leq   w_{b_2} \left(\sum_{k=1}^{|\edges|} \rho_k - b_2\right) = 0.
\end{align*}
\rev{By} Farkas' lemma, this implies that there exists an attack strategy $\sigma^2 \in \Delta(\mathcal{A}_2)$ such that \rev{for all} $e \in \edges, \ \rho_{\sigma^2}(e) = \rho_e$.
%
\hfill
\Halmos
\endproof

\proof{Proof of Theorem~\ref{Constant}.}

Consider \defender's amount of resources $b_1$. If $b_1 \geq \smsc$, then \rev{for every} $b_2 < \semm$ \rev{and every} $\sigma^* \in \nash(b_1,b_2)$, $r(\sigma^*) = 1$. Henceforth, we assume that $b_1 < \smsc$. Recall from \rev{Theorem}~\ref{all_resources} that the NE of $\Gamma$ can be obtained by solving \lpob and \lptb. 

\begin{enumerate}
\item[(i.a)]
First, we show  that for $b_2 = 1$, there exists an optimal solution of \lptp, $\sigma^{2^*} \in \Delta(\overline{\mathcal{A}_2})$, that satisfies \rev{for all} $e \in \edges, \ \rho_{\sigma^{2^*}}(e) \leq \frac{1}{\semm}.$ Since $b_2 = 1$, then $\overline{\mathcal{A}_2} = \edges$, and \rev{for every} $\sigma^2 \in \Delta(\edges)$ \rev{and every} $e \in \edges, \ \sigma^2_e = \rho_{\sigma^2}(e)$.

\begin{enumitemize}
\item[--] Consider an optimal solution of \lptp, $\sigma^{2^*} \in \Delta(\edges)$, and assume on the contrary that \rev{there exist} $e^\prime \in \edges$ \rev{and} $\varepsilon >0$ \rev{such that} $\rho_{\sigma^{2^*}}(e^\prime) = \frac{1}{\semm} + \varepsilon.$ Let $\Def^* \in \rev{\argmin_{\Def \in \overline{\mathcal{A}_1}} U(\Def,\sigma^{2^*})}$. 
From \rev{Proposition}~\ref{best_set_cover}, we know that: 
\vspace{-0.0cm}
\begin{align}
\rev{\mathbb{E}_{\sigma^{2^*}}[\valdet{\Def^*}{\att}] \overset{\eqref{payoff2}}{=} - U(\Def^*,\sigma^{2^*}) + 1 }\leq -\max\{ 1 - \frac{b_1}{\semm},0\} + 1 \leq \frac{b_1}{\semm}.\label{One of the very last}
\end{align}

\vspace{-0.0cm}

Therefore, we can show that \rev{there exists} $i^\prime \in \Def^*$ \rev{such that} $\rev{\mathbb{E}_{\sigma^{2^*}}[\valdet{\Def^*}{\att}]  - \mathbb{E}_{\sigma^{2^*}}[\valdet{\Def^*\backslash\{i^\prime\}}{\att}] } \leq \frac{1}{\semm}$. Indeed, if \rev{for all} $i \in \Def^*, \ \rev{\mathbb{E}_{\sigma^{2^*}}[\valdet{\Def^*}{\att}]  - \mathbb{E}_{\sigma^{2^*}}[\valdet{\Def^*\backslash\{i\}}{\att}] }  > \frac{1}{\semm}$, then we obtain the following contradiction:
\vspace{-0.0cm}
\begin{align*}
\rev{(b_1  -1)\mathbb{E}_{\sigma^{2^*}}[\valdet{\Def^*}{\att}]} & \rev{\overset{\eqref{nice_detection_inequality}}{\leq}\sum_{i \in \Def^*} \mathbb{E}_{\sigma^{2^*}}[\valdet{\Def^*\backslash\{i\}}{\att}]  <b_1\mathbb{E}_{\sigma^{2^*}}[\valdet{\Def^*}{\att}] - \frac{b_1}{\semm} }\\
&\rev{\overset{\eqref{One of the very last}}{\leq} (b_1 - 1) \mathbb{E}_{\sigma^{2^*}}[\valdet{\Def^*}{\att}].}
\end{align*}

\vspace{-0.0cm}

\item[--] This implies that $e^\prime \in \set{\Def^*}$: If  $e^\prime \notin \set{\Def^*}$ instead, then repositioning the detector on node $i^\prime$ to a node that can monitor $e^\prime$ improves \defender's payoff by at least $\varepsilon$ and contradicts the definition of $\Def^*$. 

\item[--] Then, we show that at least $b_1$ components are monitored by $\Def^*$. 
%
%
%
If we assume that $|\set{\Def^*}| < b_1$, then at least one detector in $\Def^*$ can be removed without changing \defender's payoff. Let $i_0 \in \Def^*$ denote the location of that detector. Now, we can show that \rev{there exists} $e \in \edges\backslash\{\set{\Def^*}\}$ \rev{such that} $\sigma_e^{2^*} >0$: if on the contrary, we had \rev{for all} $e \in \edges\backslash\{\set{\Def^*}\}, \ \sigma_e^{2^*} =0$, then we would obtain \rev{$U(\Def^*,\sigma^{2^*}) = 0$}, which is \rev{a} contradiction: Since $b_1 < \smsc$, \rev{the value of the game} must be strictly \rev{positive}. Let $e \in \edges\backslash\{\set{\Def^*}\}$ \rev{such that} $\sigma_e^{2^*} >0$ and let $i \in \nodes$ \rev{such that} $\valdet{i}{e}=1$. Repositioning the detector from node $i_0$ to node $i$ will \rev{improve} \defender's payoff by at least $\sigma_e^{2^*}$, which contradicts the definition of $\Def^*$. Thus, $|\set{\Def^*}| \geq b_1$.

\item[--] Now, we show that we can construct $\sigma^{2^\prime}$ which is the same probability distribution as $\sigma^{2^*}$ except that it reallocates $\varepsilon$ probability from $e^\prime$ to a subset of components $\att^1$ monitored by $\Def^*$ while ensuring that the  attack probability of each component in $\att^1$ is not above $\frac{1}{\semm}$. Let us split $\mathcal{C}_{\Def^*}$ into $\{e^\prime\}$, $\att^1 \coloneqq \{e \in \mathcal{C}_{\Def^*} \backslash\{e^\prime\} \ | \ \sigma^{2^*}_e <\frac{1}{\semm}\}$, and  $\att^2 \coloneqq \{e \in \mathcal{C}_{\Def^*} \backslash\{e^\prime\} \ | \ \sigma^{2^*}_e \geq\frac{1}{\semm}\}$.
%
Now, we have:
\begin{align*}
\varepsilon =   \rev{\mathbb{E}_{\sigma^{2^*}}[\valdet{\Def^*}{\att}] }  - \sum_{\mathclap{e \in \att^1\cup \att^2}}\sigma^{2^*}_e - \frac{1}{\semm}\overset{\eqref{One of the very last}}{\leq}   \frac{|\set{\Def^*}| - |\att^2| - 1}{\semm}  - \sum_{e \in \att^1}\sigma^{2^*}_e = \sum_{e \in \att^1} (\frac{1}{\semm} - \sigma^{2^*}_e),
\end{align*}
%
%
which implies that $\att^1 \neq \emptyset$, and that it is possible to allocate $\varepsilon$ additional probability to components in $\att^1$ so that the attack probability of each component in $\att^1$ is not above $\frac{1}{\semm}$. Thus, $\sigma^{2^\prime}$ can be constructed. It satisfies \rev{for all} $e \in \edges \backslash \{\att^1 \cup\{e^\prime\}\}, \ \sigma^{2^\prime}_e = \sigma^{2^*}_e$, $\sigma^{2^\prime}_{e^\prime} = \sigma^{2^*}_{e^\prime} - \varepsilon$, and $\sum_{e \in \att^1} (\sigma^{2^\prime}_e - \sigma^{2^*}_e) = \varepsilon$.


\item[--] Now, consider \rev{$\Def^\prime \in \argmin_{\Def \in \overline{\mathcal{A}_1}} U(\Def,\sigma^{2^\prime})$.} The goal of this step is to show (by contradiction) that $e^\prime \in \mathcal{C}_{\Def^\prime}$. First, we derive the following calculations:
%
%
%
\begin{align}
&\rev{\mathbb{E}_{\sigma^{2^\prime}}[\valdet{\Def^\prime}{\att}] } = (\sigma^{2^*}_{e^\prime} - \varepsilon)\valdet{\Def^\prime}{e^\prime} +\sum_{\mathclap{e \in \edges\backslash\{\att^1\cup\{e^\prime\}\}}}\sigma^{2^*}_e  \valdet{\Def^\prime}{e} +\sum_{e \in \att^1}\underset{\geq 0}{\underbrace{(\sigma^{2^\prime}_e - \sigma^{2^*}_e)}}\underset{\leq 1}{\underbrace{\valdet{\Def^\prime}{e}}} + \sum_{e \in \att^1}\sigma^{2^*}_e \valdet{\Def^\prime}{e}\nonumber\\
& \leq \sum_{e \in \edges}\sigma^{2^*}_e  \valdet{\Def^\prime}{e} +\sum_{e \in \att^1}(\sigma^{2^\prime}_e - \sigma^{2^*}_e)  -\varepsilon\valdet{\Def^\prime}{e^\prime}  = \rev{\mathbb{E}_{\sigma^{2^*}}[\valdet{\Def^\prime}{\att}] } + \varepsilon(1 - \valdet{\Def^\prime}{e^\prime}).\label{Good epsilon}
\end{align}

Thus, if $e^\prime \notin \mathcal{C}_{\Def^\prime}$, then $\rev{\mathbb{E}_{\sigma^{2^\prime}}[\valdet{\Def^\prime}{\att}] } \leq \rev{\mathbb{E}_{\sigma^{2^*}}[\valdet{\Def^\prime}{\att}] }+ \varepsilon$. Let  $i^* \in \nodes \backslash\Def^\prime$ \rev{such that} $e^\prime \in \mathcal{C}_{i^*}$. Then, we have:
\vspace{-0.0cm}
\begin{align}
\forall i \in \Def^\prime, \ \rev{\mathbb{E}_{\sigma^{2^*}}[\valdet{\Def^*}{\att}]}   & \rev{\overset{\eqref{payoff2}}{=} 1 - U(\Def^*,\sigma^{2^*})  \geq 1 - U(\{i^*\} \cup \Def^\prime\backslash\{i\},\sigma^{2^*}) } \nonumber\\
&\rev{\overset{\eqref{payoff2}}{=} \mathbb{E}_{\sigma^{2^*}}[\valdet{\{i^*\} \cup \Def^\prime\backslash\{i\}}{\att}]} \overset{\eqref{monotonicity_detection}}{\geq}\frac{1}{\semm} + \varepsilon + \rev{\mathbb{E}_{\sigma^{2^*}}[\valdet{\Def^\prime\backslash\{i\}}{\att}]}.\label{Almost...}
\end{align}
\vspace{-0.0cm}

Next, we can derive the following calculations:
%
\vspace{-0.0cm}
\begin{align}
&\rev{\mathbb{E}_{\sigma^{2^*}}[\valdet{\Def^\prime}{\att}] } \overset{\eqref{nice_detection_inequality}}{\leq} \frac{1}{b_1-1}\sum_{i \in \Def^\prime}\rev{\mathbb{E}_{\sigma^{2^*}}[\valdet{\Def^\prime\backslash\{i\}}{\att}] } \overset{\eqref{One of the very last},\eqref{Almost...}}{\leq} \rev{\mathbb{E}_{\sigma^{2^*}}[\valdet{\Def^*}{\att}] } - \frac{b_1}{b_1-1}\varepsilon. \label{Out of ideas}
\end{align}

\vspace{-0.0cm}

Combining everything together, we obtain the following contradiction:
\vspace{-0.0cm}
\begin{align*}
\rev{\min_{\Def \in \overline{\mathcal{A}_1}}U(\Def,\sigma^{2^\prime}) \overset{\eqref{payoff2},\eqref{Good epsilon},\eqref{Out of ideas}}{\geq} U(\Def^*,\sigma^{2^*}) + \frac{1}{b_1-1}\varepsilon  >  U(\Def^*,\sigma^{2^*}) =  \max_{\sigma^2 \in \Delta(\edges)} \min_{\Def \in \overline{\mathcal{A}_1}}U(\Def,\sigma^{2}).}
\end{align*}

\vspace{-0.0cm}

\item[--] Therefore, we showed that $e^\prime \in \mathcal{C}_{\Def^\prime}$. This implies that $\varepsilon(1-\valdet{\Def^\prime}{e^\prime}) =0$, and we obtain: \rev{$\min_{\Def \in \overline{\mathcal{A}_1}}U(\Def,\sigma^{2^\prime})  \overset{\eqref{payoff2},\eqref{Good epsilon}}{\geq} U(\Def^\prime,\sigma^{2^*})  \geq  \min_{\Def \in \overline{\mathcal{A}_1}} U(\Def,\sigma^{2^*})  = \max_{\sigma^2 \in \Delta(\edges)} \min_{\Def \in \overline{\mathcal{A}_1}} U(\Def,\sigma^2).$}
Thus, $\sigma^{2^\prime}$ is also an optimal solution of \lptp. Therefore, if  an optimal solution of \lptp is such that at least one component is targeted with probability more than $\frac{1}{\semm}$, we can create another optimal solution of \lptb with one less component targeted with probability more than $\frac{1}{\semm}$. We can then repeat this process until all attack probabilities are no more than $\frac{1}{\semm}$. 
\end{enumitemize}

\item[(i.b)] Given $b_1 < \smsc$, let $z^*(b_2)$ denote the optimal value of \lptb for any $b_2 <\semm$. Now, consider $b_2 < \semm$, and let $\sigma^{2^*} \in \Delta(\mathcal{A}_2)$ be an optimal solution of \lptp. Since $\sum_{e \in \edges}\frac{\rho_{\sigma^{2^*}(e)}}{b_2} =1$, we can construct an attack strategy $\sigma^{2^\prime} \in \Delta(\edges)$  such that \rev{for all} $e \in \edges, \ \sigma^{2^\prime}_e = \rho_{\sigma^{2^\prime}}(e) =  \frac{\rho_{\sigma^{2^*}}(e)}{b_2}$. Then, the additivity of $\detect$ gives: \rev{$z^*(b_2) = \min_{\Def \in \overline{\mathcal{A}_1}} (b_2 - \sum_{e \in \edges} \valdet{\Def}{e}\rho_{\sigma^{2^*}}(e)) = b_2\min_{\Def \in \overline{\mathcal{A}_1}}U(\Def,\sigma^{2^\prime}) \leq b_2 z^*(1).$} 

Now, consider $\widetilde{\sigma}^2 \in \Delta(\edges)$ which is an optimal solution of \lptp (where the number of attack resources is 1) with the additional property that \rev{for all} $e \in \edges, \ \rho_{\widetilde{\sigma}^2}(e) \leq \frac{1}{\semm}$. Then, given $b_2 <\semm$, since \rev{for all} $e \in \edges, \ b_2\rho_{\widetilde{\sigma}^2}(e) \leq 1$ and $\sum_{e \in \edges} b_2\rho_{\widetilde{\sigma}^2}(e) = b_2$, there exists a probability distribution $\widehat{\sigma}^{2} \in \Delta(\mathcal{A}_2)$ that satisfies $\rho_{\widehat{\sigma}^{2}}(e) = b_2  \rho_{\widetilde{\sigma}^2}(e)$ \rev{for all} $e \in \edges$ (Lemma~\ref{Big Support}). Then, by additivity of $\detect$,  we obtain: \rev{$z^*(1) =\min_{\Def \in \overline{\mathcal{A}_1}} (1- \sum_{e \in \edges} \valdet{\Def}{e}\rho_{\widetilde{\sigma}^{2}}(e)) = \frac{1}{b_2} \min_{\Def \in \overline{\mathcal{A}_1}}(b_2-\sum_{e \in \edges} \valdet{\Def}{e}\rho_{\widehat{\sigma}^{2}}(e)) \leq \frac{1}{b_2} z^*(b_2).$}
%

Thus, \rev{for every} $b_2 < \semm, \ z^*(b_2) = b_2z^*(1).$ Therefore, we can conclude that \rev{for every} $\sigma^* \in \nash(b_1,b_2), \ r(\sigma^*) \overset{\eqref{payoff2},\eqref{Exp_detection_rate},\eqref{res2}}{=} \rev{1 - \frac{U(\sigma^*)}{b_2} = 1 - \frac{z^*(b_2)}{b_2} = 1 - z^*(1)} \eqqcolon r^*_{b_1},$ which does not depend on $b_2$.

\item[(ii)] Given \defender's resources $b_1 < \smsc$, let $\sigma^{1^*} \in \Delta(\overline{\mathcal{A}_1})$ be an inspection strategy in equilibrium of the game $\Gamma(b_1,1)$. From \rev{Theorem}~\ref{all_resources}, we know that $\sigma^{1^*}$ is an optimal solution of \lpob for $b_2 = 1$. Now, consider $b_2 < \semm$. We can derive the following inequality:
\begin{align*}
\rev{\forall \att \in \overline{\mathcal{A}_2}, \ U(\sigma^{1^*},\att) \overset{\eqref{detection_disruptions}}{=} \sum_{e \in \att}U(\sigma^{1^*},e) \leq  \sum_{e \in \att} \underset{= 1- r^*_{b_1}}{\underbrace{\max_{e^\prime \in \edges} U(\sigma^{1^*},e^\prime)}} =  b_2(1-r^*_{b_1}) = z^*(b_2).}
\end{align*}
Since this inequality is valid for any $\att \in \overline{\mathcal{A}_2}$, we deduce that \rev{$\max_{\att \in \overline{\mathcal{A}_2}} U(\sigma^{1^*},\att) \leq z^*(b_2) = \min_{\sigma^{1} \in \Delta(\overline{\mathcal{A}_1})} \max_{\att \in \overline{\mathcal{A}_2}}U(\sigma^{1},\att)$}. Therefore, $\sigma^{1^*}$ is an optimal solution of \lpob (when the number of attack resources is $b_2$), and is an inspection strategy in equilibrium of the game $\Gamma(b_1,b_2)$.
\hfill
\Halmos
%
\end{enumerate}

\endproof

\section{\rev{Additional Illustrations}}

\subsection{\rev{Illustrative Example}}\label{Sec:Additional_Ex}
\begin{figure}[htbp]
\centering
\begin{subfigure}[t]{0.5\textwidth} \centering
        \begin{tikzpicture}
              \node[inner sep = 0pt] (Image) at (0,0) {\includegraphics[trim = 40mm 70mm 10mm 70mm, clip,scale = 0.5]{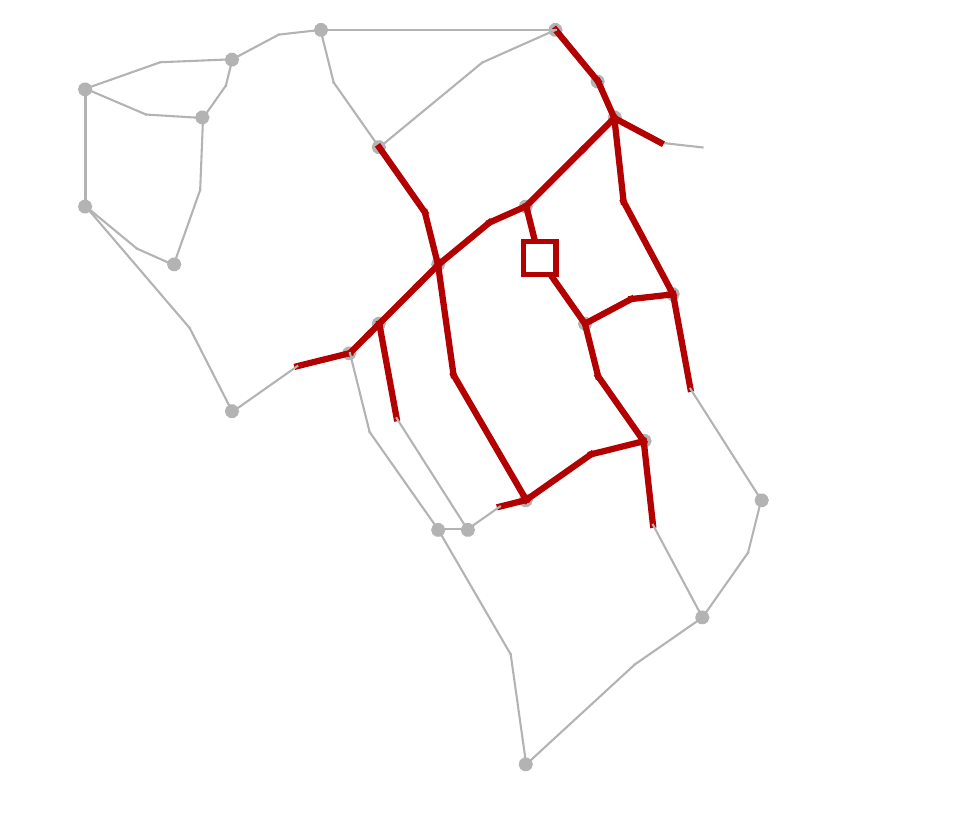}}; 
              \node (hey) at (2.7,1.75) {{\small \textcolor{red!75!black}{pipeline break}}};
         \draw[->,thick, red!75!black] (1.55,1.70) -- (0.6,1.35);     
                    \end{tikzpicture}
		\caption{\rev{Network layout. The colored region indicates the locations from where a particular break event can be detected.}}    \label{fig:net1_burst} 
        \end{subfigure}
        ~
 \begin{subfigure}[t]{0.45\textwidth} \centering
\includegraphics[scale=0.26]{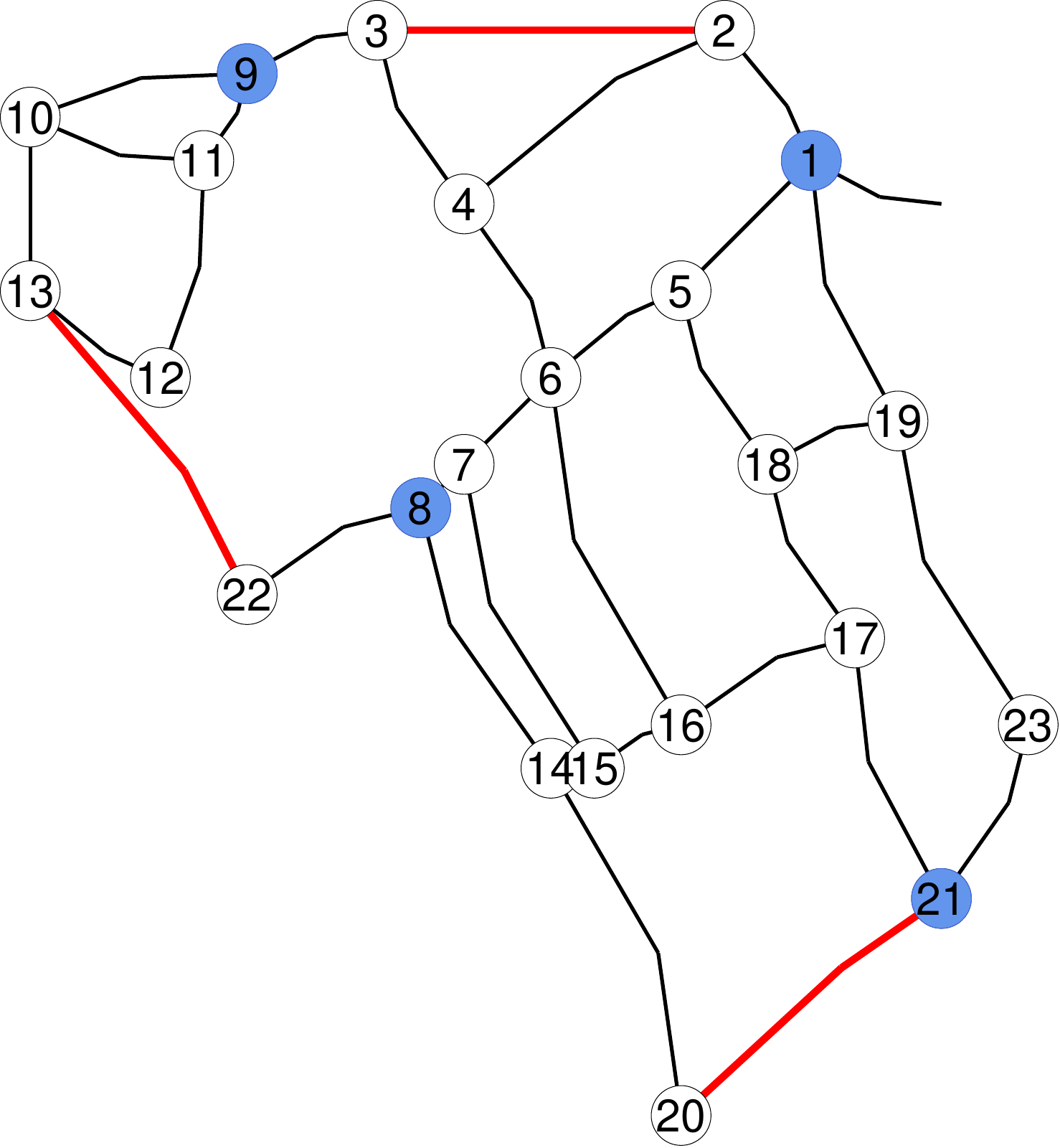}

\caption{\rev{MSC (blue nodes) and MSP (red pipes).}}\label{Apulian_MSC}
\end{subfigure}
        ~ 
        \begin{subfigure}[t]{0.5\textwidth} \centering
              \begin{tikzpicture}[scale=0.90]
\begin{axis}[
  xmin=0,
  xmax=4,
  ymin=0,
  ymax=1,
  xtick={0,1,2,3,4},
  font = \small,
  legend pos=south east,
	legend cell align={left},
  legend style={draw=none, font = \scriptsize},
  grid=major, 
  grid style={dashed,gray!30}, 
  xlabel=Number of detectors: \ $b_1$ ,
  ylabel=Equilibrium detection rate: \ $\sdet{*}$,
]

\addplot[dashdotted, color = red,very thick] table[x=b1, y=ub, col sep=comma, comment chars={\%}] {Evaluation_2.csv};
\addplot[solid, color = black,very thick,mark=*] table[x=b1, y=ne, col sep=comma, comment chars={\%}] {Evaluation_2.csv};
 \addplot[dashed, color = blue,very thick] table[x=b1, y=lb, col sep=comma, comment chars={\%}] {Evaluation_2.csv};
\legend{MSP-based upper bound, NE, MSC-based lower bound}

 \addplot[dotted, color = green!0!black,thick] table[x=b1, y=alpha, col sep=comma, comment chars={\%}] {Evaluation_2.csv};

\end{axis}
 \node at (-0.3,4.27) {$\alpha$};

\end{tikzpicture}

		\caption{\rev{Equilibrium detection rate with respect to the number of detectors $b_1$.}}   \label{fig:net1_15}  
        \end{subfigure}
~
 \begin{subfigure}[t]{0.45\textwidth} \centering
\includegraphics[scale=0.26]{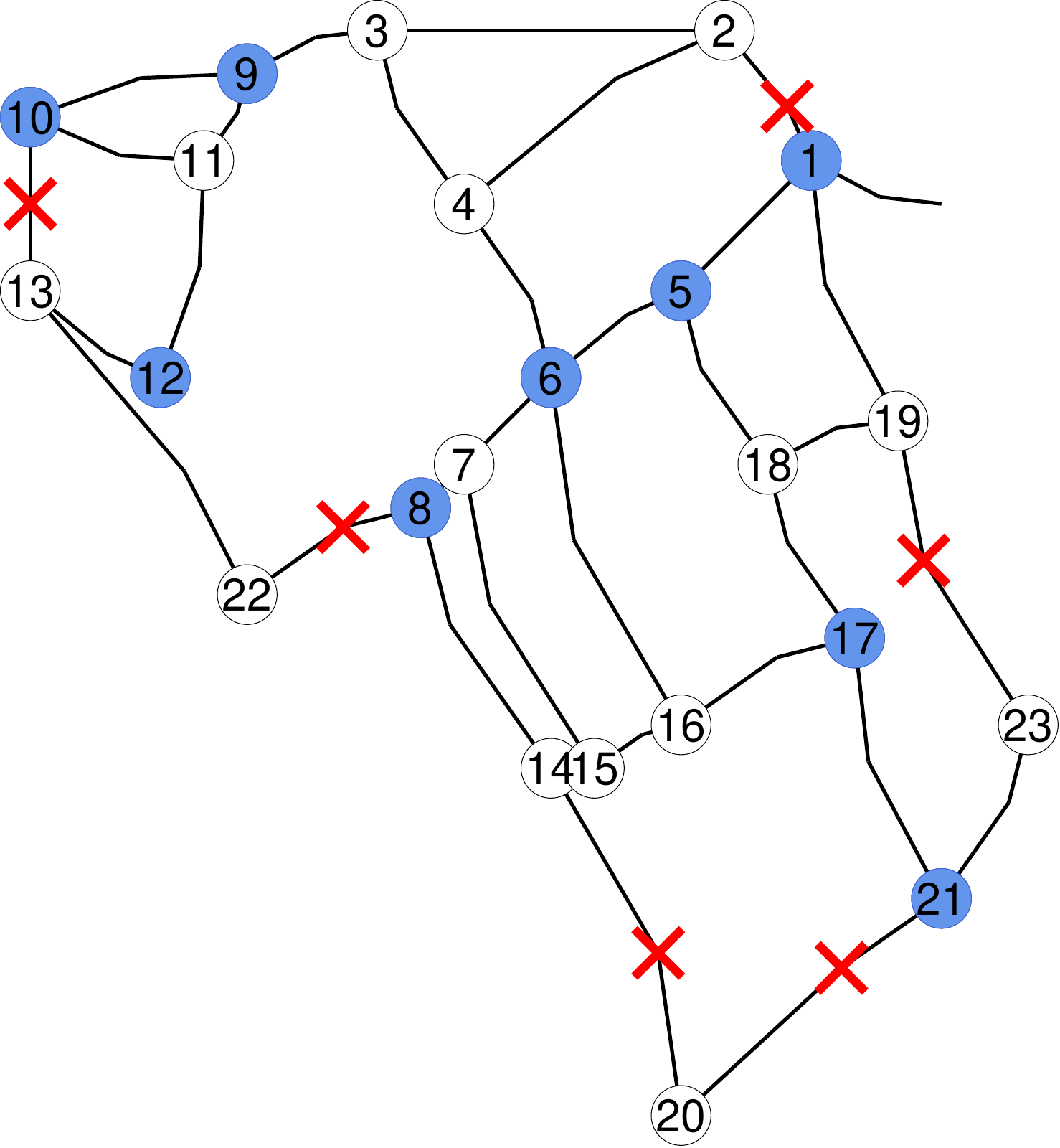}
\caption{\rev{Node basis (blue nodes) and component basis (red crosses) in equilibrium of $\Gamma(3,b_2)$ for $b_2 \leq 3$.}}\label{Apulian_Opt}
\end{subfigure}

\caption{\rev{Solving $(\mathcal{P})$ for the Apulian benchmark network facing adversarial pipeline break events.}}\label{Apulian_Net}
\end{figure}

\rev{We} illustrate \rev{the} solution approach \rev{outlined in Section~\ref{sec:Answer}} on a small-sized pipeline network of 23 nodes and 34 pipelines \citep{apulian}, as shown in \rev{Figure}~\ref{fig:net1_burst}. 
\rev{Consider the network inspection problem \ECOP, where the defender wants to ensure an expected detection rate of $\alpha = 0.75$ against an adversary with $b_2 \leq \semm$ resources.
First, we determine the detection model $\mathcal{G}$: The set of monitoring locations $\nodes$ is given by the set of network nodes, and the set $\edges$ of critical components is the set of pipes. Then, we determine the monitoring sets $\set{i}$ for detecting break events from each location $i \in \nodes$, as illustrated in Figure~\ref{fig:net1_burst}.}



\rev{\textbf{MSC/MSP-based solution:} We solve (\hyperlink{(MSC)}{$\mathcal{I}_{\text{MSC}}$}) to compute an MSC $\Def^{min} = \{1,8,9,21\}$ of size $\smsc  = 4$, as shown in Figure~\ref{Apulian_MSC}. From \eqref{Eq:Bounds_r}, we know that with $\lceil \alpha \smsc \rceil = 3$ detectors, the expected detection rate in any equilibrium is at least $\alpha$. Using Lemma~\ref{algebra2}, we construct the inspection strategy $\sold{\Def^{min}}{3}{}$, which uniformly randomizes over $\supp(\sold{\Def^{min}}{3}{}) =  \{\{1,8,9\},\{8,9,21\},\{9,21,1\},\{21,1,8\}\}$. Then, we deduce from \eqref{Eq:detect_guarantee} that $(3,\sold{\Def^{min}}{3}{})$ is a feasible solution of \ECOP.}
Next, we solve (\hyperlink{(MSP)}{$\mathcal{I}_{\text{MSP}}$}) to compute an MSP $\att^{max}$, which is of size $\semm = 3$, \rev{as illustrated in Figure~\ref{Apulian_MSC}}. From \eqref{Eq:Bounds_r}, we deduce that the optimality gap associated with $(3,\sold{\Def^{min}}{3}{})$ is given by $\lceil \alpha \smsc\rceil - \lceil \alpha \semm \rceil = \lceil 3\rceil - \lceil 2.25 \rceil = 0$. \rev{Therefore, $(3,\sold{\Def^{min}}{3}{})$ is an optimal solution of \ECOP. Although $\sold{\Def^{min}}{3}{}$ satisfies constraints \eqref{all_NE}, we note that its detection performance is lower than the detection performance in equilibrium: The upper bound on the corresponding loss in detection performance derived from \eqref{Eq:detect_guarantee} is given  by $1 - \frac{\max\{\lceil \alpha \smsc \rceil,\semm\}}{\smsc} = 25\,\%$.} 
\rev{\rev{Figure}~\ref{fig:net1_15} illustrates the bounds on the equilibrium expected detection rate derived in \eqref{Eq:Bounds_r}.}

\rev{\textbf{Refinement Procedure:} Next, we improve our MSC-based inspection strategy using the refinement procedure described in Section~\ref{conclusions}. We first select $\mathcal{I} = \{\{1,8,9\},\{8,9,21\},$ $\{9,21,1\},\{21,1,8\}\}$ as subset of indices for the restricted master problem \pcg to warm-start the column generation algorithm. After 17 iterations of \aref{primal_dual_alg}-\aref{end_CG}, we obtain an equilibrium inspection strategy $\sigma^{1^*}$ that uniformly randomizes over $\supp(\sigma^{1^*}) = \{\{1,8,9\},$ $\{8,9,21\},\{21,1,8\},\{5,12,17\},\{1,10,17\},\{6,9,21\}\}$. The corresponding node basis is given by $\nodes_{\sigma^{1^*}} = \{1,5,6,8,9,10,12,17,21\}$, as illustrated in Figure~\ref{Apulian_Opt}. The equilibrium detection rate is $\frac{5}{6}$, which improves the detection performance of $\sold{\Def^{min}}{3}{}$ by $11.11\,\%$. Finally, we find that the optimal dual variables of \lpob, obtained at termination of the column generation algorithm, imply that an attack strategy in equilibrium of $\Gamma(3,1)$ uniformly randomizes over the 6 pipes highlighted in Figure~\ref{Apulian_Opt}. Thus, for every $b_2 \leq \semm = 3$, an attack strategy in equilibrium of $\Gamma(3,b_2)$ is such that each of these 6 pipelines is targeted with probability $\frac{b_2}{6}$.}

\subsection{\rev{Additional Computational Results of Column Generation}}\label{Add:Figures}

\rev{In Figure~\ref{Fig:Others}, we illustrate the results of the column generation algorithm \aref{alg:CG} for solving \lpob with the optimal number of detectors in networks ky5, ky13, and ky2.}
\begin{figure}
\begin{subfigure}[t]{0.99\textwidth} \centering
\caption*{\normalsize{Network ky5 with $b_1 = 14$}}

\ifarXiv

\else
\vspace{-0.2cm}
\fi

              \begin{tikzpicture}[scale=0.90]
\begin{semilogxaxis}[
  grid=major, 
  grid style={dashed,gray!30}, 
  xlabel=Runtime {[s]} ,
  ylabel=Worst-case detection rate: \ $\min r(\sigma^1{,}\cdot)$,
   font=\small,
  xmin=1,
  xmax=100,
  ymin=0,
  ymax=1,
 ytick={0,0.25, 0.5,0.75,1},
  legend pos=south east,
	legend cell align={left},
  legend style={draw=none, font = \scriptsize},
]


  \addplot[color = red, very thick] table[x=Time2, y=Objective2, col sep=comma, comment chars={\%}] {Net3_ky5.csv};
  
 \addplot[color = blue, very thick] table[x=Time, y=Objective, col sep=comma, comment chars={\%}] {Net3_ky5.csv};

\legend{No warm-start, MSC warm-start}
 
\end{semilogxaxis}

\end{tikzpicture}       
~
 \begin{tikzpicture}[scale=0.90]
\begin{semilogxaxis}[
  grid=major, 
  grid style={dashed,gray!30}, 
  xlabel=Runtime {[s]} ,
  ylabel=Node basis size: \ $|\nodes_{\sigma^1}|$,
   font=\small,
  xmin=1,
  xmax=100,
  ymin=0,
  ymax=200,
 ytick={0,40,80,120,160,200},
  legend pos=south east,
	legend cell align={left},
  legend style={draw=none, font = \scriptsize},
]


  \addplot[color = red, very thick] table[x=Time2, y=Node_Basis_Size2, col sep=comma, comment chars={\%}] {Net3_ky5.csv};

 \addplot[color = blue, very thick] table[x=Time, y=Node_Basis_Size, col sep=comma, comment chars={\%}] {Net3_ky5.csv};

\legend{No warm-start, MSC warm-start}
 
\end{semilogxaxis}

\end{tikzpicture}

        \end{subfigure}
        ~
        \begin{subfigure}[t]{0.99\textwidth} \centering
        
        \ifarXiv

\else
\vspace{0.1cm}
\fi
        \caption*{\normalsize{Network ky13 with $b_1 = 22$}}
        
        \ifarXiv

\else
\vspace{-0.2cm}
\fi

              \begin{tikzpicture}[scale=0.90]
\begin{semilogxaxis}[
  grid=major, 
  grid style={dashed,gray!30}, 
  xlabel=Runtime {[s]} ,
  ylabel=Worst-case detection rate: \ $\min r(\sigma^1{,}\cdot)$,
   font=\small,
  xmin=1,
  xmax=1000,
  ymin=0,
  ymax=1,
 ytick={0,0.25, 0.5,0.75,1},
  legend pos=south east,
	legend cell align={left},
  legend style={draw=none, font = \scriptsize},
]


  \addplot[color = red, very thick] table[x=Time2, y=Objective2, col sep=comma, comment chars={\%}] {Net7_ky13_b1_22.csv};

 \addplot[color = blue, very thick] table[x=Time, y=Objective, col sep=comma, comment chars={\%}] {Net7_ky13_b1_22.csv};

\legend{No warm-start, MSC warm-start}
 
\end{semilogxaxis}

\end{tikzpicture}       
~
              \begin{tikzpicture}[scale=0.90]
\begin{semilogxaxis}[
  grid=major, 
  grid style={dashed,gray!30}, 
  xlabel=Runtime {[s]} ,
  ylabel=Node basis size: \ $|\nodes_{\sigma^1}|$,
   font=\small,
  xmin=1,
  xmax=1000,
  ymin=0,
  ymax=375,
 ytick={0,75,150,225,300,375},
  legend pos=south east,
	legend cell align={left},
  legend style={draw=none, font = \scriptsize},
]


  \addplot[color = red, very thick] table[x=Time2, y=Node_Basis_Size2, col sep=comma, comment chars={\%}] {Net7_ky13_b1_22.csv};

 \addplot[color = blue, very thick] table[x=Time, y=Node_Basis_Size, col sep=comma, comment chars={\%}] {Net7_ky13_b1_22.csv};

\legend{No warm-start, MSC warm-start}
 
\end{semilogxaxis}

\end{tikzpicture}

%
%
%

        \end{subfigure}
        ~
        \begin{subfigure}[t]{0.99\textwidth} \centering
        
        \ifarXiv

\else
\vspace{0.1cm}
\fi

        \caption*{\normalsize{Network ky2 with $b_1 = 14$}}

        \ifarXiv

\else
\vspace{-0.2cm}
\fi

              \begin{tikzpicture}[scale=0.90]
\begin{semilogxaxis}[
  grid=major, 
  grid style={dashed,gray!30}, 
  xlabel=Runtime {[s]} ,
  ylabel=Worst-case detection rate: \ $\min r(\sigma^1{,}\cdot)$,
   font=\small,
  xmin=1,
  xmax=200,
  ymin=0,
  ymax=1,
 ytick={0,0.25, 0.5,0.75,1},
  legend pos=south east,
	legend cell align={left},
  legend style={draw=none, font = \scriptsize},
]


  \addplot[color = red, very thick] table[x=Time2, y=Objective2, col sep=comma, comment chars={\%}] {Net8_ky2_b1_14.csv};

 \addplot[color = blue, very thick] table[x=Time, y=Objective, col sep=comma, comment chars={\%}] {Net8_ky2_b1_14.csv};

\legend{No warm-start, MSC warm-start}
 
\end{semilogxaxis}

\end{tikzpicture}       
        ~
              \begin{tikzpicture}[scale=0.90]
\begin{semilogxaxis}[
  grid=major, 
  grid style={dashed,gray!30}, 
  xlabel=Runtime {[s]} ,
  ylabel=Node basis size: \ $|\nodes_{\sigma^1}|$,
   font=\small,
  xmin=1,
  xmax=200,
  ymin=0,
  ymax=250,
 ytick={0,50,100,150,200,250},
  legend pos=south east,
	legend cell align={left},
  legend style={draw=none, font = \scriptsize},
]


  \addplot[color = red, very thick] table[x=Time2, y=Node_Basis_Size2, col sep=comma, comment chars={\%}] {Net8_ky2_b1_14.csv};

 \addplot[color = blue, very thick] table[x=Time, y=Node_Basis_Size, col sep=comma, comment chars={\%}] {Net8_ky2_b1_14.csv};

\legend{No warm-start, MSC warm-start}
 
\end{semilogxaxis}

\end{tikzpicture}

        \end{subfigure}
         \caption{\rev{Results of column generation applied to ($\overline{\text{LP}_1}$) for networks ky5, ky13, and ky2 with and without the MSC-based warm-start.}} \label{Fig:Others}
        \end{figure}

\section{\rev{Case When $b_2 \geq \semm$}}\label{Other Case}

As argued in Section~\ref{2 Comb Pbs}, the network inspection problem \ECOP  and \rev{the game $\Gamma(b_1,b_2)$} when $b_2 \geq \semm$ are of limited practical interest. However, for the sake of completeness, we now briefly discuss this case. 
\rev{To solve problem \ECOP, we}
%
derived in Section~\ref{Main_case} equilibrium properties of the game $\Gamma(b_1,b_2)$ that hold when $b_1 < \smsc$ and $b_2 < \semm$. Note that all these properties, except \rev{Proposition}~\ref{all_edges}, also hold when $b_1 < \smsc$ and $b_2 = \semm$. This implies that \rev{our approach to solve \ECOP is still valid when $b_2 = \semm$: the optimality gap and detection guarantees of the MSC/MSP-based solution hold, and can be improved with the refinement procedure.}


However, most of the equilibrium properties \rev{described in Section~\ref{Main_case} are not always} satisfied by the NE of $\Gamma(b_1,b_2)$ when $b_1 < \smsc$ and $b_2 > \semm$, as discussed in the following example:
%
%
%
%
%
\begin{example}\label{example_other_case}

Consider the detection model $\mathcal{G} = (\nodes,\edges,\{\set{i}, \ i \in \nodes\})$ defined as follows: Let $\nodes = \rev{\{i_1,\dots,i_{n}\}}$, with $n \in \mathbb{N}$, and let $\edges = \edges_1 \cup \{e_1,\dots,e_n\}$, where $\edges_1$ is a discrete set. Then, the monitoring sets are defined by \rev{$\set{i_k} = \edges_1 \cup \{e_k\}$, for every $k \in \llbracket 1,n\rrbracket$}.
In this example, $\Def^{min} = \{i_1,\dots,i_n\}$ is an MSC, and  $\att^{max} = \{e_1,\dots,e_n\}$ is an MSP; so $\smsc = \semm = n$. Given any $b_1 < n$ and any $b_2 \in \llbracket n, |\edges_1| + n\rrbracket$, one can check that \rev{for all} $\att \in \mathcal{A}_2$ \rev{such that} $\att^{max} \subseteq \att, \ (\{i_1,\dots,i_{b_1}\},\att)$ is a pure NE, whose node basis is not a set cover. Therefore, \rev{Proposition}~\ref{all_edges} does not hold anymore.

Now, if we consider $b_2 = |\edges_1| + n$, we just showed that we could construct NE where \attacker can use $n,n+1,\dots$, or $n+|\edges_1| = b_2$ resources. Thus, \eqref{res2} in \rev{Theorem}~\ref{all_resources} does not hold anymore. This also implies that the expected detection rate is not constant in equilibrium anymore: We found equilibria where the equilibrium detection rates are equal to $\frac{b_1}{n},\frac{b_1+1}{n+1},\dots,\frac{b_1+|\edges_1|}{n+|\edges_1|}$, which violates \rev{Theorem}~\ref{Constant}. Since $\frac{b_1+|\edges_1|}{n+|\edges_1|} \underset{|\edges_1| \to +\infty}{\longrightarrow} 1$, the \rev{MSP-based} upper bound on the expected detection rates given in~\eqref{Eq:Bounds_r} is violated. Furthermore, the \rev{upper bound on the relative loss of performance derived from~\eqref{Eq:detect_guarantee}} is not valid anymore. By choosing $\sold{\Def^{min}}{b_1}{}$, the expected detection rate may be arbitrarily far from an equilibrium expected detection rate. 
%
\hfill $\triangle$


\end{example}


Still, some results remain valid when $b_1 < \smsc$ and $b_2 > \semm$: In \rev{Theorem}~\ref{all_resources}, property \eqref{res1} still holds. In~\eqref{Eq:Bounds_r}, the \rev{MSC-based} lower bound on the equilibrium expected detection rates is still valid. \rev{If \defender selects the strategy $\sold{\Def^{min}}{b_1}{}$,} the expected detection rate is guaranteed to be at least $\frac{b_1}{\smsc}$, regardless of \attacker's strategy.
From these remaining results, we can \rev{deduce} that $(\lceil \alpha \smsc \rceil,\sold{\Def^{min}}{\lceil \alpha \smsc \rceil}{})$ is a feasible solution to \ECOP when $b_2 > \semm$.  \rev{However, the optimality gap derived in Section~\ref{Sec:MSC/MSP} does not hold anymore.}

\ifadditional
\section{A Simple Example}\label{example_detection_sensing}


\def \sizefig {1.4cm}
\def \scalefig {\small}
\def \innersepfig {0.07cm}

Consider the problem \ECOP with the target expected detection rate $\alpha = 0.75$ and the number of attack resources $b_2 = 2$. For this small-sized problem, we can solve  \lpo and \lpt to compute the NE of $\Gamma$, and obtain  the expected detection rate in any NE for each $b_1 \in \mathbb{N}$; see Table~\ref{r_sigma_example}. Notice that for a given $b_1$, the expected detection rate is the same in any NE; this property is shown in Thm.~\ref{bounds_game}.
\begin{table}[htbp]
\centering
\caption{Expected detection rate in equilibrium for every $b_1 \in \mathbb{N}$.}
\def\arraystretch{\spacingtable}
\begin{tabular}{|c|c|c|c|c|c|}\hline $b_1$ & 0 & 1 & 2 & 3 & $\geq 4$ \\\hline $\sdet{*}$ & 0 & $\frac{2}{7}$ & $\frac{4}{7}$ & $\frac{6}{7}$ & 1 \\\hline \end{tabular}
\label{r_sigma_example}
\end{table}

From Table~\ref{r_sigma_example}, we conclude that the optimal value of (\hyperlink{(P)}{$\mathcal{P}$}) is $b_1^\dag = 3$, i.e., with  $3$ sensors, \defender is capable of detecting $\frac{6}{7} \geq \alpha$ of the failure events in equilibrium. However, as mentioned in Section~\ref{sec:Formal problem}, this method does not extend to larger networks, as it requires solving large \lpo and \lpt for each $b_1 \in \mathbb{N}$ and checking if the constraint on the expected detection rate in any equilibrium \eqref{all_NE} is met.



Now consider another approach that is based on our results in Sections~\ref{general} and \ref{sec:Answer}: First, by solving (\hyperlink{(MSC)}{$\mathcal{I}_{\text{MSC}}$}), we obtain an MSC given by  $\Def^{min} = \{i_3,i_4,i_6,i_8\}$; thus $\smsc = 4$. Then, from Thm.~\ref{bounds_game},  we know that the expected detection rate in any equilibrium is lower bounded by $\frac{b_1}{\smsc}$. Thus, with $b_1^\prime:=\lceil \alpha \smsc \rceil = 3$ sensors, the expected detection rate in any equilibrium is at least $\alpha$.

Secondly, by solving (\hyperlink{(MSP)}{$\mathcal{I}_{\text{MSP}}$}), we obtain an MSP given by $\att^{max} = \{e_3,e_4,e_8\}$; thus $\semm = 3$. Then, from Thm.~\ref{bounds_game}, we know that the expected detection rate in any equilibrium is upper bounded by $\frac{b_1}{\semm}$, which implies that if $b_1< \lceil \alpha \semm \rceil$, the equilibrium constraints \eqref{all_NE} are not satisfied. This enables us  to deduce that an optimality gap associated with $b_1^\prime$ is given by $\lceil \alpha \smsc\rceil - \lceil \alpha \semm \rceil$ (Proposition~\ref{Prop_interim}). In fact, for this example, we obtain that the optimality gap is $\lceil 3\rceil - \lceil 2.25 \rceil = 0$. Therefore, by solving (\hyperlink{(MSC)}{$\mathcal{I}_{\text{MSC}}$}) and (\hyperlink{(MSP)}{$\mathcal{I}_{\text{MSP}}$}), we can estimate the number of sensing resources to ensure that the equilibrium constraints in \ECOP are satisfied; furthermore, we can verify that this number is optimal for this example, i.e., $b_1^\prime = b_1^\dag = 3$.

Thirdly, given $b_1^\prime=3$ and $b_2 =2$, we can use $\Def^{min}$ and $\att^{max}$ to construct an approximate NE. Using Lemma~\ref{algebra2}, we construct a strategy profile ${\sigma} = (\sigma^1,{\sigma}^2)$, where ${\sigma}^1_{\{i_3,i_4,i_6\}} = {\sigma}^1_{\{i_8,i_3,i_4\}}  =  {\sigma}^1_{\{i_6, i_8,i_3\}}  = {\sigma}^1_{\{i_4,i_6,i_8\}}  = \frac{1}{4}$, and ${\sigma}^2_{\{e_3,e_4\}} = {\sigma}^2_{\{e_8,e_3\}} = {\sigma}^2_{\{e_4,e_8\}}  =  \frac{1}{3}$; $\sigma$ is illustrated in Fig.~\ref{Strategies}. From Thm.~\ref{MSC-MSP-eNE}, we obtain that the above-constructed strategy profile ${\sigma}$ is an $\epsilon-$NE, and provides each player a payoff that is $\epsilon-$close to their equilibrium payoff, with $\epsilon=  b_1^\prime b_2\left( \frac{1}{\max\{b_1^\prime,\semm\}} - \frac{1}{\smsc}\right) = \frac{1}{2}$. Indeed, from \lpo and \lpt, we can deduce that \defender and \attacker's equilibrium payoffs in the game $\Gamma$ are $\frac{12}{7}$ and $\frac{2}{7}$ respectively, while our strategy profile $\sigma$ provides them with the respective payoffs $\frac{3}{2}$ and $\frac{1}{2}$. One can easily check that $\sigma$ gives each player a payoff that is $\frac{3}{14} \ (\leq \epsilon)$ close to their equilibrium payoff.

\begin{figure}[ht]
\centering
\begin{subfigure}[b]{0.23\textwidth}
\centering
{\scalefig
\begin{tikzpicture}[-,auto,x=\sizefig, y=\sizefig,
  thick,main node/.style={circle,draw,inner sep = \innersepfig,black},main node2/.style={circle,black,draw,fill=blue!60,inner sep = \innersepfig},flow_a/.style ={gray!100}]
\tikzstyle{edge} = [draw,thick,-,black]
\tikzstyle{cut} = [draw,very thick,-]
\tikzstyle{match} = [draw,very thick,black,-, dashed]
\tikzstyle{flow} = [draw,line width = 1.5pt,->,gray!100]

	\node[main node] (1) at (0,2) {$i_1$};
	\node[main node] (2) at (1,2) {$i_2$};
	\node[main node2] (3) at (2,2) {$i_3$};
	\node[main node2] (4) at (1,1) {$i_4$};
	\node[main node] (5) at (2,1) {$i_5$};
	\node[main node2] (6) at (0,0) {$i_6$};
	\node[main node] (7) at (1,0) {$i_7$};
	\node[main node] (8) at (2,0) {$i_8$};
	
	\path[edge]	
	(1) edge node[left]{$e_3$} (6)
	(4) edge node {$e_6$} (5)
	(2) edge  node {$e_2$} (3)
	(2) edge node[left] {$e_4$} (4)
	(6) edge node[below] {$e_9$} (7)
	(7) edge node[below] {$e_{10}$} (8)
	(5) edge node {$e_8$} (8);

	\path[edge]
	(1) edge node {$e_1$} (2)
	(3) edge node {$e_5$} (5)
	(4) edge node[left] {$e_7$} (7);

\end{tikzpicture}}

\caption{${\sigma}^1_{\{i_3,i_4,i_6\}} = \frac{1}{4}$.} 
\end{subfigure}
~
\begin{subfigure}[b]{0.23\textwidth}
\centering
{\scalefig
\begin{tikzpicture}[-,auto,x=\sizefig, y=\sizefig,
  thick,main node/.style={circle,draw,inner sep = \innersepfig,black},main node2/.style={circle,black,draw,fill=blue!60,inner sep = \innersepfig},flow_a/.style ={gray!100}]
\tikzstyle{edge} = [draw,thick,-,black]
\tikzstyle{cut} = [draw,very thick,-]
\tikzstyle{match} = [draw,very thick,black,-, dashed]
\tikzstyle{flow} = [draw,line width = 1.5pt,->,gray!100]

	\node[main node] (1) at (0,2) {$i_1$};
	\node[main node] (2) at (1,2) {$i_2$};
	\node[main node2] (3) at (2,2) {$i_3$};
	\node[main node2] (4) at (1,1) {$i_4$};
	\node[main node] (5) at (2,1) {$i_5$};
	\node[main node] (6) at (0,0) {$i_6$};
	\node[main node] (7) at (1,0) {$i_7$};
	\node[main node2] (8) at (2,0) {$i_8$};
	
	\path[edge]	
	(1) edge node[left]{$e_3$} (6)
	(4) edge node {$e_6$} (5)
	(2) edge  node {$e_2$} (3)
	(2) edge node[left] {$e_4$} (4)
	(6) edge node[below] {$e_9$} (7)
	(7) edge node[below] {$e_{10}$} (8)
	(5) edge node {$e_8$} (8);

	\path[edge]
	(1) edge node {$e_1$} (2)
	(3) edge node {$e_5$} (5)
	(4) edge node[left] {$e_7$} (7);

\end{tikzpicture}}

\caption{${\sigma}^1_{\{i_8,i_3,i_4\}} = \frac{1}{4}$.} 
\end{subfigure}
~
\begin{subfigure}[b]{0.23\textwidth}
\centering
{\scalefig
\begin{tikzpicture}[-,auto,x=\sizefig, y=\sizefig,
  thick,main node/.style={circle,draw,inner sep = \innersepfig,black},main node2/.style={circle,black,draw,fill=blue!60,inner sep = \innersepfig},flow_a/.style ={gray!100}]
\tikzstyle{edge} = [draw,thick,-,black]
\tikzstyle{cut} = [draw,very thick,-]
\tikzstyle{match} = [draw,very thick,black,-, dashed]
\tikzstyle{flow} = [draw,line width = 1.5pt,->,gray!100]

	\node[main node] (1) at (0,2) {$i_1$};
	\node[main node] (2) at (1,2) {$i_2$};
	\node[main node2] (3) at (2,2) {$i_3$};
	\node[main node] (4) at (1,1) {$i_4$};
	\node[main node] (5) at (2,1) {$i_5$};
	\node[main node2] (6) at (0,0) {$i_6$};
	\node[main node] (7) at (1,0) {$i_7$};
	\node[main node2] (8) at (2,0) {$i_8$};
	
	\path[edge]	
	(1) edge node[left]{$e_3$} (6)
	(4) edge node {$e_6$} (5)
	(2) edge  node {$e_2$} (3)
	(2) edge node[left] {$e_4$} (4)
	(6) edge node[below] {$e_9$} (7)
	(7) edge node[below] {$e_{10}$} (8)
	(5) edge node {$e_8$} (8);

	\path[edge]
	(1) edge node {$e_1$} (2)
	(3) edge node {$e_5$} (5)
	(4) edge node[left] {$e_7$} (7);

\end{tikzpicture}}

\caption{${\sigma}^1_{\{i_6,i_8,i_3\}} = \frac{1}{4}$.} 
\end{subfigure}
~
\begin{subfigure}[b]{0.23\textwidth}
\centering
{\scalefig
\begin{tikzpicture}[-,auto,x=\sizefig, y=\sizefig,
  thick,main node/.style={circle,draw,inner sep = \innersepfig,black},main node2/.style={circle,black,draw,fill=blue!60,inner sep = \innersepfig},flow_a/.style ={gray!100}]
\tikzstyle{edge} = [draw,thick,-,black]
\tikzstyle{cut} = [draw,very thick,-]
\tikzstyle{match} = [draw,very thick,black,-, dashed]
\tikzstyle{flow} = [draw,line width = 1.5pt,->,gray!100]

	\node[main node] (1) at (0,2) {$i_1$};
	\node[main node] (2) at (1,2) {$i_2$};
	\node[main node] (3) at (2,2) {$i_3$};
	\node[main node2] (4) at (1,1) {$i_4$};
	\node[main node] (5) at (2,1) {$i_5$};
	\node[main node2] (6) at (0,0) {$i_6$};
	\node[main node] (7) at (1,0) {$i_7$};
	\node[main node2] (8) at (2,0) {$i_8$};
	
	\path[edge]	
	(1) edge node[left]{$e_3$} (6)
	(4) edge node {$e_6$} (5)
	(2) edge  node {$e_2$} (3)
	(2) edge node[left] {$e_4$} (4)
	(6) edge node[below] {$e_9$} (7)
	(7) edge node[below] {$e_{10}$} (8)
	(5) edge node {$e_8$} (8);

	\path[edge]
	(1) edge node {$e_1$} (2)
	(3) edge node {$e_5$} (5)
	(4) edge node[left] {$e_7$} (7);

\end{tikzpicture}}

\caption{${\sigma}^1_{\{i_4,i_6,i_8\}} = \frac{1}{4}$.} 
\end{subfigure}

\medskip

\begin{subfigure}[b]{0.3\textwidth}
\centering
{\scalefig
\begin{tikzpicture}[-,auto,x=\sizefig, y=\sizefig,
  thick,main node/.style={circle,draw,inner sep = \innersepfig,black},main node2/.style={circle,black,draw,fill=Green,inner sep = 0.08cm},flow_a/.style ={gray!100}]
\tikzstyle{edge} = [draw,thick,-,black]
\tikzstyle{cut} = [draw,very thick,-]
\tikzstyle{match} = [draw,very thick,black,-, dashed]
\tikzstyle{flow} = [draw,line width = 1.5pt,->,gray!100]

	\node[main node] (1) at (0,2) {$i_1$};
	\node[main node] (2) at (1,2) {$i_2$};
	\node[main node] (3) at (2,2) {$i_3$};
	\node[main node] (4) at (1,1) {$i_4$};
	\node[main node] (5) at (2,1) {$i_5$};
	\node[main node] (6) at (0,0) {$i_6$};
	\node[main node] (7) at (1,0) {$i_7$};
	\node[main node] (8) at (2,0) {$i_8$};
	
	\path[edge]	
	(1) edge node[left = 0.1cm]{$e_3$} (6)
	(4) edge node {$e_6$} (5)
	(2) edge  node {$e_2$} (3)
	(2) edge node[left = 0.1cm] {$e_4$} (4)
	(6) edge node[below] {$e_9$} (7)
	(7) edge node[below] {$e_{10}$} (8)
	(5) edge node {$e_8$} (8);

	\path[edge]
	(1) edge node {$e_1$} (2)
	(3) edge node {$e_5$} (5)
	(4) edge node[left] {$e_7$} (7);

	\node(12) at (0,1) {\textcolor{Red}{\Huge{\Cross}}};
	\node(13) at (1,1.5) {\textcolor{Red}{\Huge{\Cross}}};

\end{tikzpicture}}

\caption{${\sigma}^2_{\{e_3,e_4\}} = \frac{1}{3}$.}
\end{subfigure}
~
\begin{subfigure}[b]{0.3\textwidth}
\centering
{\scalefig
\begin{tikzpicture}[-,auto,x=\sizefig, y=\sizefig,
  thick,main node/.style={circle,draw,inner sep = \innersepfig,black},main node2/.style={circle,black,draw,fill=Green,inner sep = 0.08cm},flow_a/.style ={gray!100}]
\tikzstyle{edge} = [draw,thick,-,black]
\tikzstyle{cut} = [draw,very thick,-]
\tikzstyle{match} = [draw,very thick,black,-, dashed]
\tikzstyle{flow} = [draw,line width = 1.5pt,->,gray!100]

	\node[main node] (1) at (0,2) {$i_1$};
	\node[main node] (2) at (1,2) {$i_2$};
	\node[main node] (3) at (2,2) {$i_3$};
	\node[main node] (4) at (1,1) {$i_4$};
	\node[main node] (5) at (2,1) {$i_5$};
	\node[main node] (6) at (0,0) {$i_6$};
	\node[main node] (7) at (1,0) {$i_7$};
	\node[main node] (8) at (2,0) {$i_8$};
	
	\path[edge]	
	(1) edge node[left = 0.1cm]{$e_3$} (6)
	(4) edge node {$e_6$} (5)
	(2) edge  node {$e_2$} (3)
	(2) edge node[left] {$e_4$} (4)
	(6) edge node[below] {$e_9$} (7)
	(7) edge node[below] {$e_{10}$} (8)
	(5) edge node[right = 0.1cm] {$e_8$} (8);

	\path[edge]
	(1) edge node {$e_1$} (2)
	(3) edge node {$e_5$} (5)
	(4) edge node[left] {$e_7$} (7);
\node(12) at (0,1) {\textcolor{Red}{\Huge{\Cross}}};
	\node(13) at (2,0.5) {\textcolor{Red}{\Huge{\Cross}}};
\end{tikzpicture}}

\caption{${\sigma}^2_{\{e_8,e_3\}} = \frac{1}{3}$.} 
\end{subfigure}
~
\begin{subfigure}[b]{0.3\textwidth}
\centering
{\scalefig
\begin{tikzpicture}[-,auto,x=\sizefig, y=\sizefig,
  thick,main node/.style={circle,draw,inner sep = \innersepfig,black},main node2/.style={circle,black,draw,fill=Green,inner sep = 0.08cm},flow_a/.style ={gray!100}]
\tikzstyle{edge} = [draw,thick,-,black]
\tikzstyle{cut} = [draw,very thick,-]
\tikzstyle{match} = [draw,very thick,black,-, dashed]
\tikzstyle{flow} = [draw,line width = 1.5pt,->,gray!100]

	\node[main node] (1) at (0,2) {$i_1$};
	\node[main node] (2) at (1,2) {$i_2$};
	\node[main node] (3) at (2,2) {$i_3$};
	\node[main node] (4) at (1,1) {$i_4$};
	\node[main node] (5) at (2,1) {$i_5$};
	\node[main node] (6) at (0,0) {$i_6$};
	\node[main node] (7) at (1,0) {$i_7$};
	\node[main node] (8) at (2,0) {$i_8$};
	
	\path[edge]	
	(1) edge node[left]{$e_3$} (6)
	(4) edge node {$e_6$} (5)
	(2) edge  node {$e_2$} (3)
	(2) edge node[left = 0.1cm] {$e_4$} (4)
	(6) edge node[below] {$e_9$} (7)
	(7) edge node[below] {$e_{10}$} (8)
	(5) edge node[right = 0.1cm] {$e_8$} (8);

	\path[edge]
	(1) edge node {$e_1$} (2)
	(3) edge node {$e_5$} (5)
	(4) edge node[left] {$e_7$} (7);
\node(12) at (2,0.5) {\textcolor{Red}{\Huge{\Cross}}};
	\node(13) at (1,1.5) {\textcolor{Red}{\Huge{\Cross}}};
\end{tikzpicture}}

\caption{${\sigma}^2_{\{e_4,e_8\}} = \frac{1}{3}$.}
\end{subfigure}

\caption{\defender's sensing strategy ${\sigma}^1$ (top) and \attacker's attack strategy ${\sigma}^2$ (bottom) for the example network in Fig.~\ref{Example1}.} \label{Strategies}
\end{figure}

Finally, we give an upper bound on the loss in detection performance by choosing our sensing strategy $\sigma^1$ (see Fig.~\ref{Strategies} (top)), instead of choosing an equilibrium strategy for \defender.
It turns out that, with 3 sensors, $\sigma^1$ detects at least $\frac{3}{4}$ of the failures regardless of \attacker's strategy (Proposition~\ref{robust_r}), while in equilibrium $\frac{6}{7}$ of the failures are detected (Table~\ref{r_sigma_example}). Thus,  the loss in detection performance is $12.5\,\%$. This exact calculation is possible only because, for this  example, we can solve \lpo and \lpt and compute the value of the expected detection rate in equilibrium. In contrast,  Proposition~\ref{robust_r} provides  an upper bound on the loss in detection performance \emph{without} solving \lpo and \lpt, but instead by computing $\smsc$ and $\semm$ from \eqref{IP1} and \eqref{IP2}. This upper bound is given  by $1 - \frac{\max\{b_1^\prime,\semm\}}{\smsc} = 25\,\%$. In the next two sections, we formalize these results.

\else
\fi

\ifadditional
\section{Example}

Fig. \ref{fig:net1_msc_emm} shows the location of the $\Def^{min}$ nodes and the $\att^{max}$ pipelines for the break (left) and the contamination (right) scenarios.

\begin{table}[htbp] \footnotesize
  \centering
  \caption{NE solution for the Apulian network with $b_1 = 2, b_2 = 2$.}
    \begin{tabular}{|ccc|cc|ccc|cc|}
\hline
\multicolumn{5}{|c|}{\textbf{Operator strategy}} &\multicolumn{5}{c|}{\textbf{Attacker strategy}}\\
	\hline
    \multicolumn{2}{|c}{\textbf{Support}} & \textbf{Probability}  & \textbf{Node basis} & \textbf{Probability} & \multicolumn{2}{c}{\textbf{Support}} & \textbf{Probability}  & \textbf{Edge basis} & \textbf{Probability} \\
 \multicolumn{2}{|c}{\boldmath{$\supp(\sigma^1)$}} & \boldmath{$\sigma^1_\Def$}  & \boldmath{$s_{\sigma^1}$} & \boldmath{$\nprob$} &  \multicolumn{2}{c}{\boldmath{$\supp(\sigma^2)$}} & \boldmath{$\sigma^2_\Def$}  & \boldmath{$s_{\sigma^2}$} & \boldmath{$\eprob$}  \\
\hline
    1     & 8         & 0.286   & 1     & 0.286 &   5     	& 34    & 0.143  		&5     	& 0.143\\
    6     & 21         & 0.036  & 6     & 0.071 & 10     & 31   & 0.143 		& 10     & 0.143\\
    6     & 23         & 0.036  & 7     & 0.214 & 13     & 15   & 0.286  		& 13    & 0.286\\
    7     &9        & 0.071  & 8    & 0.286 & 28     & 30 	 & 0.286  	& 15     & 0.286\\
    7     & 11         & 0.071   & 9    & 0.214 & 29    	& 32       & 0.143  	& 28  	& 0.286 \\
    7     & 17        & 0.071   & 10    & 0.107 & &       	&   			& 29    & 0.143 \\
    9     & 23        & 0.143   & 11    & 0.071 & &         	&   			& 30    & 0.286\\
    10     & 17        & 0.107   &  12     &  0.107 & &          &       		& 31    & 0.143 \\
    12     & 17      & 0.107  &  13     &  0.071 & &          &       		& 32    & 0.143\\
    13     & 21        & 0.071  &  17     &  0.286 &  &           &       		& 34    & 0.143 \\
          &       &       &        21     &  0.107 &&&&&\\
          &       &       &       23     &  0.179 &&&&& \\ 
\hline
    \end{tabular}\\
  \label{tab:1}%
\end{table}%

\begin{table}[htbp] \footnotesize
  \centering
  \caption{Equilibrium properties for the Apulian network with $b_1 = 2, b_2 = 2$.}
    \begin{tabular}{|ccc|cc|ccc|cc|}
\hline
\multicolumn{5}{|c|}{\textbf{Operator strategy}} &\multicolumn{5}{c|}{\textbf{Attacker strategy}}\\
	\hline
    \multicolumn{2}{|c}{\textbf{Support}} & \textbf{Probability}  & \textbf{Node basis} & \textbf{Probability} & \multicolumn{2}{c}{\textbf{Support}} & \textbf{Probability}  & \textbf{Link basis} & \textbf{Probability} \\
 \multicolumn{2}{|c}{\boldmath{$\supp(\sigma^1)$}} & \boldmath{$\sigma^1_\Def$}  & \boldmath{$s_{\sigma^1}$} & \boldmath{$\nprob$} &  \multicolumn{2}{c}{\boldmath{$\supp(\sigma^2)$}} & \boldmath{$\sigma^2_\Def$}  & \boldmath{$s_{\sigma^2}$} & \boldmath{$\eprob$}  \\
\hline
     1     & 8         & 0.5   & 1     & 0.5 &   13     	& 30    & 0.333  		&13     	& 0.667\\
    10     & 23         & 0.5  & 8     & 0.5 & 13     & 34   & 0.333 		& 30     & 0.667 \\
    &          &   & 10     & 0.5 & 30     & 34   & 0.333  		& 34   & 0.667\\
    &      &  & 23    & 0.5 &&&&&\\
\hline
    \end{tabular}\\
  \label{tab:2}%
\end{table}%

\begin{figure}
        \centering
        \begin{subfigure}[b]{0.45\textwidth} \centering
              \includegraphics[trim = 60mm 30mm 50mm 30mm, clip,scale = 0.4]{or_fig35burst.pdf}
       \caption{Pipeline break}  \label{fig:net1_msc_emm_burst}
        \end{subfigure}
        ~ 
        \begin{subfigure}[b]{0.45\textwidth} \centering
               \includegraphics[trim = 60mm 30mm 50mm 30mm, clip,scale = 0.4]{or_fig35contamination.pdf}       
		\caption{Contaminant intrusion}   \label{fig:net1_msc_emm_cont}  
        \end{subfigure}
       \caption{$\Def^{min}$ and $\att^{max}$ for the Apulian water network. The labeled links indicate the links in a $\att^{max}$ and the labeled nodes indicate the nodes in a $\Def^{min}$.}   \label{fig:net1_msc_emm}

\end{figure}

\tr{Hence, to solve \ECOP  we compute a NE of the game $\Gamma$ for each $b_1 \in \mathbb{N}$, and compute the expected detection rate $\sdet{*}$. One NE solution of $\Gamma$ given $b_1 = 2, b_2 = 2$ for the pipeline break scenario is listed in Table \ref{tab:1}. 
Table \ref{tab:1} lists the supports, the bases, and their corresponding probability for the operator (columns one and two) and for the attacker (columns three and four). Result show that the node basis spans 12 nodes, which implies that the operator will have to shift between 12 different locations in the network.  }

We use the solutions to (\hyperlink{(MSC)}{$\mathcal{I}_{\text{MSC}}$}) and (\hyperlink{(MSP)}{$\mathcal{I}_{\text{MSP}}$}) to derive an approximate solution of \ECOP according to Proposition~\ref{Prop_interim}. In the pipeline break scenario, we have $\smsc  = 4 > \semm = 3$. Restricting support of the operator to $\Def^{min}$ and the support of the attacker to $\att^{max}$, the strategy profile is the solution to the two relaxed linear programs, $LP_{S'}$ and $LP_{T'}$ (Propositions \ref{best_set_cover} and \ref{best_maximum_matching}). The solution of the two linear programs is listed in Table \ref{tab:2}, showing the node and component bases and the corresponding sensing probability. For example, the operator should equally shift between sensing nodes $\{1, 8\}$ and nodes $\{10, 23\}$. The attacker should equally shift between pipelines $\{13,20\}$, $\{13,34\}$, and $\{30,34\}$. Under this strategy, all the nodes in the MSC are monitored with equal probability, $\rho_{\sigma^{\Def}}(i) = \frac{b_1}{n}=\frac{2}{4}=0.5$, as well as all the pipes in MSP are disrupted with equal probability, $\rho_{\sigma^{\att}}(e) = \frac{b_2}{m}=\frac{2}{3}= 0.667$ (see Lemma \ref{algebra1}). 
Using Theorem \ref{bounds_game}, we evaluate the upper and lower bounds of the expected detection rate as a function of $b_1$ (see Fig. ~\ref{fig:net1_15}) of our approximate NE. Our estimate of the number of sensors required to achieve a targeted detection rate is close to the \textit{optimal} value.

\else
\fi



\end{document}